\documentclass[fisica]{tesis}
\usepackage{tesis}
\usepackage[utf8]{inputenc}
\usepackage[spanish]{babel}
\usepackage{verbatim}
\usepackage{subfig}

\title{Dinámica y termodinámica del modelo d-HMF de campo medio fuera del equilibrio.}
\author{Mg. Boris Atenas Núñez}
\profguia{Sergio Curilef Huichalaf}
\date{2021}

\begin{document}
\maketitle
%

%
\begin{agradecimientos}

Difícil es encontrar las palabras justas para agradecer a quienes me han apoyado a seguir mis sueños y moldear quien soy hoy en día. Cómo dijo Albert Einstein: "\textit{vivimos en un mundo hecho por otros}"; comemos comida hecha por otros, vestimos ropa fabricada por otros, nos transportamos en vehículos hechos por otros, dormimos en camas construidas por otros y pocas veces nos detenemos para agradecer y pensar que todo a nuestro alrededor es posible gracias a otros. Nuestra compleja sociedad y todas las comodidades que tenemos hoy en día, no serían posibles si no fuera por el arduo trabajo y dedicación de quienes han vivido antes que nosotros. Por ello pienso que insuficientes serán mis palabras para agradecer a todas y a cada una de las personas con las que he compartido a lo largo de mi vida, sin embargo, haré mi mejor esfuerzo.

Quisiera comenzar agradeciendo en especial al profesor Sergio Curilef, mi profesor guía, por creer en mí y apoyarme en mi formación. Desde el comienzo creyó en mis capacidades y buscó la manera de apoyarme cada vez que necesité ayuda, por su confianza para contarle mis problemas personales y por guiarme y aconsejarme en el ámbito profesional y personal. Agradezco también al profesor Luis Del Pino que en paz descanse, quién fue también mi profesor guía y me apoyó en mis primeras publicaciones. Agradezco también al profesor Luisberis, quien  me ha apoyado de manera desinteresada en mi formación como científico durante mi doctorado, le agradezco en particular por su gran compromiso y apoyo incondicional cada vez que necesité su ayuda, y por la confianza que me ha entregado en estos últimos años.

Agradezco además a nuestra casa de estudios y al departamento de Física por las facilidades e infraestructura que me permitieron llevar a cabo mis investigaciones de manera óptima. Pero no sólo por ello, pues en esta universidad me formé profesionalmente. Desde que era estudiante de pregrado soñaba con hacer clases en alguna universidad, y en la Universidad Católica del Norte tuve la oportunidad de cumplir uno de mis sueños; enseñar y formar estudiantes, mientras continuaba perfeccionándome como científico con mi magíster y doctorado. 

Como no agradecer a todos y cada uno de los profesores que me han hecho clases en la vida, a los profesores del departamento de física y matemática que me han formado, con los que he compartido pues cada uno de ellos ha sido una interacción que me ha transformado y me ha hecho ser quien soy hoy en día.

Se agradece también a los fondos UCN, a Becas ANID por la Beca de Doctorado nacional 2020 Conicyt Folio: 21202551. Al apoyo computacional de Alessandro Pluchino. Al grupo Powered@NLHPC: que apoyó esta investigación parcialmente por la infraestructura de supercómputo del NLHPC (ECM-02). Al sistema de cómputo HPC de ciencias de la Universidad Católica del Norte y el financiamiento parcial del proyecto FONDECYT-Chile 1170834.

Quiero también agradecer a mis amigos por darse el tiempo para estar conmigo cada vez que los necesito. A mis padres y hermanas por preocuparse siempre de mí, apoyarme, educarme, entregarme cariño y dejarme soñar. Y por último, agradecer al motor de mi vida, mi esposa Verónica, por ser mi fiel compañera todos estos años, por su paciencia, por motivarme y darme ánimo para seguir adelante en los momentos difíciles, por hacer todavía más felices mis momentos de gloria y felicidad, por hacerme reír con sus bromas y por enseñarme cada  día a ser una mejor persona, a superarme, a conocer mejor a los demás y mí mismo, por enseñarme a disfrutar de cada momento con mis amigos, familia y mi trabajo, por recordarme siempre que la vida es una sola, que hay que disfrutarla y aprovecharla para dejar nuestra huella en la sociedad para que \textit{otros} tengan la posibilidad de seguir creciendo con nuestros aprendizajes. 
\end{agradecimientos}

\begin{resumen}

En la presente investigación, se estudia el modelo \textbf{d-HMF} propuesto por Curilef y  Atenas \cite{ATENAS3}, un modelo de campo medio con interacciones de largo alcance inspirado en la interacción dipolo-dipolo, cuyo nombre se debe a su similitud con el modelo HMF (\emph{Hamiltonian Mean Field model}). Dentro de los desafíos de esta tesis se destaca la resolución del modelo d-HMF en los ensambles canónico y microcanónico y la descripción analítica y numérica de la función de distribución del sistema. Este modelo ha sido estudiado tanto en el equilibrio como fuera de este. En el equilibrio se han encontrado soluciones analíticas para la energía interna por partícula y la temperatura mediante el empleo de los procedimientos estándar de la mecánica estadística como lo son el cálculo de la función partición en el ensamble canónico y el cálculo del número de microestados accesibles en el ensamble microcanónico. Los resultados indican que existe una equivalencia de ensambles entre el canónico y microcanónico.
También, se calcula la función de distribución de equilibrio de Boltzmann-Gibbs (BG) de este sistema, la cual coincide con los resultados analíticos de los ensambles canónico y microcanónico, por lo que el estado de equilibrio del modelo ha sido descrito completamente. 
A pesar de que el estado de equilibrio presenta varias anomalías discutidas en la literatura reciente, nuestro propósito se centra en describir los estados de estacionarios de evolución de cortos tiempos comparados con los tiempos que involucra el equilibrio.
En cuanto a la dinámica del sistema fuera del equilibrio, se estudian los estados cuasi-estacionarios \textit{Quasi-Stationary-States} (QSS) presentes en este sistema. Mediante simulaciones de dinámica molecular se encuentran dos tipos de estados QSS fuera del equilibrio. Para su descripción, se utiliza una combinación de dos técnicas; los métodos de la dinámica molecular y soluciones estacionarias de la ecuación de Vlasov asociadas a las ecuaciones de movimiento del sistema. A partir de las simulaciones de dinámica molecular, se obtienen datos relevantes de la dinámica como la energía interna, cinética y potencial por partícula, la magnetización, las distribuciones marginales en los momentos y en las orientaciones. Además, se obtiene una ley de potencia para el tiempo de duración del segundo estado QSS. Desde el punto de vista analítico se encuentran soluciones estacionarias de la ecuación de Vlasov para describir los estados QSS observados con la dinámica molecular. Las soluciones estacionarias ensayadas en esta tesis, corresponden a funciones de distribución del tipo q-exponencial. En particular, se encuentra que una de estas soluciones describe de manera muy precisa las distribuciones marginales de los momentos y las orientaciones de uno de los estados QSS hallados con la dinámica molecular. Asimismo, se establece una transformación matemática que vincula los parámetros de la función q-exponencial utilizada como solución de la ecuación de Vlasov con los parámetros de la estadística de Tsallis. Por último, se encuentran las distribuciones marginales analíticas generales en los momentos y las orientaciones, siendo la distribución marginal en las orientaciones una expresión integral sin primitiva, mientras que la distribución marginal en los momentos tiene una expresión matemática definida.

\bigskip
\end{resumen}

\begin{agradecimientos2}

En el capítulo \ref{Intro}, se describe el marco teórico que sustenta esta tesis y se da una visión general de la metodología utilizada en mecánica estadística y en particular en esta tesis. En el capítulo \ref{modeloHMF}, mostramos los resultados más interesantes del modelo HMF, tanto en el equilibrio (soluciones analíticas) como fuera del equilibrio (presencia de estados QSS), que vienen de la ecuación de Vlasov, y las estadísticas de Tsallis y Lynden-Bell.
En el capítulo \ref{dhmfmodel}, definimos el modelo d-HMF, derivamos las ecuaciones de movimiento y mostramos los resultados de simulaciones de dinámica molecular. Asimismo, se describe la dinámica de este sistema fuera del equilibrio, en donde se observa la presencia de los estados QSS.
En el capítulo \ref{cap2}, se muestran en detalle los cálculos analíticos en los ensambles canónico y microcanónico de las soluciones de equilibrio del sistema, cuya obtención se basa en los procedimientos estándar de la mecánica estadística: el cálculo de la función de partición canónica y el conteo de microestados accesibles microcanónico.
En el capítulo \ref{cap3}, se obtiene la función de distribución de equilibrio de BG y de la ecuación cinética de Vlasov se obtienen las expresiones analíticas de las funciones de distribución que describen los estados cuasi-estacionarios presentes en el modelo. Por último, se obtiene una transformación que permite conectar la estadística de Tsallis con las soluciones de la ecuación de Vlasov.
Finalmente, en el capítulo \ref{conclu}, se comentan los principales resultados de este trabajo resaltando las principales conclusiones y posibles trabajos futuros a desarrollar a raíz de esta tesis.
\end{agradecimientos2}

\tableofcontents
\listoffigures

\chapter{Introducción}\label{Intro}
\section{Comentarios generales y motivación}

En la física teórica se proponen modelos matemáticos con el fin de explicar la realidad. Algunos son tan simplificados que sólo pretenden capturar aspectos significativos de alguna realidad física; sobre todo, cuando se trata de analizar procesos y situaciones complejas. Modelos que no pretenden capturar la realidad con algún grado de aproximación, pueden ser llamados modelos matemáticos; pero no de la física o alguna ciencia práctica.

Los sistemas físicos reales constituidos por un gran número de partículas suelen albergar un número considerable de interacciones que lo forman, las cuales se pueden clasificar según la distancia en que actúan, en interacciones de corto y de largo alcance.

Las interacciones de corto alcance suelen aparecer con mayor abundancia en sistemas nanoscópicos mientras que las de largo alcance actúan principalmente en sistemas a escala astrofísica, sin embargo, a escala mesoscópica estas interacciones suelen competir fuertemente para gobernar la dinámica y termodinámica del sistema dando origen a las propiedades macroscópicas de los materiales.

En esta tesis estamos interesados en estudiar las propiedades termoestadísticas de un tipo de sistema cuyas interacciones son de largo alcance, el así llamado modelo d-HMF propuesto por Curilef y Atenas \cite{ATENAS3}. 

El modelo d-HMF es el primer modelo que considera el potencial dipolar eléctrico para describir una dinámica de muchos dipolos. Si bien es un modelo simple, este presenta una transición de fase continua del tipo para-ferromagnética. Su estudio aún es preliminar, ya que hasta el momento solo se ha resuelto la versión unidimensional en el marco de la aproximación de campo medio.

Debido a que el modelo d-HMF se inspira directamente en sistemas reales, y presenta un comportamiento macroscópico no trivial, podría ser utilizado en diversas aplicaciones, ya que tanto en el área de la biología, química como de la tecnología, existen sistemas formados por dipolos eléctricos (enlaces iónicos) tales como NaCl, HCl, LiF, CaNO3$^+$, MgO, SiO2, NO3$^-$ etc, los cuales, además son abundantes en la naturaleza y esenciales tanto para el metabolismo de los seres vivos, así como en la elaboración de materiales de uso industrial.
En los últimos años la caracterización de rotores y motores moleculares ha causado gran interés en la comunidad científica, pues constituye un aporte sustancial en la comprensión del funcionamiento de los organismos microscópicos que permiten la vida. Esta es una de las motivaciones del presente estudio. Recientemente se han realizado varios trabajos que concluyeron en la creación de los primeros \emph{motores moleculares artificiales} \cite{KOMURA, KOTTAS, ZHANG, ERBAS}, en los que mediante campos eléctricos y magnéticos se ha logrado generar movimientos de traslación y rotación de moléculas. Con la tecnología actual por ejemplo, se pueden controlar bancos de células (bacterias) mediante el uso de estas técnicas \cite{MARTEL}.

Asimismo, varios estudios teóricos clásicos han sido desarrollados para modelar el comportamiento de cargas eléctricas sometidas a campos eléctricos y/o magnéticos \cite{CLARO, TRONCOSO, ESCOBAR, ATENAS1, ATENAS2, DELPINO}, los cuales nos han mostrado la diversidad de movimientos posibles para distintas condiciones iniciales. En el ámbito de la tecnología, estos estudios se han utilizado para crear un aparato llamado \textbf{cortina eléctrica} (\textit{Electric curtain device}) \cite{PURSEY, CLARO, TRONCOSO, ESCOBAR, ATENAS1, ATENAS2, DELPINO}, que sirve para aislar pequeñas partículas en su interior mediante el uso de campos eléctricos. De ahí la importancia de estudiar sistemas formados por dipolos eléctricos en presencia de campos eléctricos y/o magnéticos. Por lo tanto, los resultados de esta investigación pueden ser considerados en las aplicaciones antes mencionadas como es el caso de los motores moleculares y los aparatos de cortina eléctrica, los cuales están relacionados con el estudio de estos sistemas formados por dipolos eléctricos.

\section{Soluciones analíticas y métodos numéricos en física estadística}

La mecánica estadística junto con la teoría cinética, constituyen herramientas muy poderosas para estudiar la dinámica y la termodinámica de sistemas formados por muchas partículas dentro y fuera del equilibrio a partir de una configuración inicial dada. A lo largo de la historia se han desarrollado diversos modelos  teóricos que intentan describir situaciones físicas reales, tales como Van der Waals, Curie-Weiss, Einstein, Brag-Williams, Bethe-Peierls, Ising, Potts, Chandrasekhar, Hydrodynamics, Self-gravitating, HMF \cite{VAN, WEISS, EINSTEIN, ISING, POTTS, CHANDRASEKHAR, FINE, KONISHI, ANTONI, TATEKAWA} e incluso con aplicaciones sociales \cite{STAUFFER}.  Estos modelos han sido popularizados, resueltos y discutidos en libros tales como \cite{KAC2, REICHL, HUANG2, LEBELLAC, STANLEY, PATHRIA, SOTO, LUO, CAMPA}. Sin embargo, muchos de ellos aún no tienen solución analítica conocida \footnote{Solución analítica se refiere al cálculo exacto de la función de partición y la energía libre de Helmholtz (si se utiliza el ensamble canónico), el número de microestados accesibles y la entropía en el caso microcanónico, etc.}, como es el caso del modelo de Ising en tres dimensiones. Esto debido a la dificultad a la hora de resolver las integrales multidimensionales. Por esto, es que se hace necesario resolver el sistema numéricamente. En la literatura podemos encontrar algunos mecanismos de carácter determinista y otros estocásticos los cuales se detallan a continuación:

\begin{enumerate}
\item \textbf{Dinámica molecular (DM):} existen dos versiones clásica y cuántica. La versión cuántica es conocida como ab initio, y se basa en la resolución de la ecuación de Shrödinger de un sistema de partículas. La versión clásica (utilizada en esta tesis) es determinista y está basada en la integración directa de las ecuaciones de movimiento clásicas, es decir, las ecuaciones de Hamilton, o bien las de Langrange o Newton, según el esquema a utilizar. De esta simulación se obtiene la posición y velocidad de cada partícula que conforma el sistema en cada instante a partir de unas condiciones iniciales. Con esta información es posible obtener los promedios estadísticos correspondientes a cantidades termodinámicas de interés como la magnetización, el calor específico, energía cinética promedio, entre otras.   

\item \textbf{Ecuaciones cinéticas:} otra forma de estudiar los sistemas de partículas es mediante ecuaciones cinéticas \footnote{Si bien estas ecuaciones han sido planteadas aquí como un método numérico, estas pueden tener soluciones analíticas conocidas, pero no necesariamente serán equivalentes a la solución analítica obtenida por la mecánica estadística, ya que son aproximaciones.}. Estas ecuaciones son aproximaciones útiles, cuyas soluciones permiten obtener una función de distribución de probabilidad del sistema con la que se calculan los promedios estadísticos. Ejemplos de ellas tenemos: las ecuaciones de Boltzmann, Lorentz, Vlasov, Poisson-Boltzmann, etc. Aunque de manera más general, la única dinámica exacta para funciones de distribución es la ecuación de Liouville basada en la función de distribución de todas las partículas que forman el sistema, esto es $F(\vec{r}_1,\vec{p}_1,...,\vec{r}_N,\vec{p}_N,t)$. Esta función contiene toda la información del sistema y nos dice como están distribuidas las partículas en el espacio de fases. A partir de ella se pueden calcular todas las cantidades termodinámicas tales como la energía cinética promedio, energía interna, entropía, etc. Sin embargo, a pesar de que la ecuación de Liouville describe de manera exacta la dinámica, ésta no parece aclarar cómo los sistemas con interacciones no lineales evolucionan hacia el equilibrio \cite{Gallavotti1999}. Por otro lado, conocer la función de distribución de todas las partículas $F(\vec{r}_1,\vec{p}_1,...,\vec{r}_N,\vec{p}_N,t)$ es computacionalmente inviable cuando el número de partículas es grande, por eso se suelen utilizar funciones de distribución reducidas que involucran menos partículas. Estas funciones reducidas omiten parte de la información del sistema a cambio de expresiones más manejables. Así el caso más sencillo consiste en la función reducida de una sola partícula esto es $f(\vec{r},\vec{p}, t)$.

Como la función de distribución de una partícula es una función de tres variables, la evolución temporal de ésta nos da como resultado la famosa ecuación de Boltzmann,
    \begin{equation}
    \frac{df}{dt}=\frac{\partial f}{\partial \vec{r}}  \frac{d\vec{r}}{dt}+\frac{\partial f}{\partial\vec{p}}\frac{d\vec{p}}{dt} +\frac{\partial f}{\partial t}.
    \end{equation}
La expresión da cuenta que el cambio de la función de distribución $f$ en el tiempo equivale al ritmo con el que entran y salen partículas en una parte infinitamente pequeña del espacio de fases $d^3\vec{r}d^3\vec{p}$. Este término es llamado operador de colisión $I(f)=\frac{df}{dt}$, ya que los eventos de colisión entre las partículas determinarán este flujo.

En esta tesis, utilizaremos soluciones de la ecuación de Vlasov, la cual es útil para describir estados fuera del equilibrio. Esta ecuación no es más que la ecuación de Boltzmann sin efectos correlativos, esto quiere decir que el operador de colisión es idénticamente cero. En los siguientes capítulos se muestra una deducción formal de esta ecuación a partir de la Jerarquía BBGKY (Bogoliubov–Born–Green–Kirkwood–Yvon) \cite{SOTO} de funciones reducidas.

Por otro lado, cuando no se conocen las interacciones que gobiernan el sistema, existe otro tipo de ecuaciones llamadas ecuaciones maestras, que incluyen términos probabilísticos similares al operador de colisión, ejemplos de ellas son las ecuaciones de Langevin, Difusión, Fokker-Planck, Chapman-Kolmogorov, etc. \cite{SOTO}; todas intentan obtener información del sistema por medio de fuerzas de carácter aleatorio, lo que permite entre otras cosas, describir por ejemplo el movimiento Browniano. 

\item \textbf{Métodos Monte Carlo (MC):} en física estadística, estos métodos utilizan probabilidades y ecuaciones maestras para simular un sistema de partículas. En general, estos simulan la situación de equilibrio en el ensamble canónico, aunque más recientemente estos métodos han sido ampliados para simular otros ensambles estadísticos y estados fuera del equilibrio. En estas simulaciones, la dinámica es ficticia, pues mediante números aleatorios se generan nuevas configuraciones del sistema hasta llegar a una configuración de equilibrio. Los cálculos estadísticos del tipo función de partición y los promedios de  observables macroscópicos convergen más rápido que realizando la integración numérica directa (por cuadraturas como los métodos de trapecio o Simpson), por lo que el costo computacional de estos métodos es mucho menor que el de la integración directa para el cálculo de la función partición, lo que acelera los cálculos a la hora de obtener diagramas de fase donde se observa el comportamiento general de un sistema como lo son los cambios de fase. En esta tesis se discute un modelo específico desde el punto de vista determinista, eliminando cualquier aproximación implícita en la aleatoriedad.

\end{enumerate}
La utilización de alguna de estas herramientas depende en concreto de lo que se desee estudiar. Si bien es posible utilizar todas, cada una de ellas entregará, más o menos, la misma información del equilibrio, pero si queremos estudiar lo que sucede fuera del equilibrio, lo mejor sería utilizar la dinámica molecular y alguna ecuación cinética apropiada.

Lo interesante del estudio fuera del equilibrio es la presencia de estados \textbf{cuasi-estacionarios}, que corresponden a estados en los que el sistema queda atrapado por un período de tiempo considerable y que crece conforme aumenta el tamaño del sistema. Estos estados pueden ser estados de equilibrio estables o inestables y suelen aparecer en regiones cercanas a la zona de transición de fase del sistema. La conexión de estos estados con la termodinámica es que están representados por valores anómalos en las funciones respuesta, como por ejemplo capacidades caloríficas negativas \footnote{Capacidad calorífica negativa quiere decir que conforme aumenta la energía interna se produce una disminución en la energía cinética promedio.}. Hasta el momento los estados cuasi-estacionarios han sido observados tan sólo en el ensamble microcanónico y bajo ciertas condiciones iniciales, mediante dinámica molecular y la ecuación cinética de Vlasov. Asimismo, han sido observados experimentalmente en plasmas \cite{HUANG}.

Es importante destacar que en la literatura también se mencionan los así llamados estados \textbf{metaestables}, que aunque en la literatura muchas veces suelen ser tratados como sinónimos de los estados cuasi-estacionarios, no son lo mismo. Se habla de estados metaestables cuando un sistema posee más de un estado de equilibrio estable, y dentro de estos estados se les llama metaestables a los de mayor energía, porque ante pequeñas perturbaciones el sistema transita hacia el estado de equilibrio con menor energía. En estos estados también se obtienen valores anómalos de las funciones respuesta que aparecen durante las transiciones de fase discontinuas, donde las distribuciones de los observables macroscópicos fundamentales son multimodales.

Si bien la utilización de alguna de estas tres herramientas depende fuertemente del sistema a estudiar, se debe tener en cuenta qué es lo que se desea estudiar, si sólo queremos describir el sistema en el equilibrio, una dinámica de Monte Carlo es más que suficiente, pero si se desea estudiar al sistema fuera del equilibrio, deben aplicarse los métodos de la dinámica molecular y/o los tratamientos ofrecidos por las ecuaciones cinéticas o métodos especiales para Monte Carlo fuera del equilibrio.

Otro aspecto importante a la hora de elegir alguna de estas herramientas, es la naturaleza del propio sistema. En la literatura existen diversos algoritmos para tratar los sistemas según su naturaleza. Por ejemplo, los algoritmos de clústeres utilizados en los métodos Monte Carlo suelen ser mejores en sistemas cuyas interacciones conllevan fuertes correlaciones. En dinámica molecular, si bien existen algoritmos generales para la integración de los sistemas de ecuaciones diferenciales como Runge-Kutta, la naturaleza de un sistema de ser hamiltoniano (conservativo) permite utilizar métodos de integración más adecuados basados en algoritmos simplécticos, como Verlet o Leapfrog \cite{YOSHIDA}.

Debido a su naturaleza, los sistemas formados por dipolos eléctricos y en particular el modelo d-HMF se enmarcan dentro de los sistemas con interacciones de largo alcance los cuales serán descritos en la siguiente sección.

\section{Interacciones de largo alcance}\label{LRI}

Los sistemas se pueden clasificar según su interacción en dos grupos, los sistemas con interacciones de corto alcance, y los sistemas con interacciones de largo alcance \cite{CAMPA} (aunque, también existen sistemas cuyas interacciones tienen un comportamiento marginal entre estos dos grupos, como los propios sistemas 3D con interacciones dipolares).


Para saber a qué grupo pertenece una interacción, consideremos un potencial de interacción entre partículas, de la forma,
\begin{equation}\label{potinter}
  V=\frac{A}{r^{\alpha}},
\end{equation}
donde $r$ es el módulo de la distancia entre partículas y $A$ es un factor de acoplamiento que se considera constante para $r\gg1$ y depende de $r$ cuando $r\ll1$. Si $\alpha\leq d$ la interacción es de largo alcance, donde $d$ es la dimensión del espacio en la que se encuentra el sistema. Por el contrario, si $\alpha > d$, la interacción es de corto alcance.
\begin{figure}
    \centering
    \includegraphics[width=7cm]{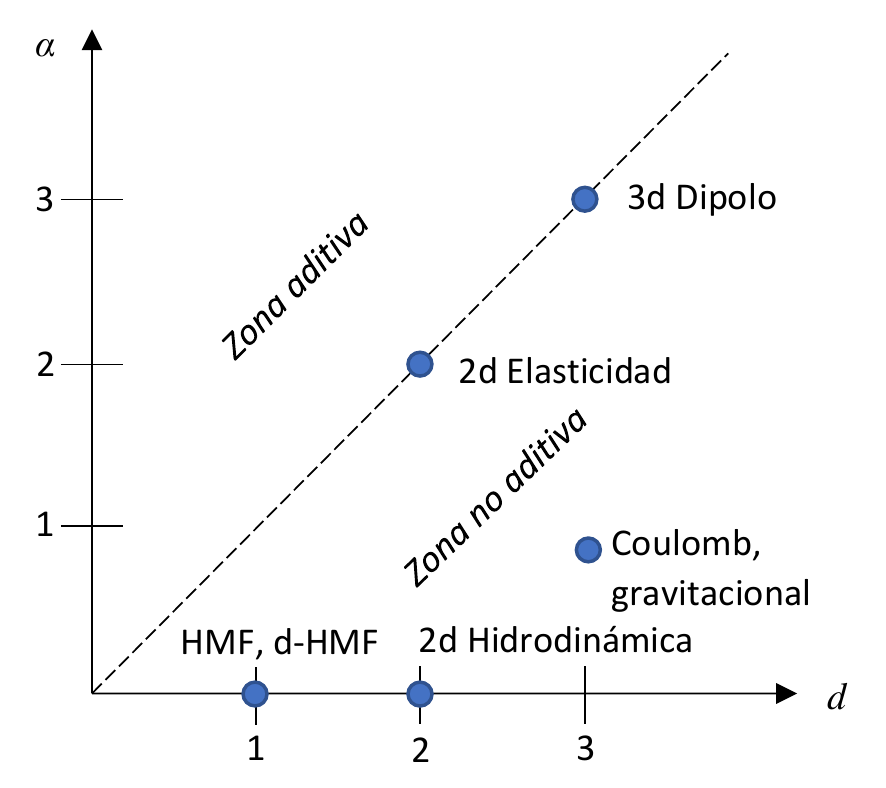}
    \caption{La figura muestra la separación de diferentes sistemas en las zonas aditivas y no aditivas de la energía. El modelo d-HMF está en la zona no aditiva.}
\end{figure}
Esto puede evidenciarse al tomar la energía potencial de una partícula $\varepsilon$ interactuando con el potencial de la ec.( \ref{potinter}), situada en el centro de una distribución homogénea de partículas en una esfera de radio $R$ de dimensión $d$, con $\alpha \neq d$,
\begin{equation}\label{long}
\varepsilon=\int_{\delta}^R \rho \frac{A}{r^{\alpha}} d^dr= A \rho\Omega_d \int_{\delta}^R  r^{d-1-\alpha} dr = \frac{ \rho A \Omega_d}{d-\alpha} \left(R^{d-\alpha}-\delta^{d-\alpha}\right),
\end{equation}
\begin{figure}[b]
    \centering
    \includegraphics[width=4cm]{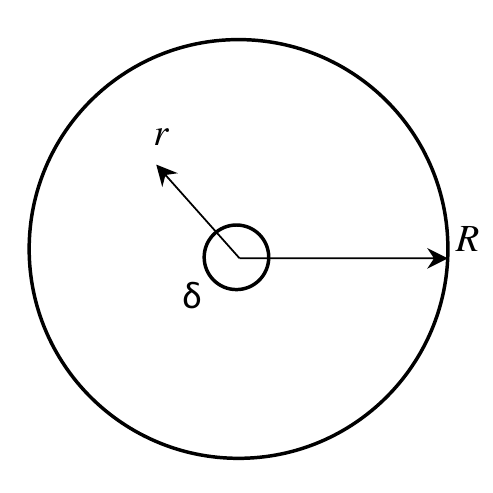}
    \caption{Esquema para la evaluación de la energía $\varepsilon$.}
\end{figure}
donde $\delta$ es un pequeño radio introducido para evitar la divergencia del potencial con $\delta \ll R$, el cual no está relacionado ni con la naturaleza de las interacciones de largo alcance, ni con la constante de integración. $\Omega_d$ es el volumen angular en dimensión $d$, y $\rho$ es la densidad de carga, masa etc., que hemos considerado constante.
Para el caso marginal $\alpha=d$, se tiene una divergencia logarítmica de la forma,
\begin{equation}
\varepsilon\backsim \log(R)-\log(\delta)
\end{equation}
Si $\alpha > d$ de la ecuación \ref{long} vemos que la energía permanece finita y proporcional al volumen para $A$ constante, algo natural en los sistemas con interacciones de corto alcance, donde el número de interacciones es proporcional a $N$. Por el contrario, si $\alpha \leq d$ con $A$ constante, la energía diverge como una ley de potencias de $R$, ya que se vuelve superlineal conforme crece el tamaño del sistema, esto es, $\varepsilon\propto {\cal V}^{2-\alpha/d}$, donde $\cal{V}$ es el volumen del sistema, esto produce la pérdida de extensividad \footnote{Extensividad se refiere a la escalabilidad de las magnitudes macroscópicas en relación con el tamaño del sistema, esto incluye magnitudes no observables como la entropía. Por ejemplo, cuando la energía escala proporcional al tamaño del sistema, se dice que la energía del sistema es extensiva, si ésta escala de manera superlineal (como en este caso), entonces la energía del sistema no es extensiva.}.
Otra característica de los sistemas con interacciones de largo alcance, es que en ellos se presenta la pérdida de aditividad ya sea en la energía o en la entropía. Supongamos que dividimos un sistema en dos o más partes y que la suma de las energías de cada subsistema no es igual la energía total del sistema,  entonces se dice que la energía no es aditiva, lo mismo puede ocurrir con la entropía del sistema, en tal caso si la entropía del sistema no es igual a la suma de las entropías de los subsistemas la entropía no es aditiva \footnote{El surgimiento de esta pérdida en la aditividad de la energía o de la entropía se debe al efecto de las correlaciones, donde la longitud de correlación es comparable con el tamaño del sistema.}. \cite{DAUXOIS,LEVIN,DELPINO2}.

Sin embargo, para los modelos de campo medio (donde $\alpha=0$), es posible introducir un factor, de manera de recuperar la extensividad como veremos más adelante.

La pérdida de aditividad ya sea en la energía o en la entropía, está frecuentemente acompañada de la pérdida de extensividad en estas magnitudes.
La extensividad es sutilmente diferente a la aditividad, porque pueden existir sistemas donde la energía es extensiva pero no aditiva. La extensividad indica como dicha magnitud física escala proporcional al tamaño del sistema, por ejemplo, en un gas ideal, si hay más partículas, éstas ocupan un tamaño mayor acorde a este aumento y es de esperar que la energía del sistema crezca si tenemos más partículas. Pero puede ocurrir que al aumentar el número de partículas de un sistema no aumente de energía o que aumente muy poco y no de manera proporcional. Si pensamos en los microestados posibles de un sistema, puede que al aumentar el número de partículas, éstas se muevan a regiones de menor energía, incluso energía negativa, disminuyendo la energía total del sistema, o simplemente se agrupen en niveles de energía de valor cero, lo que no aumentaría la energía del sistema \footnote{Es importante señalar que en el caso de las interacciones de corto alcance la entropía de Boltzmann es aditividad y extensiva, sin embargo, la energía interna solo es aditiva para el caso de un gas ideal, mientras que para el resto de sistemas es siempre extensiva. Para sistemas con interacciones de largo alcance lo que se busca es que la entropía sea extensiva porque esto preserva la estructura de Legendre de la termodinámica y para conseguirlo debe utilizarse una entropía no aditiva.}. Para ilustrar como se produce la pérdida de aditividad en la energía cuanto se tienen correlaciones de largo alcance, utilizaremos el modelo generalizado de \textbf{Curie-Weiss}, el cual es un sistema tipo Ising pero con una variación que envuelve interacciones de largo y corto alcance dado por,
\begin{equation}
E \propto \sum_{i<j}^{N} \frac{{s}_i{s}_j}{|i-j|^\alpha},
\end{equation}
donde $s_i = \pm 1$ es el espín, $\forall i$ y  $\alpha$ es un parámetro cuyo valor permite considerar casos desde interacciones de corto alcance cuando $\alpha\rightarrow\infty$ a interacciones de largo alcance cuando $\alpha$ se hace más pequeño.

\begin{figure}
    \centering
    \includegraphics[width=7cm]{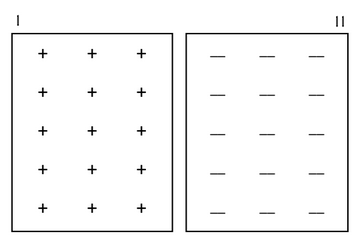}
    \caption{Separación del sistema en los subsistemas I y II.}\label{aditivi}
\end{figure}

Ahora dividamos el sistema en dos partes I y II, cada una compuesta por $N/2$ sitios, donde todos los espines del subsistema I se encuentran en el estado arriba (up), mientras que todos los del subsistema II se encuentran en el estado abajo (down), como se muestra en la Fig. \ref{aditivi}.

\begin{figure}[hbt!]
    \centering
    \includegraphics[width=15cm]{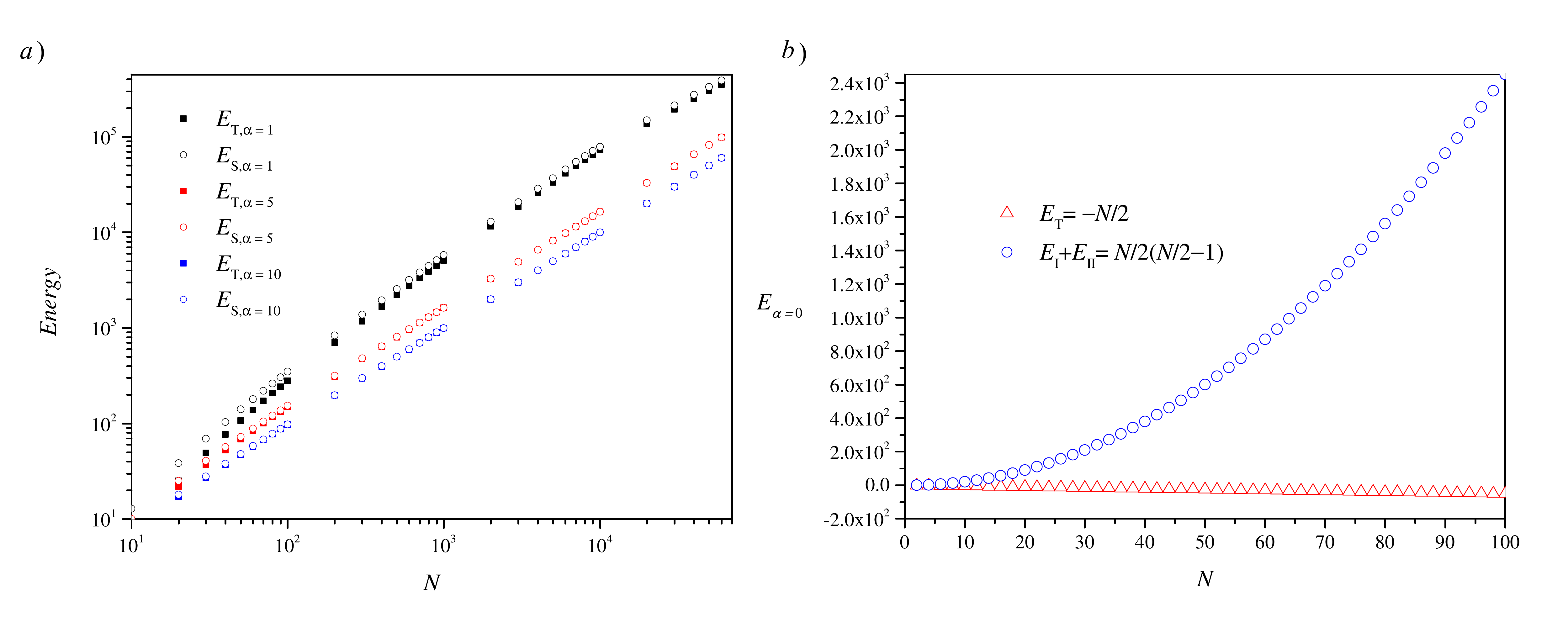}
    \includegraphics[width=7cm]{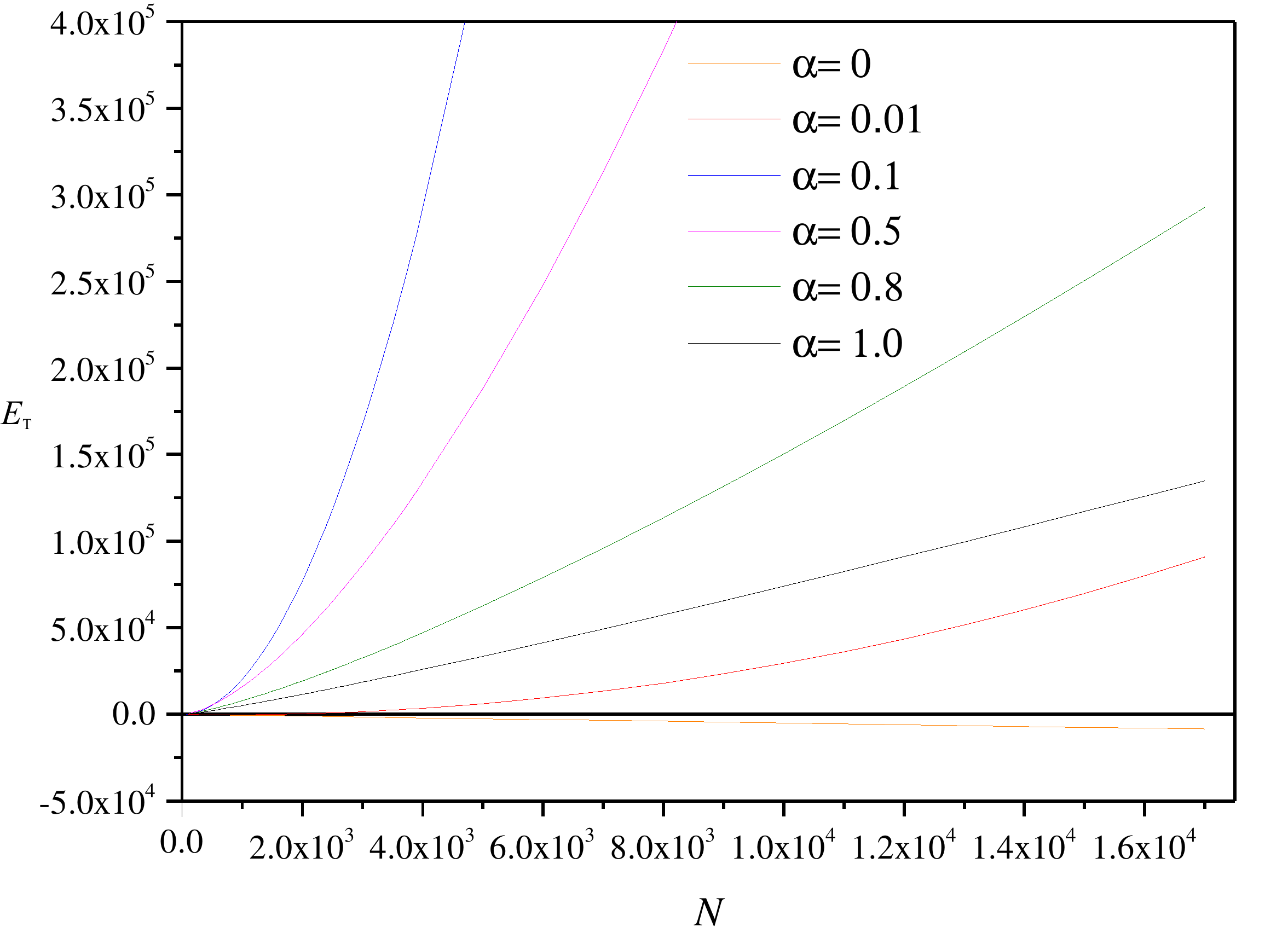}\includegraphics[width=7.1cm]{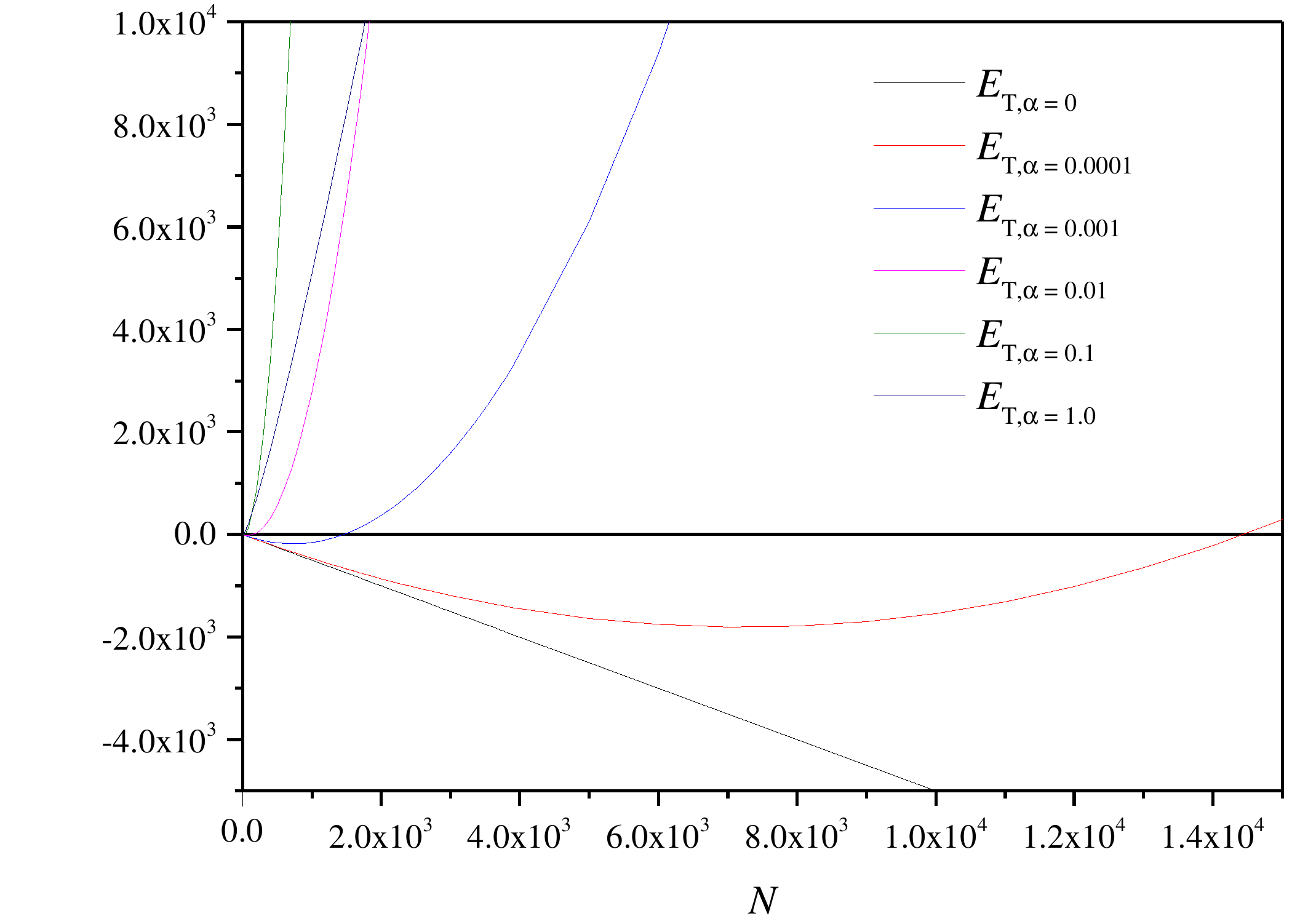}
    \caption{Energía total en función del número de partículas del sistema. En el panel superior izquierdo la energía para diferentes valores de $\alpha$. En el panel superior derecho la energía total \textit{versus} la suma de energías de los subsistemas I y II. En el panel inferior izquierdo la energía total para diferentes valores del parámetro  $0<\alpha<1$. En el panel inferior derecho un acercamiento para valores de $\alpha$ $0<\alpha<0$.$1$.} \label{ExtensividadAditividad}
\end{figure}

En la Fig. \ref{ExtensividadAditividad} se muestra la energía total del sistema para la misma situación anterior con la mitad de los espines arriba y la otra mitad con espines abajo. En el panel superior izquierdo se muestra el efecto del tamaño, se observa que a medida que el número de partículas $N$ crece, la energía total del sistema $E_T$ (cuadrados rellenos) se acerca a la suma de las energías $E_S$ para valores de $\alpha$ mayores, mientras que para $\alpha=1$ se observa una gran diferencia,  claramente cuando $\alpha$ crece, las interacciones son de corto alcance y por tanto la aditividad se cumple. Para interacciones de largo alcance, como el caso $\alpha=1$ y el caso $\alpha=0$ se observa una discrepancia entre $E_T$ y $E_S$. En el panel superior derecho se muestra como para $\alpha=0$ la diferencia entre $E_T$ y $E_S$ es muy grande y a medida que el tamaño del sistema crece, la discrepancia observada es todavía mayor.
En el panel inferior izquierdo podemos ver el comportamiento de la energía total para diferentes valores de $\alpha$ $0<\alpha<1$, mientras que en el inferior derecho se muestra un acercamiento para valores de $\alpha$ $0<\alpha<$0.$1$.

A pesar de que estos sistemas producen pérdida de aditividad acompañada de pérdida de extensividad, es posible recuperar ésta última mediante la prescripción de Kac \cite{KAC} al  introducir un factor $1/N$ en el potencial, lo que vuelve el sistema extensivo, pero no aditivo, esto es,
\begin{equation}
H =-\frac{J}{2N}\sum_{i,j}{s}_i{s}_j.
\end{equation}
Con esto la energía escala proporcional al número de partículas $N$.
Este tipo de escalamiento presenta una estructura termodinámica estándar porque preserva las ecuaciones de Euler y las relaciones de Gibbs-Duhem \cite{DELPINO2} recuperando la linealidad de las propiedades termodinámicas de esta clase de sistemas \footnote{Si bien la prescripción de Kac es un procedimiento ampliamente utilizado, existen otros escalamientos no extensivos que preservan relaciones de Gibbs-Duhem asociadas \cite{VELAZQUEZ2016}. }. 



A pesar de la abundancia de resultados teóricos y numéricos, la descripción termodinámica estadística en los sistemas con interacciones de largo alcance aún plantea muchos desafíos. Primero, por la gran cantidad de correlaciones y acoplamientos se traducen en no linealidades que crean serias dificultades tanto analíticas como numéricas. Además, las suposiciones comunes en la base misma de la mecánica estadística como el estado de equilibrio térmico con un baño se basan en la localidad de corto alcance de la interacción (lo que permite separar el sistema en partes independientes), pues la propiedad de aditividad es utilizada para establecer las condiciones para el equilibrio \cite{CALLEN, PATHRIA}. Los ensambles canónico, gran canónico, el carácter intensivo de la temperatura, el contacto con un foco térmico entre sistemas, etc. son todos derivados para sistemas extensivos. Por otro lado, la termo-estadística pretende reconstruir la termodinámica macroscópica a partir de representaciones microscópicas del sistema mediante los métodos de la matemática estadística independientemente del carácter de la interacción si es de corto o largo alcance. Para los sistemas de largo alcance, la comprensión de los fenómenos aquí mencionados no está del todo clara. La teoría de Tsallis \cite{TSALLIS}, y la superestadística de Beck-Cohen \cite{BECK}, son algunos ejemplos de esfuerzos por intentar explicar lo que sucede cuando los sistemas poseen este tipo de interacciones.

Como se mencionó anteriormente, una correcta descripción de los sistemas con interacciones de largo alcance, fuera del equilibrio, puede llevarse a cabo mediante los métodos de la dinámica molecular o la aplicación de ecuaciones cinéticas apropiadas. En lo que sigue presentamos la obtención de la ecuación de Vlasov a partir de la jerarquía BBGKY.

\section{Ecuación de Vlasov}

La descripción de $N$ partículas en el espacio de fase requiere hallar una función de distribución de probabilidad $F$ que tenga en cuenta todas las partículas del sistema, esto es $F=F_N(\vec{r}_1, \vec{p}_1, \vec{r}_2, \vec{p}_2,..., \vec{r}_N, \vec{p}_N; t)$, donde $F_N$ es la probabilidad de que en un tiempo $t$, la partícula $1$ se encuentre en el volumen infinitesimal $d^3\vec{r}_1d^3\vec{p}_1$, la partícula $2$ en $d^3\vec{r}_2d^3\vec{p}_2$, y así sucesivamente hasta $N$. Sin embargo en la práctica esto no es posible, y se suelen utilizar funciones de distribución reducidas de la jerarquía BBGKY \cite{SOTO}.
Consideremos un hamiltoniano dado por,
\begin{eqnarray}
H=\sum_{i=1}^N \frac{p_i^2}{2m}+\sum_{i=1}^N V(\vec{r}_i)+\sum_{i<j}^N\phi_{ij},
\end{eqnarray}
donde $V(\vec{r}_i)$ es la energía potencial de la partícula $i$-ésima debida a una fuerza externa $\vec{F}=-\nabla V$, y $\phi_{ij}=\phi(\vec{r}_i,\vec{r}_j)=\phi(|\vec{r}_i-\vec{r}_j|)$ es el potencial entre partículas.

Partiremos de la ecuación de Liouville \cite{SOTO},
\begin{equation}\label{liouville}
\frac{\partial F}{\partial t}= -\{H,F\},
\end{equation}
donde $\{\}$ es el paréntesis de Poisson y $F$ es la función de distribución. Asumiendo que las partículas son indistinguibles (esto es, que la función de distribución es simétrica, es decir no cambia al intercambiar $\vec{r}_i,\vec{p}_i$ por $\vec{r}_j,\vec{p}_j$), se pueden obtener las ecuaciones de la jerarquía BBGKY, que vienen dadas por,
\begin{eqnarray}
\frac{\partial F^{(n)}}{\partial t}=
-\left\{H,F^{(n)}\right\}-\sum_{i=1}^{n}\int\left\{\phi_{i,n+1},F^{(n+1)}\right\}d^3\vec{r}_{n+1}d^3\vec{p}_{n+1}\cdot\cdot\cdot d^3\vec{r}_{N}d^3\vec{p}_{N},\nonumber\\
\end{eqnarray}
donde, $F^{(n)}$ es la función de distribución reducida,
\begin{eqnarray}
F^{(n)}(\Gamma;t)=\int F(\Gamma,t)d^3\vec{r}_{n+1},\vec{p}_{n+1},...,\vec{r}_N,\vec{p}_N.
\end{eqnarray}
Aquí hemos utilizado la notación $\Gamma=\vec{r}_1,\vec{p}_1,\vec{r}_2,\vec{p}_2,...,\vec{r}_N,\vec{p}_N$.
Entonces la primera de las ecuaciones BBGKY, queda en términos de la función de distribución reducida de dos partículas,
\begin{eqnarray}
\frac{\partial F^{(1)}}{\partial t}+\frac{\vec{p}_1}{m}
\frac{\partial F^{(1)}}{\partial \vec{r}_1}+\frac{\vec{F}}{m}\cdot\frac{\partial F^{(1)}}{\partial \vec{p}_1}
=\int\frac{\partial \phi_{12}}{\partial \vec{r}_1}\left(\frac{\partial F^{(2)}}{\partial \vec{p}_1}-\frac{\partial F^{(2)}}{\partial \vec{p}_2}\right) d^3\vec{r}_2d^3\vec{p}_2,\label{BBGKY}
\end{eqnarray}
donde se ha usado el hecho que.
\begin{equation}
    \frac{\partial \phi_{12}}{\partial \vec{r}_1}=-\frac{\partial \phi_{12}}{\partial \vec{r}_2}.
\end{equation}

Esta primera ecuación de la Jerarquía BBGKY es la ecuación de Boltzmann, donde el lado derecho corresponde al término colisional.

En general, la función de distribución reducida de dos partículas lleva asociada una función de correlación en el espacio de fases; pero para la aproximación de campo medio asumimos que existe independencia estadística, de modo que $F^{2}$ puede escribirse como el producto de las funciones de distribución de las partículas individuales,
\begin{equation}
F^{(2)}(\vec{r}_1,\vec{p}_1,\vec{r}_2,\vec{p}_2,t)=f(\vec{r}_1,\vec{p}_1,t)f(\vec{r}_2,\vec{p}_2,t)\equiv f_1f_2.
\end{equation}



Si consideramos que no hay fuerzas externas, esto es $\vec{F}=0$, entonces la ec.( \ref{BBGKY}) resulta 
\begin{eqnarray}
\frac{\partial F^{(1)}}{\partial t}+\frac{\vec{p}_1}{m}\cdot
\frac{\partial F^{(1)}}{\partial \vec{r}_1}
&=&\int \frac{\partial\phi_{12}}{\partial\vec{r}_1}\left(\frac{\partial f_1f_2}{\partial \vec{p}_1}-\frac{\partial f_1f_2}{\partial \vec{p}_2}\right) d^3\vec{r}_2d^3\vec{p}_2\nonumber\\
&=&\int \frac{\partial\phi_{12}}{\partial\vec{r}_1}\left(f_2\frac{\partial f_1}{\partial \vec{p}_1}-f_1\frac{\partial f_2}{\partial \vec{p}_2}\right) d^3\vec{r}_2d^3\vec{p}_2,
\end{eqnarray}
donde el segundo término del lado derecho desaparece al integrar en $\vec{p}_2$, ya que la función de distribución se anula al evaluar en los extremos, entonces
\begin{eqnarray}
\frac{\partial F^{(1)}}{\partial t}+\frac{\vec{p}_1}{m}\cdot
\frac{\partial F^{(1)}}{\partial \vec{r}_1}
=\frac{f(\vec{r}_1,\vec{p}_1,t)}{\partial \vec{p}_1}\cdot \frac{\partial }{\partial \vec{r}_1}\int \phi(\vec{r}_1,\vec{r}_2)f(\vec{r}_2,\vec{p}_2,t) d^3\vec{r}_2d^3\vec{p}_2.
\end{eqnarray}
Recordando la definición de promedio en \cite{PATHRIA} tenemos
\begin{equation}
\langle A\rangle=\int f(\vec{r},\vec{p},t)A d^3\vec{r}d^3\vec{p},
\end{equation}
luego
\begin{eqnarray}
\int \phi(\vec{r}_1,\vec{r}_2)f(\vec{r}_2,\vec{p}_2,t)d^3\vec{r}_2d^3\vec{p}_2=\langle\phi(\vec{r}_1,t)\rangle,
\end{eqnarray}
por lo tanto, la ecuación resultante es,
\begin{eqnarray}
\frac{\partial F^{(1)}}{\partial t}+\frac{\vec{p}_1}{m}\cdot
\frac{\partial F^{(1)}}{\partial \vec{r}_1}
-\frac{f(\vec{r}_1,\vec{p}_1,t)}{\partial \vec{p}_1}\cdot \frac{\partial \langle\phi(\vec{r}_1)\rangle}{\partial \vec{r}_1}=0,
\end{eqnarray}
pero $F^{(1)}=f(\vec{r}_1,\vec{p}_1,t)\equiv f(\vec{r},\vec{p},t)$, por lo que podemos escribir,
\begin{eqnarray}
\frac{\partial f}{\partial t}+\frac{\vec{p}}{m}\cdot
\frac{\partial f}{\partial \vec{r}}
-\frac{\partial \langle\phi\rangle}{\partial \vec{r}}\cdot\frac{\partial f}{\partial \vec{p}} =0. \label{vlasovecc0}
\end{eqnarray}

\'Esta es la ecuación de Vlasov, conocida también como la ecuación de Boltzmann sin colisiones, pero en vez de una fuerza externa en el tercer término, se tiene una fuerza de campo medio $\vec{F}_{mf}=-\partial \langle\phi\rangle/\partial \vec{r}$, que viene de la interacción entre las partículas. Aunque ese término parezca colisional, en realidad la ecuación de Vlasov simplemente establece que, en ausencia de colisiones, la función de distribución $f$ se conserva mediante la evolución temporal en el espacio de fase, lo que significa $df/dt = 0$. 

Vamos a aplicar ahora la ecuación de Vlasov en un modelo unidimensional con interacciones de largo alcance de campo medio, cuyo hamiltoniano viene dado por,
\begin{equation}
H=\sum_{i=1}^N\frac{p_i^2}{2}+\sum_{i<j}U(\theta_i-\theta_j),
\end{equation}
entonces la ecuación de Vlasov se reduce a
\begin{eqnarray}
\frac{\partial f}{\partial t}+p
\frac{\partial f}{\partial \theta}
-\frac{\partial \langle U\rangle}{\partial \theta}\frac{\partial f}{\partial p} =0,\label{vlasoveccuni}
\end{eqnarray}
donde la energía potencial promedio viene dada por,
\begin{equation}
\langle U(\theta,t)\rangle=\int d\theta'dp' U(\theta-\theta')f(\theta',p',t).\label{potenergyHMF}
\end{equation}
aquí $U(\theta-\theta')$ ya ha sido reescalada por la prescripción de Kac, por lo que la expresión corresponde justamente al promedio.

Como se planteó anteriormente, la dinámica molecular y la ecuación de Vlasov, son las indicadas para describir lo que sucede fuera del equilibrio, y en efecto se sabe que para una amplia gama de potenciales de campo medio (de largo alcance), el teorema de Braun-Hepp \cite{BRAUN}\footnote{En la comunidad científica existe actualmente un debate respecto a este teorema, el cual no considera que en el espacio de fases exista una medida de Lebesge tal que permita  integrar. Si en en el espacio de fases la estructura no es suave (por ejemplo el caso de un fractal), no se puede integrar, por lo que este teorema necesitaría, en sus hipótesis, indicar que se está considerando el caso en que es posible integrar, de otra manera no sería posible pasar de la ecuación de Liouville a ninguna otra ecuación cuya distribución sea de una sola partícula.}, demuestra rigurosamente que en el límite continuo $N\rightarrow\infty$ las soluciones de las ecuaciones de movimiento convergen a las soluciones de la ecuación de Vlasov.

En la sección siguiente veremos la aplicación de estas dos técnicas al modelo HMF.

\chapter{Modelo HMF}\label{modeloHMF}
\section{Aspectos generales}
En la literatura se puede encontrar una cantidad extensa de trabajos relacionados al modelo HMF \cite{KONISHI, ANTONI}, el cual permite describir el complejo comportamiento de los sistemas con interacciones de largo alcance fuera del equilibrio. En base a este modelo, han surgido importantes contribuciones, como por ejemplo la exhibición de dos tipos de relajación, que ha sido observada en sistemas estelares, plasmas, vórtices $2$-D, etc. La primera conocida como relajación violenta sin colisiones, donde el sistema rápidamente transita desde una condición inicial a un aparente estado de equilibrio conocido como QSS. Continuando con su evolución, el sistema sufre una lenta relajación (con colisiones o correlaciones) que lo lleva a alcanzar el estado de  equilibrio estadístico descrito por la distribución de Maxwell-Boltzmann-Gibbs (MGB) \footnote{Esto no ocurre en todos los sistemas con interacciones de largo alcance. Por ejemplo en un sistema de osciladores, el equilibrio BG nunca es alcanzado.}. El tiempo de relajación colisional crece conforme crece el tamaño del sistema, por lo que la duración de un estado QSS teóricamente se vuelve infinita en el límite termodinámico. 
Esto ha creado un gran debate en la comunidad de la mecánica estadística. Por un lado, inspirándose en el trabajo de Tsallis, se ha intentado dar una explicación a estos estados a partir de una forma de entropía generalizada,  e inspirándose en el trabajo de Lynden-Bell, se ha propuesto interpretar estos estados QSS por medio de la ecuación de Vlasov, que corresponde a la ecuación de Boltzmann sin colisiones, y que describe por tanto, el régimen estacionario tras la relajación violenta. 

Desde el punto de vista físico, la distribución de Lynden-Bell es una solución al problema variacional, mientras que las distribuciones de leyes de potencia corresponden a ajustes de parámetros que después de ciertos esfuerzos se tratan de conectar la termodinámica, haciendo de ésta una rama conocida como termodinámica no extensiva y/o no aditiva. 


En la perspectiva de Lyndenbell, la idea consiste en buscar el estado más probable del sistema resultado de la mezcla de fases, compatible con todas las restricciones impuestas por la dinámica de Vlasov, en el que se asume que el sistema está \textbf{bien mezclado} y que por tanto, se cumple la hipótesis de ergodicidad. Si la distribución inicial sólo toma dos valores (ejemplo water-bag), Lynden-Bell predice para el estado QSS, una distribución similar a la de Fermi-Dirac. 

Antoniazzi et.al.  \cite{Antoniazzi2007}, observó que en el modelo HMF para una magnetización crítica inicial $M_{0c}\approx 0$.$897$, el sistema queda atrapado en estados QSS con diferentes magnetizaciones, de los cuales tan solo aquellos con magnetización inicial menor a la crítica $M_{x}<M_{0c}$, pueden ser descritos por la distribución de Lynden-Bell, ya que la distribución no depende de las orientaciones.

Por otro lado Tsallis et.al \cite{Tsallis2008}, propuso tres tipos de eventos asociados a  estados QSS diferentes para el caso en que la magnetización inicial $M_0\approx 1$. Cuantificando su frecuencia de exhibición en las simulaciones, observaron diferentes tipos de comportamiento del teorema del límite central en cada uno de ellos, en los que destaca la distribución de Tsallis como un posible ajuste para la distribución en los momentos. 
La función q-exponencial de Tsallis, es también una solución estacionaria de la ecuación de Vlasov \cite{Tsallis2008}, sin embargo, en la literatura sólo se encuentran ajustes en la distribución de velocidades (o momentos).

Diversos estudios han sido llevados a cabo con este modelo, análisis difusivo \cite{LATORA}, caoticidad de las condiciones iniciales (análisis de los exponentes de Lyapunov) \cite{DAUXOIS, LATORA2, ANTUNES, MANOS, MIRITELLO}, verificación del primer principio de la termodinámica \cite{MOYANO}, verificación de la propiedad de ergodicidad y del teorema del límite central \cite{MUKAMEL, PLUCHINO2, BENETTI, RIBEIRO, FIGUEIREDO}, etc. En la siguiente sección se presenta formalmente el modelo HMF algunos resultados importantes y estudios realizados que fueron aplicados al modelo d-HMF en esta tesis.

\section{Definición y resultados generales}

El modelo HMF es un modelo donde las partículas interactúan a través de su espín y cuyo valor es continuo, es decir, pueden estar orientados en cualquier dirección del plano. El modelo presenta además una transición de fase de segundo orden del tipo para-ferromagnético. Su hamiltoniano viene dado por la expresión
\begin{equation}
 H \!= \!\!\sum_{i=1}^{N}\!\frac{p_i^2}{2}\!+\! \frac{\lambda}{2N} \!\sum_{i, j=1}^N (1-\cos(\theta_i\!-\!\theta_j)).\label{hamiltonianoHMF}
\end{equation}

Este modelo se ha resuelto analíticamente en los ensambles canónico y microcanónico \cite{ANTONI,CAMPA2}, se han realizado simulaciones de dinámica molecular y se ha resuelto numéricamente mediante la ecuación de Vlasov por varios autores \cite{DAUXOIS, PLUCHINO, BUYL}. En estos trabajos se ha observado la presencia de estados QSS cuando la energía interna por partícula toma valores alrededor de $\varepsilon \approx 0$.$69$.

Para hallar estos estados se han utilizado  condiciones iniciales llamadas water-bag (WBIC), que consisten en distribuciones constantes del espacio de fases, donde las orientaciones y los momentos  están distribuidos de la forma,

\begin{figure}
\centering
    \includegraphics[width=0.8\textwidth]{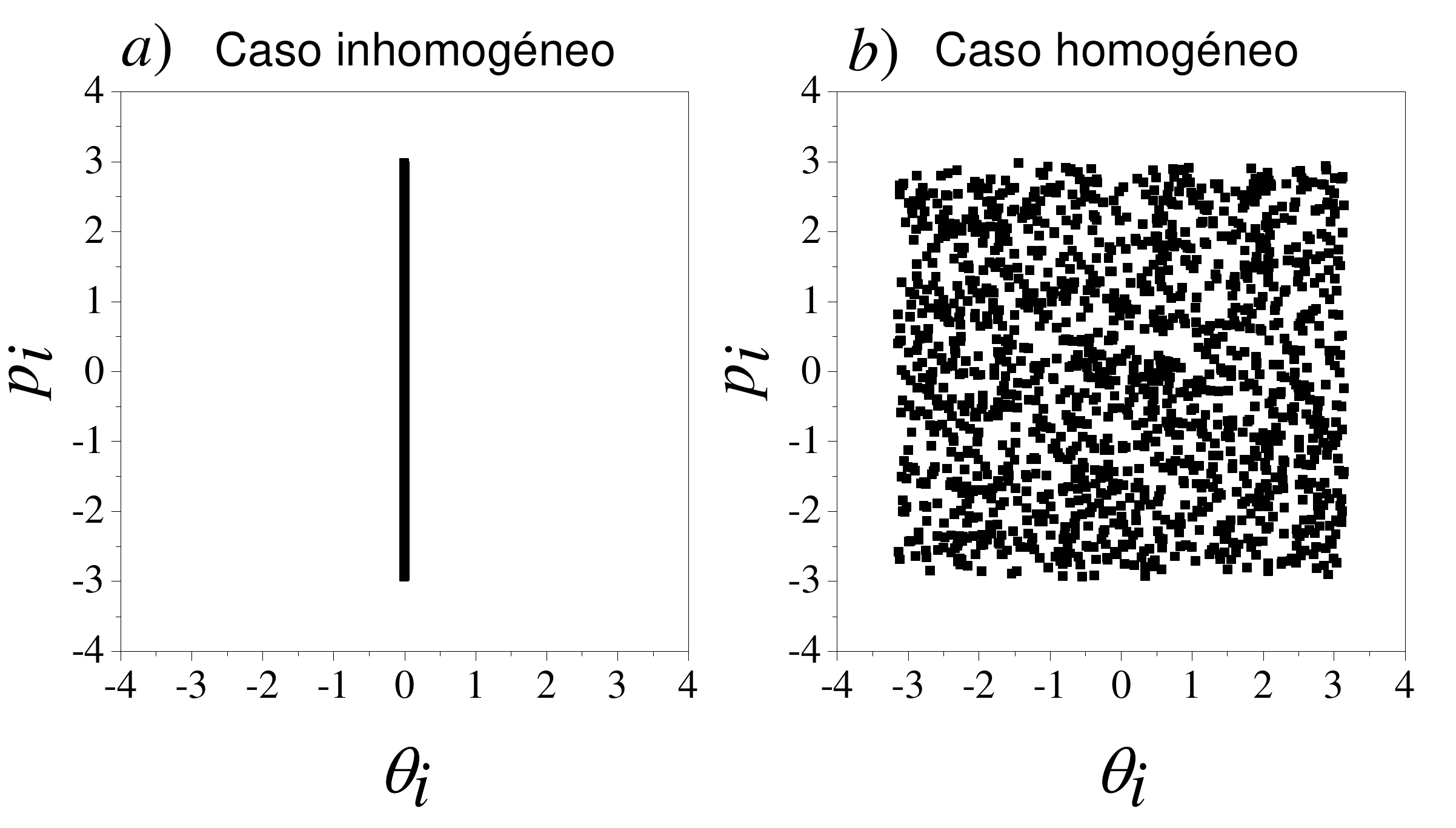}
\caption{En la izquierda las condiciones iniciales water-bag inhomogéneas. En el panel derecho las homogéneas.}\label{WBIC}
\end{figure}

\begin{figure}
\centering
    \includegraphics[width=1.0\textwidth]{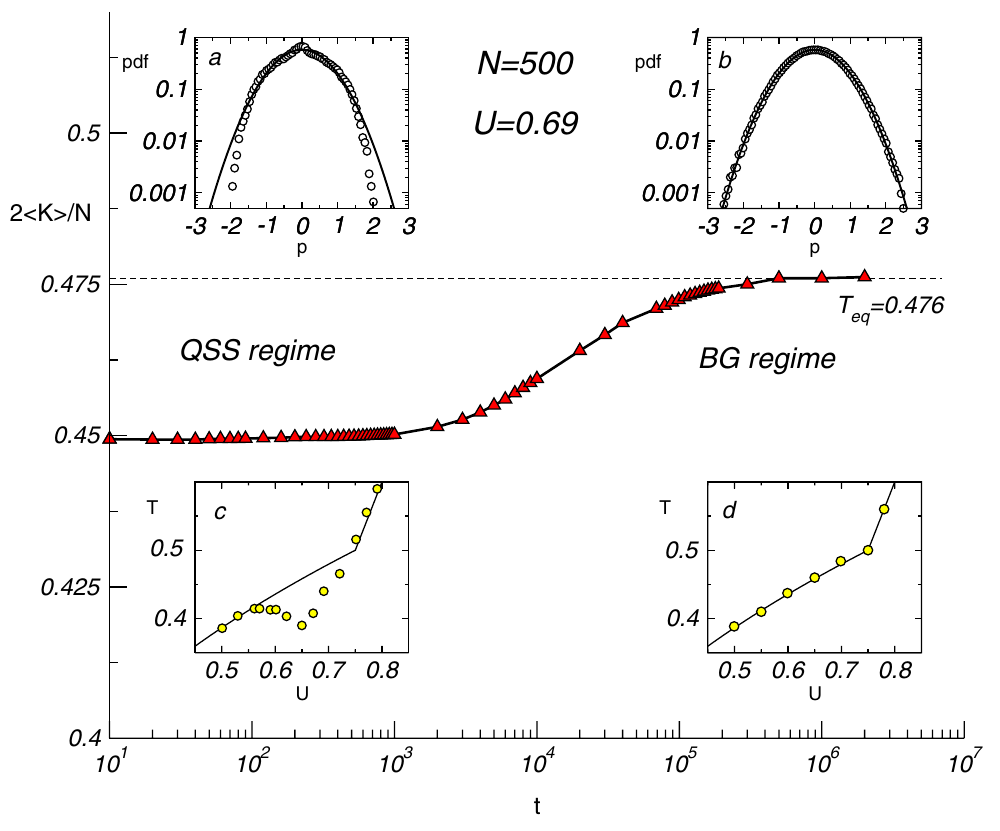}
\caption{Simulaciones numéricas en el ensamble microcanónico para $N=500$ y energía interna por partícula $\varepsilon = 0$.$69$. En la parte superior el perfil de distribución de los momentos $p$, en a) para el estado QSS y en b) para el equilibrio BG. En el centro se muestra la evolución de la energía cinética promedio por partícula (\emph{Temperatura}), habiendo dos regímenes, donde ésta permanece constante por un tiempo prolongado. En c) se muestra la curva calórica con los puntos amarillos tomados con la \emph{temperatura} del estado QSS, mientras que en d) con la temperatura del equilibrio BG (imagen obtenida de Ref. \cite{PLUCHINO}).}\label{pluchino}
\end{figure}
\begin{equation}
f(\theta,p,0)=\left\{
  \begin{array}{ll}
    \;\;\frac{1}{4\theta_0p_0}\;\;\;\;\;,|\theta|\leq \theta_0\:\: y\:\: |p|\leq p_0  \\
    \;\;\;\;\;\;0\;\;\;\;\;\;\;\;, \mbox{para otros valores de $\theta$ y $p$}. \\
  \end{array}
\right.
\end{equation}
En la Fig. \ref{WBIC} \emph{a}) se muestra un ejemplo para $p_0=\pi$ y $\theta_0=2\pi/1000\sim0$ caso inhomogéneo. En el panel \emph{b}) de la Fig. \ref{WBIC} se muestra un caso homogéneo con $p_0=\pi$ y $\theta_0=\pi$.

\begin{figure}
\centering
    \includegraphics[width=0.7\textwidth]{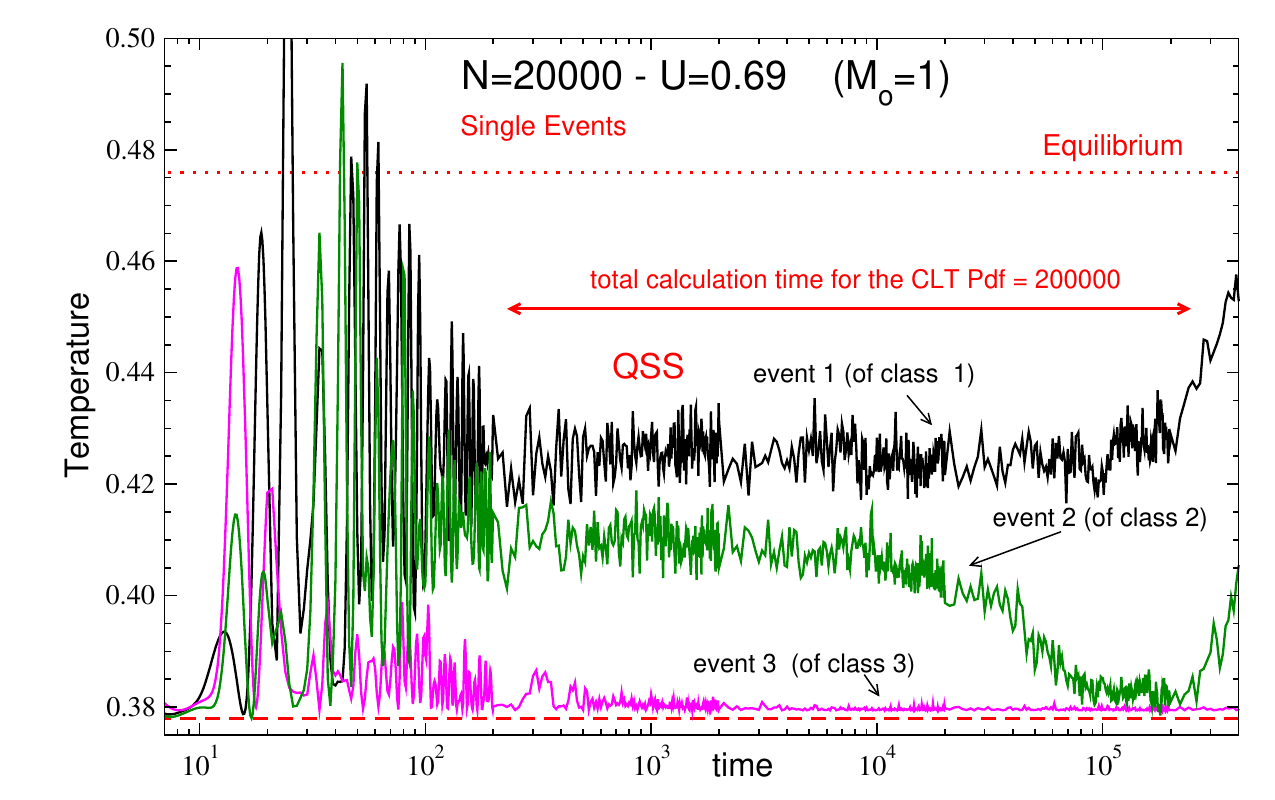}
      \includegraphics[width=0.7\textwidth]{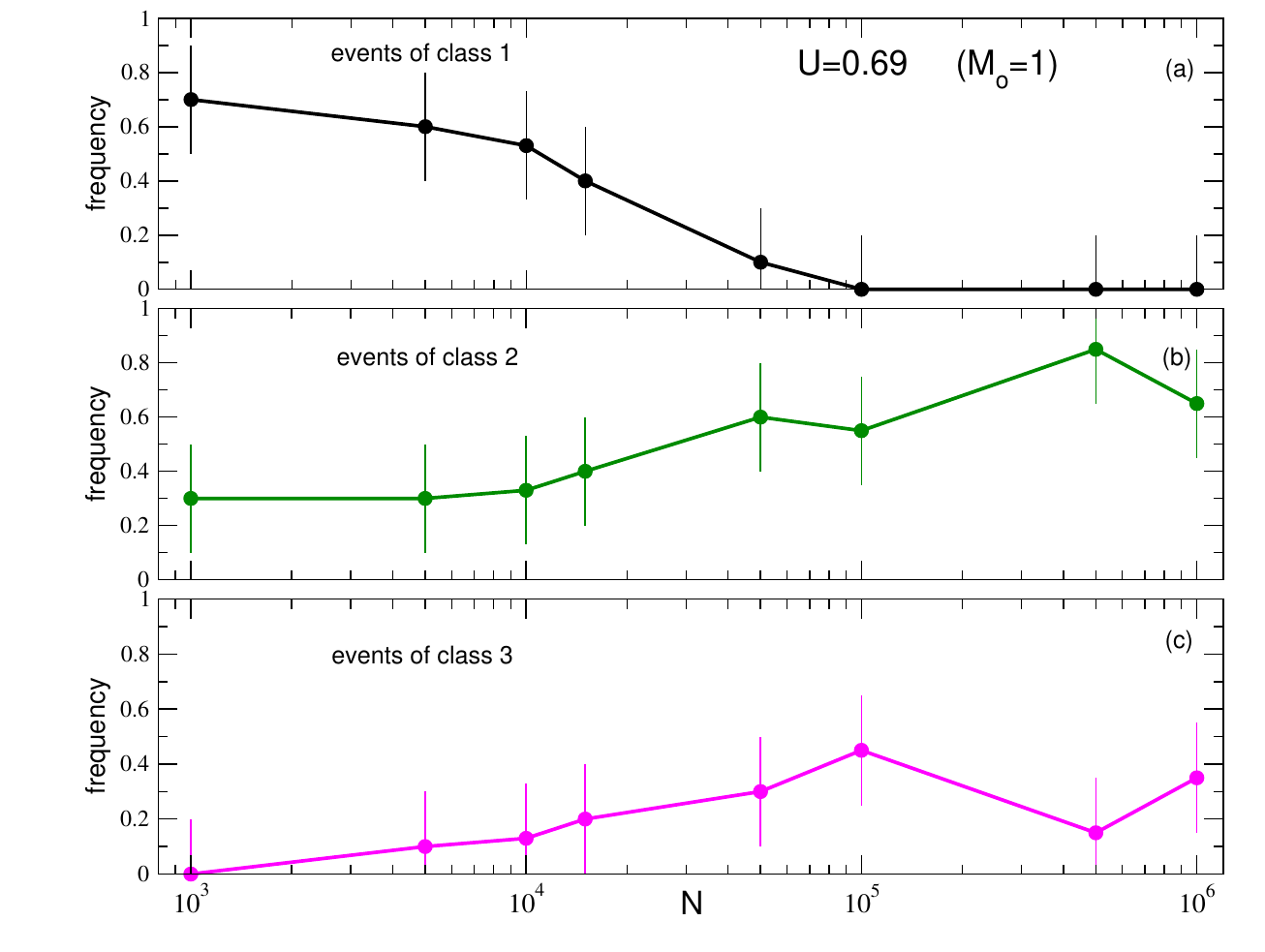}
\caption{En el panel superior se muestran tres tipos de estados cuasi-estacionarios del modelo HMF observados para la condición inicial water-bag inhomogénea, esto es, magnetización inicial casi nula $m\simeq0$. En el panel inferior la frecuencia de exhibición de los tres estados QSS descritos en el panel superior. Conforme crece el número de partículas el segundo tipo de estado QSS se vuelve más frecuente en comparación a los demás, mientras que el primero desaparece (imagen obtenida de Ref. \cite{PLUCHINO2008}).}\label{pluchino2}
\end{figure}

\begin{figure}
\centering
    \includegraphics[width=1.0\textwidth]{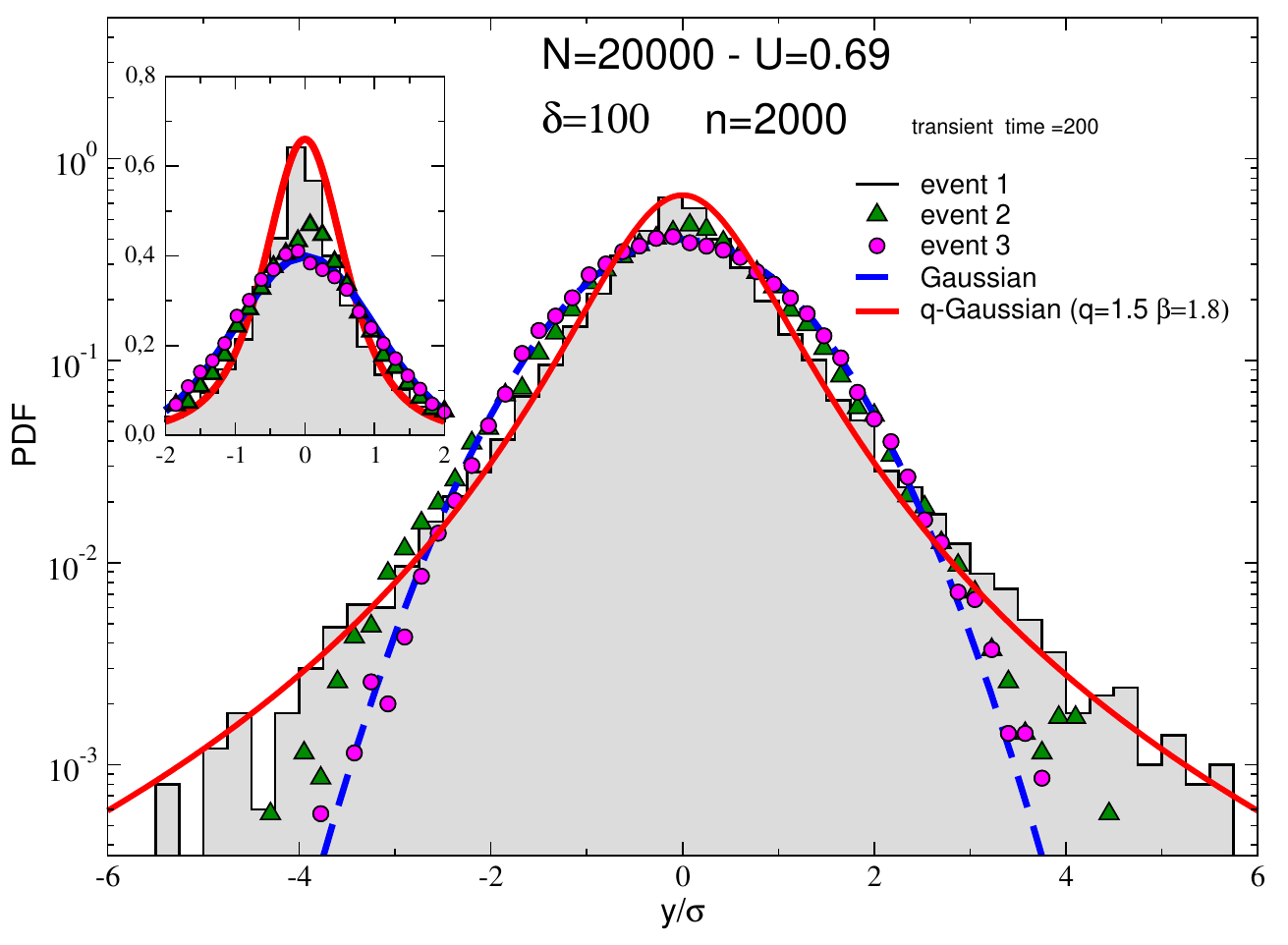}
\caption{Ajuste de Tsallis para describir los estados QSS del modelo HMF \cite{PLUCHINO2008}. Los eventos de clase 2 son los que presentan más desviación respecto a los ajustes tipo Tsallis. $M_0$ es la magnetización inicial, donde $M_0=1$ son condiciones iniciales inhomogéneas, mientras que $M_0=0$ son las homogéneas.} \label{eventos}
\end{figure}

Con estas condiciones iniciales se encontró el estado QSS descrito en \cite{PLUCHINO} como se muestra en la Fig. \ref{pluchino}, para $N=500$. Se puede apreciar que el perfil de distribución se aleja de uno gaussiano en el QSS. En el panel c) se observa una zona con capacidades caloríficas negativas, tomando en cuenta la \textit{temperatura} \footnote{Aquí el término temperatura ha sido puesto en cursiva porque corresponde a un abuso de notación. En realidad se refiere a la energía cinética promedio. La temperatura escrita sin cursiva corresponde a la del equilibrio BG derivada de la entropía.} registrada por el QSS. Las líneas continuas en los paneles $c$ y $d$ de la Fig. \ref{pluchino}  corresponden a la solución analítica del modelo HMF, encontrada por Antoni \cite{ANTONI}, dada por,
\begin{equation}
\varepsilon=\frac{1}{2\beta}+1-m^2
\end{equation}
donde $m$ es la magnetización y viene expresada por,
\begin{equation}
m=\frac{I_1(y)}{I_0(y)}
\end{equation}
donde $y=2\beta \lambda x_0$ y $x_0$ es la solución del problema extremal $x=I_1(2\beta \lambda x)/I_0(2\beta \lambda x)$. $I_0$ e $I_1$ son las funciones modificadas de Bessel de primera especie de orden cero y uno, respectivamente,
obtenidas al utilizar las transformaciones de Hubbard-Stratonovich \cite{HUBBARD, STRATONOVICH}. Un desarrollo similar y más detallado se verá en la sección siguiente para el modelo d-HMF.

Pluchino et.al Ref. \cite{PLUCHINO2008} mostraron (para el modelo HMF) que bajo condiciones iniciales tipo water-bag se observan tres clases de estados QSS, los cuales son mostrados en el panel superior de la Fig. \ref{pluchino2}, mientras que su frecuencia de exhibición es mostrada en el panel inferior. Se puede observar que conforme crece el número de partículas el primer estado desaparece, mientras que el segundo tipo se vuelve más frecuente. En la Fig. \ref{eventos}, se observa como el ajuste de Tsallis no describe correctamente la distribución del evento clase 2. En el capítulo \ref{dhmfmodel} mostramos que para el modelo d-HMF el evento clase 2 es exacerbado con las condiciones iniciales homogéneas.

\section{Dinámica de Vlasov para el modelo HMF}

Un camino para describir los estados QSS del modelo HMF es mediante soluciones estacionarias de la ecuación de Vlasov. Esta ecuación requiere conocer los momentos y las fuerzas por partícula del sistema, que pueden obtenerse  de la ec.( \ref{hamiltonianoHMF}), luego las ecuaciones de movimiento para el modelo HMF vienen dadas por,
\begin{eqnarray}
\dot{\theta}_i&=&p_i\\
\dot{p}_i&=&-M_x\sin\theta_i+M_y\cos\theta_i,
\end{eqnarray}
donde, $(M_x,M_y)=\frac{1}{N}(\sum_i\cos\theta_i, \sum_i\sin\theta_i)$.
Para describir el sistema con la dinámica de Vlasov, es necesario pasar al continuo por medio de la función de distribución,
\begin{eqnarray}
\dot{p}&=&-M_x\sin\theta+M_y\cos\theta\\
M_x&=&\int_{-\infty}^{\infty}\int_{-\pi}^{\pi}\cos\theta f(\theta,p,t)d\theta dp\\
M_y&=&\int_{-\infty}^{\infty}\int_{-\pi}^{\pi}\sin\theta f(\theta,p,t)d\theta dp,
\end{eqnarray}
donde, $\theta$ y $p$, son las coordenadas Eulerianas \cite{LUO}. Por otro lado la energía potencial de una partícula en el continuo viene dada por la ec.( \ref{potenergyHMF}), entonces para este modelo tenemos lo siguiente,
\begin{figure}
\centering
    \includegraphics[width=0.45\textwidth]{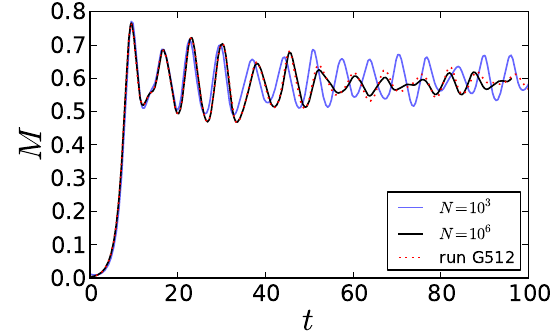}
\caption{Magnetización $M_x$ \textit{versus} tiempo. Las líneas continuas negra y azul corresponden a simulaciones por dinámica molecular. Los puntos de color rojo son de la dinámica de Vlasov (imagen obtenida de Ref. \cite{BUYL3}).}\label{magvlashmfcomp}
\end{figure}
\begin{figure}
\centering
    \includegraphics[width=0.8\textwidth]{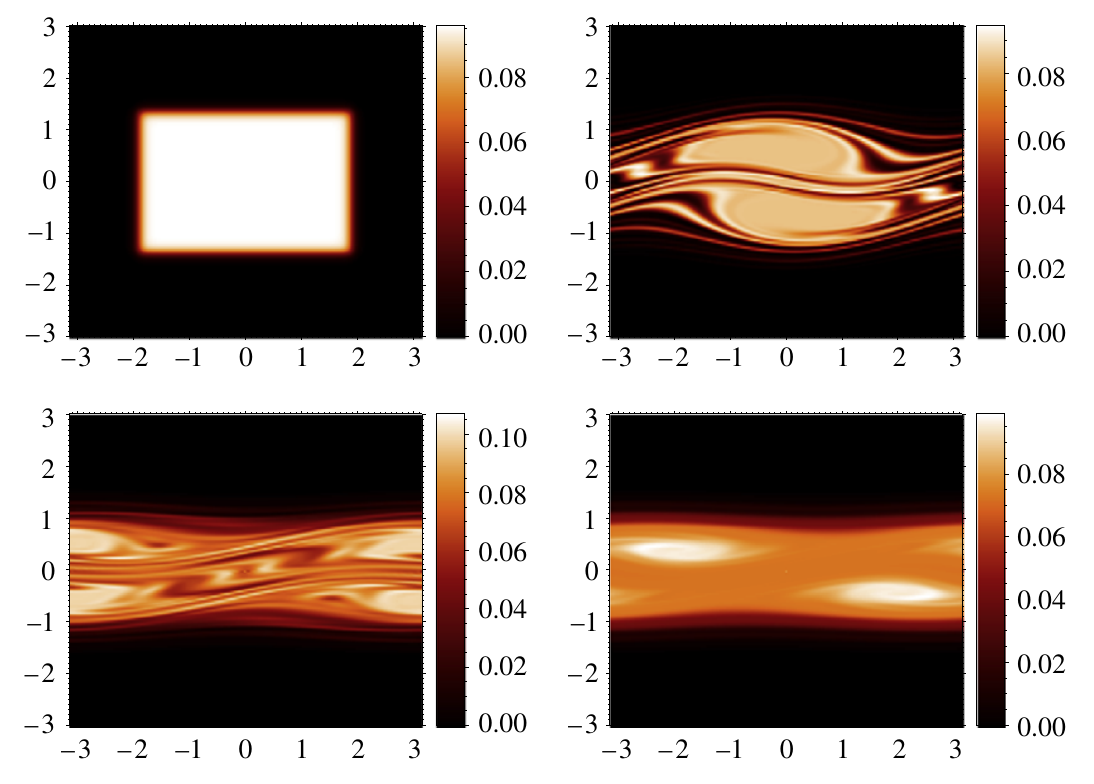}
\caption{Instantáneas (snapshots) del espacio de fases desde la condición inicial water-bag $\varepsilon=0$.$69$ y $m_0=0$.$5$. El último cuadro corresponde al estado cuasi-estacionario (imagen obtenida de Ref. \cite{CAMPA2}).}\label{qsssnaphmfvlas}
\end{figure}
\begin{eqnarray}
\langle U(\theta,t)\rangle&=&\int f(\theta',p',t)(1-\cos(\theta'-\theta))d\theta' dp'\nonumber\\
&=&1-M_x\cos\theta -M_y\sin\theta,\label{pothmfvlas}
\end{eqnarray}
entonces
\begin{eqnarray}
-\frac{\partial \langle U(\theta,t)\rangle}{\partial \theta}=-M_x\sin\theta +M_y\cos\theta
\end{eqnarray}\label{campomediohmfvlas}
Luego como $\dot{\vec{p}}=\vec{F}_{mf}=-\nabla \langle U\rangle$, la ecuación de Vlasov queda,
\begin{eqnarray}
\frac{\partial f}{\partial t}+p
\frac{\partial f}{\partial \theta}
+(-M_x\sin\theta+M_y\cos\theta)\frac{\partial f}{\partial p} =0\label{vlasoveccHMF}
\end{eqnarray}
Además, podemos calcular la energía por partícula \cite{ANTONI},
\begin{eqnarray}
e_i=\frac{p^2_i}{2}+\frac{1}{N}\sum_{j=1}^{N}\left(1-\cos(\theta_i-\theta_j)\right),
\end{eqnarray}
donde el factor $2$ en el denominador se elimina porque es la interacción de una partícula con el resto. Luego pasando al continuo tenemos
\begin{eqnarray}
\epsilon(\theta,t)=\frac{p^2}{2}+1-M_x\cos\theta-M_y\sin\theta,\label{energypphmf}
\end{eqnarray}
y la energía total específica o densidad de energía en el tiempo $t$ es
\begin{eqnarray}
e(t)&=&\frac{1}{2}\int f(\theta,p,t)p^2d\theta dp +\frac{1}{2}\int f(\theta,p,t)(1-M_x\cos\theta-M_y\sin\theta)d\theta dp\nonumber\\
&=&\frac{1}{2}\int f(\theta,p,t)p^2d\theta dp+\frac{1}{2}\left(1-M_x^2-M_y^2\right),\label{energyesphmf}
\end{eqnarray}
Como podemos apreciar, la suma de las energías por partícula de la ec.( \ref{energypphmf}) no es igual a la energía total del sistema de la ec.( \ref{energyesphmf}). Esto es un claro reflejo de que el sistema no es aditivo al tener interacciones de largo alcance.


En la Fig. \ref{magvlashmfcomp} se puede apreciar como a medida que $N$ crece, la dinámica molecular se aproxima al resultado obtenido por la dinámica de Vlasov. En la Fig. \ref{qsssnaphmfvlas} se muestran cuatro instantáneas del espacio de fases para distintos tiempos. El primero es en $t=0$ (condición inicial water-bag homogéneo), y el último es el estado cuasi-estacionario.

Otra manera de describir aproximadamente los estados QSS como soluciones de la ecuación de Vlasov es mediante la estadística de Lynden-Bell \cite{LYNDENBELL} quien desarrolló un nuevo ensamble estadístico para explicar los estados estacionarios de sistemas gravitacionales, con el que predijo la existencia de los estados QSS y además observó que cuando la distribución inicial no es una solución estacionaria de la dinámica de Vlasov, el sistema sufre fuertes oscilaciones hasta que alcanza el QSS, proceso conocido como \textbf{relajación violenta}.

\begin{figure}
\centering
    \includegraphics[width=0.8\textwidth]{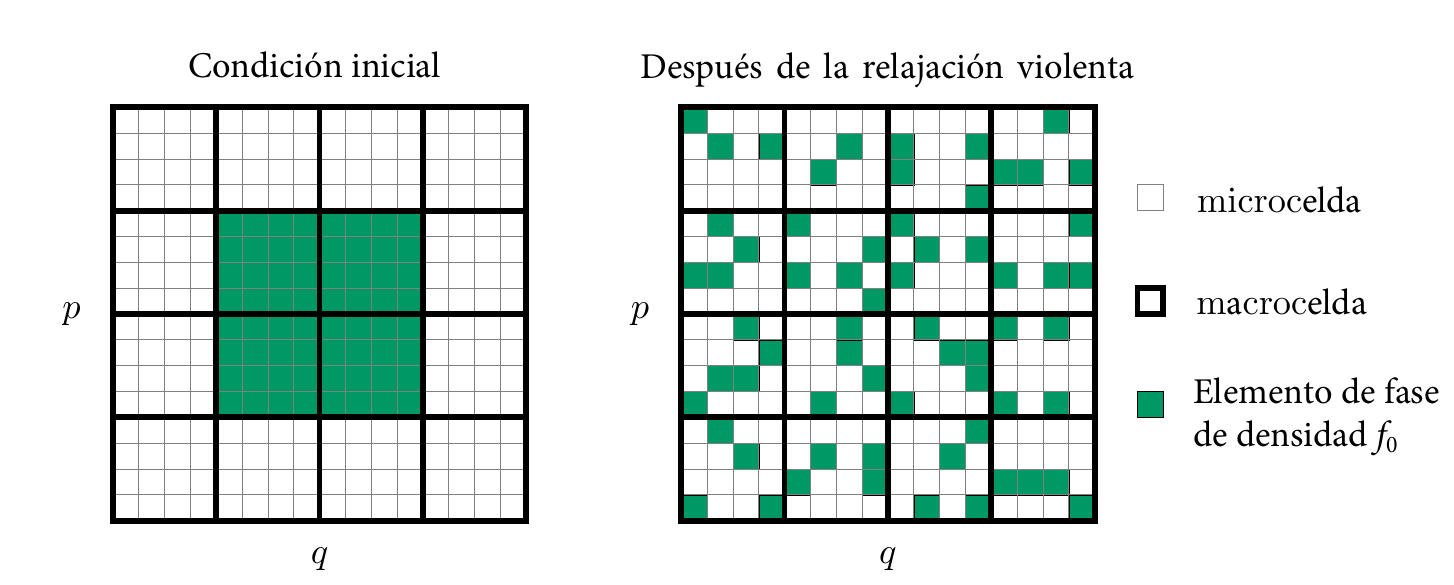}
\caption{En el panel izquierdo un water-bag. A la derecha una posible distribución de los elementos de fase tras la relajación violenta (imagen obtenida de Ref. \cite{BENETTI2}).}\label{Lyncell}
\end{figure}

En esta estadística se separa el espacio de fases en macroceldas compuestas de $\nu$ microceldas. Designamos por $\bar{f}$ a la distribución de una macrocelda de grano grueso (coarse grained) y por $f$ a la distribución microscópica de una microcelda (grano fino).
Consideremos una distribución inicial uniforme $f_0$ y sea $N$ el número de microceldas ocupadas por la distribución inicial. Este número permanece constante durante la dinámica, debido a la incompresibilidad del fluido de Vlasov. Sin embargo, la densidad en cada macrocelda no se conserva necesariamente. Un elemento de fase (cuadro pequeño de color verde) que inicialmente ocupa un lugar en una microcelda dentro de una cierta macrocelda, puede pasar a una microcelda perteneciente a otra macrocelda, como se muestra en la Fig. \ref{Lyncell}. En ella se da el ejemplo de $N=64$, es decir, $64$ microceldas ocupadas por la distribución inicial tipo water-bag. Tras la relajación violenta se distribuyen ocupando también otras macroceldas.
Bajo la dinámica sin colisiones, la función de distribución evoluciona como la densidad de un fluido incompresible. Esto significa que a medida que la distribución evoluciona, su densidad local permanece constante
a lo largo del fluido (su derivada convectiva es cero) \cite{LYNDENBELL}. A medida que la función de distribución evoluciona, se somete a un proceso de filamentación a escalas de longitud cada vez más pequeñas, hasta que finalmente la evolución ocurre en una escala de longitud tan pequeña que es imperceptible para la observación. En esta situación, se alcanza un estado estacionario macroscópico o de grano grueso, descrito por $\bar{f}$, mientras que la función de distribución microscópica $f$ continúa evolucionando.

\begin{figure}
\centering
    \includegraphics[width=0.8\textwidth]{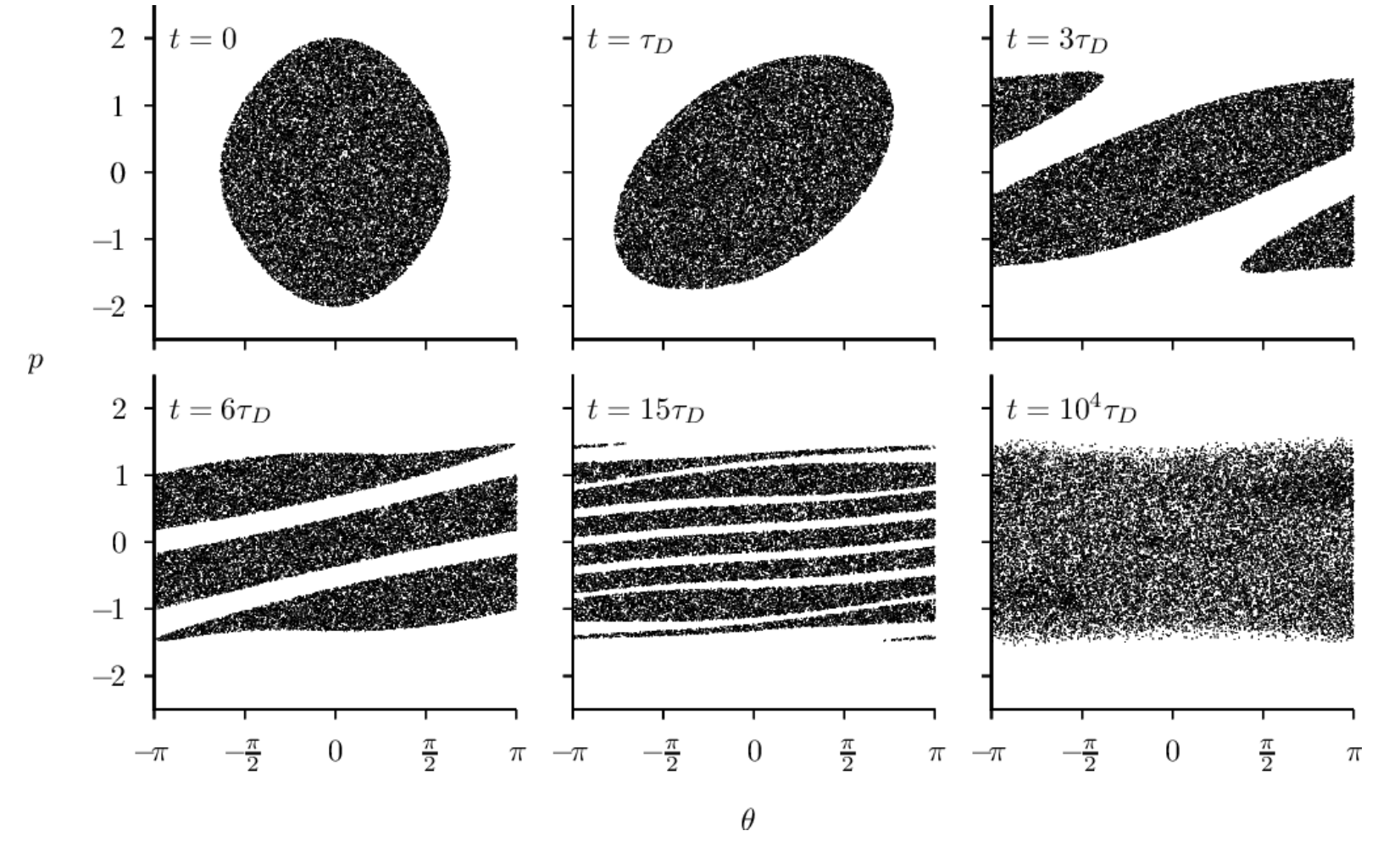}
\caption{Proceso de filamentación del modelo HMF (imagen obtenida de Ref. \cite{BENETTI2}.)}\label{filam}
\end{figure}

La Fig. \ref{filam} muestra un ejemplo del proceso de filamentación de una distribución de partículas en el espacio de fase de un sistema con interacciones de largo alcance.

Sólo la entropía de grano grueso puede aumentar; la entropía de grano fino debe conservarse. El procedimiento para obtener la entropía de grano grueso es similar al proceso de contar microestados que conduce a la entropía de Boltzmann de un gas reticular.

El número total de microestados compatibles con el macroestado donde las $n_i$ microceldas están ocupadas en la macrocelda $i$ viene dado por
\begin{eqnarray}
W_{LB}=\frac{N!}{\prod_i n_i !}\prod_i\frac{\nu!}{(\nu-n_i)!}.
\end{eqnarray}
A partir de la entropía de Boltzmann $S=\ln W$, y utilizando la aproximación de Stirling, se obtiene la entropía de la distribución de grano grueso $S_{LB}$ \cite{LYNDENBELL,CHAVANIS3}, que viene dada por la expresión
\begin{eqnarray}
S_{LB}=-\int \left(f^{*}\ln f^{*}+(1-f^{*})\ln(1-f^{*})) \right)d^3\vec{r}d^3\vec{p},\label{Lynentropy}
\end{eqnarray}
donde $f^{*}=\bar{f}/f_0$, $\bar{f}$ es la función de distribución de grano grueso y $f_0$ una distribución inicial inestable, como por ejemplo el water-bag.

\begin{figure}
\centering
    \includegraphics[width=0.8\textwidth]{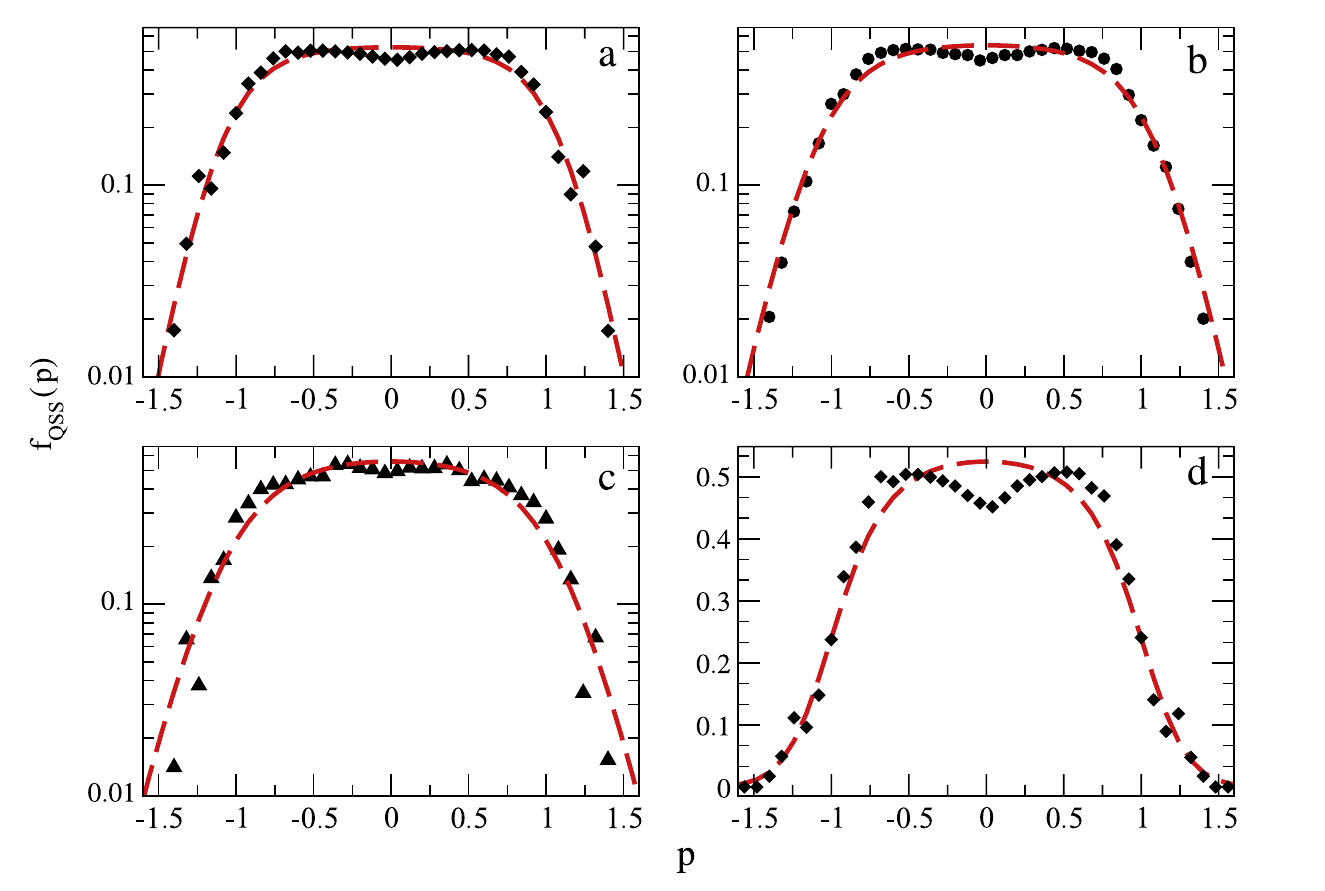}
\caption{Función de distribución de velocidades en el estado QSS para el modelo HMF con $\varepsilon=0$.$69$ y diferentes valores de $M_0$. Los puntos cuadrados se refieren a simulaciones numéricas mientras que la línea punteada es el perfil de velocidades de la ec.( \ref{Lyndenecc}). El panel (a), (b) y (c) presentan los casos, $M_0=0$.$3$, $M_0=0$.$5$ y $M_0=0$.$7$ en escala logarítmica, mientras que el panel (d) muestra el caso $M=0$.$3$ en escala lineal. La curva fue calculada de una muestra de $N=10^{7}$ en $t=100$ (imagen obtenida de Ref. \cite{CAMPA}).}\label{LyndenBelldis}
\end{figure}

La ec.( \ref{Lynentropy}) determina una expresión para la distribución de grano grueso $\bar{f}$. Esto se logra resolviendo el problema variacional,
\begin{eqnarray}
\delta S_{LB}-\beta \delta E-\alpha\delta M=0,
\end{eqnarray}
donde $\beta=1/kT$ y $\alpha=\mu/kT$ es la fugacidad. Luego obtenemos,
\begin{eqnarray}
f_{LB}\equiv\overline{f}=\frac{f_0}{1+\lambda e^{\beta(\frac{v^2}{2}+\phi)}}.\label{Lyndenecc}
\end{eqnarray}
La ec.( \ref{Lyndenecc}) ha logrado describir aproximadamente los perfiles de distribución que han sido obtenidos mediante la dinámica molecular y la ecuación de Vlasov. En la Fig. \ref{LyndenBelldis} los símbolos representan los datos obtenidos de las simulaciones numéricas y de color rojo punteado el perfil de distribución de grano grueso de la ec.( \ref{Lyndenecc}).

\chapter{Modelo d-HMF}\label{dhmfmodel}

\section{Definición del modelo}

En este apartado se presenta el modelo d-HMF \cite{ATENAS3}, el cual es una variación del modelo de Ising, pero con interacciones de largo alcance. En este modelo los espines son dipolos eléctricos que pueden orientarse en cualquier dirección. Para hacer una primera aproximación se consideran dipolos orientados en un plano pero ubicados en una recta, es decir un modelo unidimensional con condiciones de borde periódicas. Estas condiciones de borde pueden insertarse cerrando la recta en un anillo, de manera similar al modelo HMF. Al considerar el potencial dipolar eléctrico este modelo es más cercano a la realidad porque además como veremos no es isotrópico.

\begin{figure}[!h]
\centering
\includegraphics[width=1\textwidth]{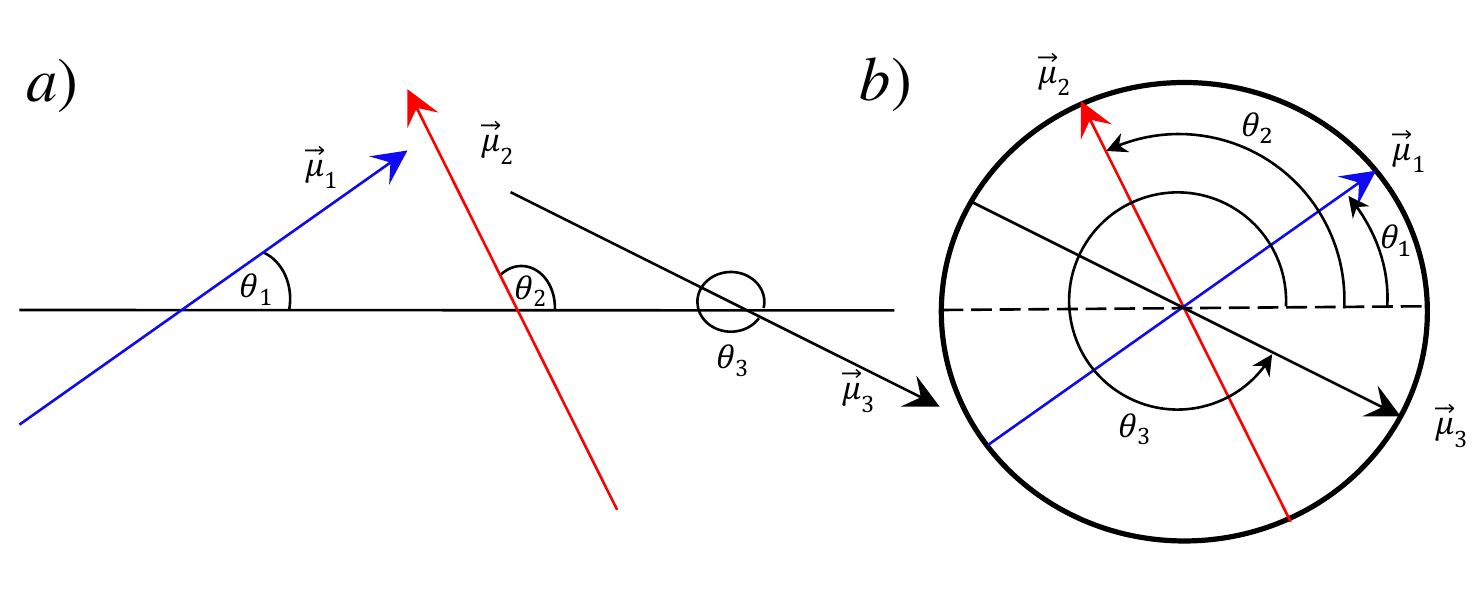}
\includegraphics[width=1\textwidth]{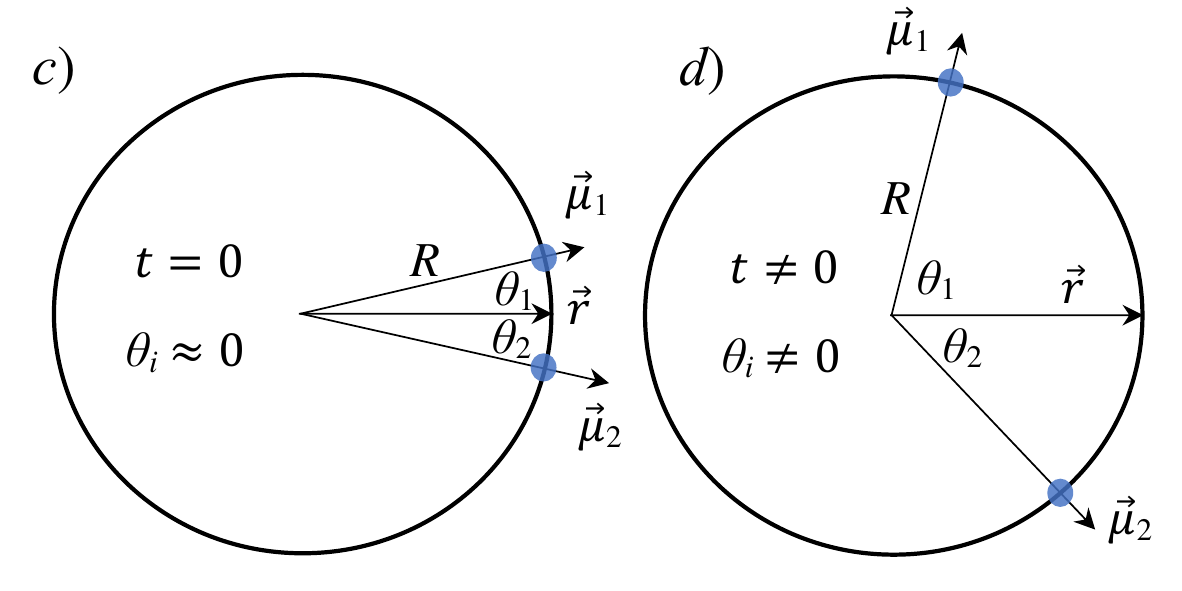}
\caption{En $a)$ una representación lineal de tres dipolos cualesquiera con orientaciones en el plano. En $b)$ una representación circular de estos mismos tres dipolos. En $c)$ se muestra un ejemplo de dos dipolos distribuidos con las condiciones iniciales inhomogéneas ($\theta_i\approx 0$) y una posible evolución para esta pareja de dipolos en un anillo en $d)$.}\label{ring}
\end{figure}

La energía potencial de la interacción entre dos dipolos $i$ y $j$ con momentos dipolares $\mu_i$ y $\mu_j$ respectivamente, viene dada por,
\begin{eqnarray}
U= - \frac{1}{4\pi\epsilon_0}\frac{3(\vec{\mu}_i\cdot \hat{r})(\vec{\mu}_j\cdot \hat{r})-\vec{\mu}_i\cdot\vec{\mu}_j}{\vert\vec{r}_i-\vec{r}_j\vert^3},
\end{eqnarray}
donde $\hat{r}$ es un vector unitario de dirección arbitraria y $\vec{r}_i$,   $\vec{r}_j$ corresponden a las posiciones de las partículas $i$, $j$.
Definimos el momento dipolar eléctrico del dipolo $i$ como $\vec{\mu}_i = e \vec{a}$ donde $e$ es el módulo de cada carga del dipolo y el módulo de $\vec{a}$ es la separación de las cargas que forman el dipolo (el vector  $\vec{a}$ describe su orientación). Luego,
$\vec{\mu}_i \cdot \hat{r} = \mu \cos {\theta_i}$ y $\vec{\mu}_i \cdot \vec{\mu}_j = \mu^2 \cos(\theta_i-\theta_j)$. Además, remarcamos que $\mu=|\vec{\mu}_i|$, los vectores $\vec{r}_i$ y $\vec{\mu}_i$ son paralelos entre ellos para todo $i$.
La Fig. \ref{ring} muestra la representación de tres dipolos en forma lineal $a)$, esos mismos tres dipolos en un anillo en $b)$, un ejemplo de dos dipolos con condiciones iniciales inhomogéneas $\theta_i \approx 0$ y una evolución en un tiempo posterior $d)$. 

Por otro lado, la distancia entre dos dipolos en un anillo de radio $R$, puede ser escrita como,
\begin{eqnarray}
r&=&\vert \vec{r}_i-\vec{r}_j \vert \nonumber\\
&=&\vert R\cos\theta_i \hat{e}_x+R\sin\theta_i \hat{e}_y-R\cos\theta_j \hat{e}_x-R\sin\theta_j \hat{e}_y\vert \nonumber\\
&=&\sqrt{2}R( 1-\cos\theta_i \cos\theta_j -\sin\theta_i \sin\theta_j +\delta)^{1/2},
\end{eqnarray}
donde $\delta$ es un parámetro de suavizamiento introducido comúnmente \cite{TATEKAWA} para evitar la divergencia del potencial a distancias cortas. Si consideramos la identidad
\begin{eqnarray}
\cos(\theta_i-\theta_j)=\cos\theta_i \cos\theta_j+\sin\theta_i \sin\theta_j,
\end{eqnarray}
entonces,
\begin{eqnarray}
r^{-3}&\approx & (2R^2)^{-3/2}(1-\cos(\theta_i-\theta_j)+\delta)^{-3/2} \nonumber\\
&\approx & (2R^2\delta)^{-3/2}\left(1-\frac{\cos(\theta_i-\theta_j)}{\delta}+\frac{1}{\delta}\right)^{-3/2}.
\end{eqnarray}
Por medio de una expansión binomial en $\frac{1}{\delta}$, la interacción entre dipolos puede ser escrita como,
\begin{equation}
U\!\approx\!  \frac{-\mu^2}{4\pi\epsilon_0(2R\delta)^{3/2}}\left(\!\frac{}{}\!3\cos\theta_i \cos\theta_j\! -\!\cos(\theta_i\! -\!\theta_j)\right) \left(1\!-\!\frac{3}{2} \frac{\cos(\theta_i\!-\!\theta_j)}{\delta}\!+\!O(\delta ^{-2})\right).
\end{equation}
Tomando el límite para $\delta\rightarrow\infty$, la aproximación de orden cero de la energía potencial de $N$ dipolos es,
\begin{equation}
U\approx \frac{\lambda}{2N} \displaystyle \sum_{i\neq j}^{N}(\cos(\theta_i -\theta_j)-3\cos\theta_i \cos\theta_j),
\end{equation}
donde $\lambda$ es una constante que da las unidades de energía y su signo define el tipo de interacción, si $\lambda>0$ el sistema es ferromagnético, y si es negativa anti-ferromagnético. El factor $1/N$ garantiza la extensividad del sistema como se mostró anteriormente en la sección \ref{LRI} para el modelo de Curie-Weiss.

Para modelar este sistema con interacciones de largo alcance consideramos el sistema de $N$ dipolos idénticos con masa igual a uno. Entonces el hamiltoniano del sistema a estudiar en aproximación de campo medio es,
\begin{equation}\label{hamiltoniano}
 H \!= \!\!\sum_{i=1}^{N}\!\frac{p_i^2}{2}\!+\! \frac{\lambda}{2N} \!\sum_{i\neq j}^N [\cos(\theta_i\!-\!\theta_j)\!-\!3\cos\theta_i \cos\theta_j\!+\!2],
\end{equation}
donde $p_i$ son los momentos lineales de cada dipolo $i$ y $\theta_i$ es su correspondiente orientación, con $i= 1,...,N$. El término $+2$ en el potencial es introducido para fijar el nivel cero de energía potencial convenientemente.

Al ser un modelo de campo medio no hay dependencia de la distancia al igual que el modelo de Curie-Weiss.

\section{Simulaciones de dinámica molecular}\label{DMdhmf}

Atenas y Curilef \cite{ATENAS3}, mostraron a través de la Dinámica Molecular, que bajo ciertas condiciones iniciales  (water-bag inhomogéneas), el modelo d-HMF presenta dos QSS. En ellos se muestra la evolución temporal de la energía cinética promedio por partícula, que en el equilibrio corresponde a la temperatura. Estos estados aparecen sólo en regiones cercanas al punto crítico evidenciando que en los sistemas con transiciones de fase continua también existen estados con capacidades caloríficas negativas, como se mostró para el modelo HMF \cite{PLUCHINO}.
En este caso los dos QSS encontrados fueron descritos mediante diversas técnicas, entre ellas, instantáneas del espacio de fases, perfil de velocidades, leyes de escalamiento y análisis difusivo.
La simulación fue hecha con una rutina de coeficientes simplécticos de cuarto orden \cite{PLUCHINO, RUTH, YOSHIDA} (ver apéndice \ref{app2}). Se utilizaron las condiciones iniciales water-bag usadas en el modelo HMF \cite{PLUCHINO}, los casos homogéneo e inhomogéneo mostrados en la Fig. \ref{WBIC}. El caso inhomogéneo es cuando todos los dipolos están orientados casi paralelos, es decir, $\theta_{0i}\approx0$. El caso homogéneo es cuando tanto en las orientaciones como en las velocidades hay una distribución homogénea.

En la Fig. \ref{magtvsu} \emph{a}) se muestra la magnetización \textit{versus} la energía interna del sistema, siendo su valor crítico en $\varepsilon_c=3/2$. En \emph{b}) se muestra la curva calórica, donde están superpuestos los datos de la simulación con el cálculo analítico mostrado en la sección anterior. De color azul están los datos de la simulación tomados en el equilibrio BG. De color rojo los datos del segundo QSS, el cual produce capacidades caloríficas negativas en la zona cercana al punto crítico.

\begin{figure}
\centering
  \subfloat{
    \includegraphics[width=0.45\textwidth]{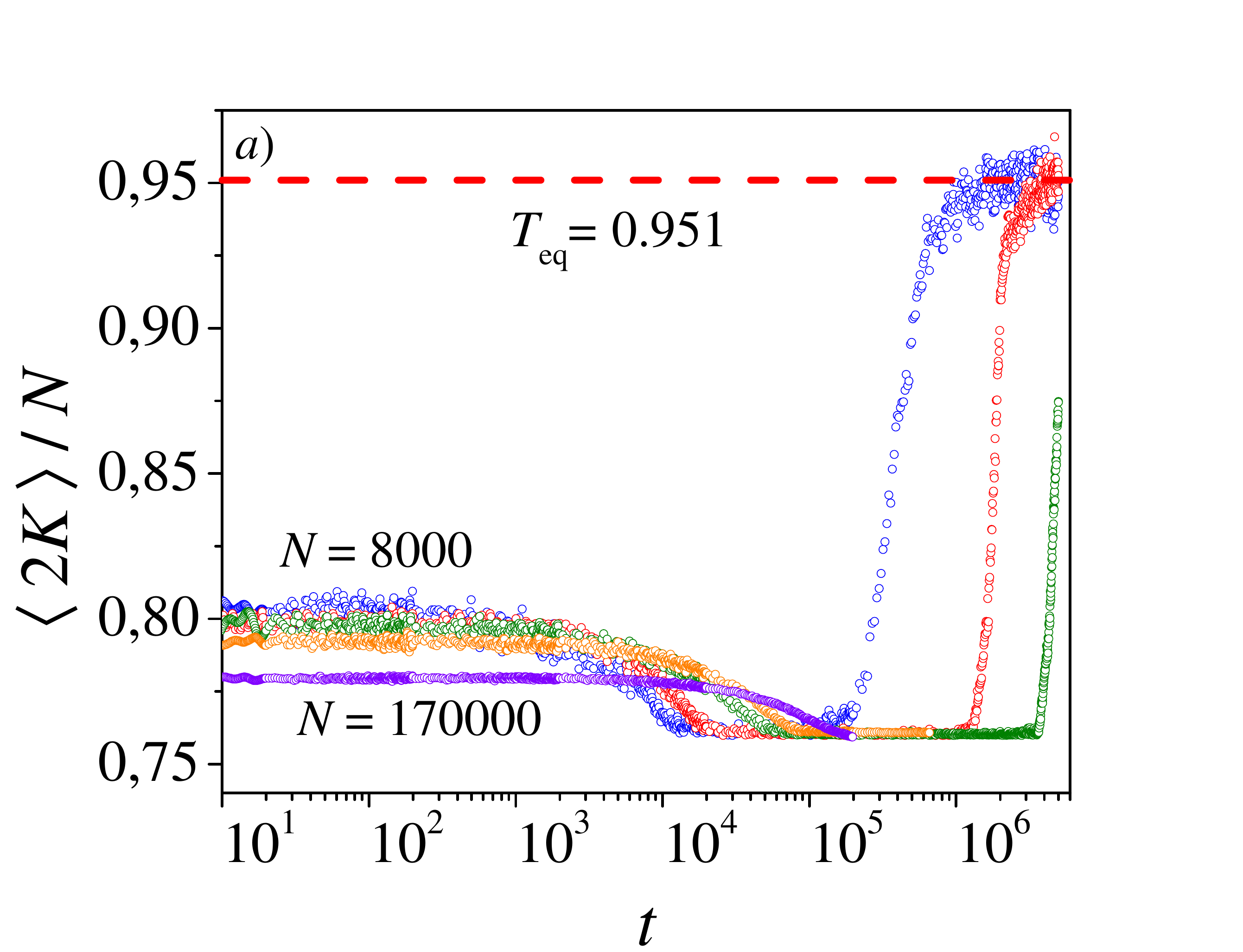}}
  \subfloat{
    \includegraphics[width=0.45\textwidth]{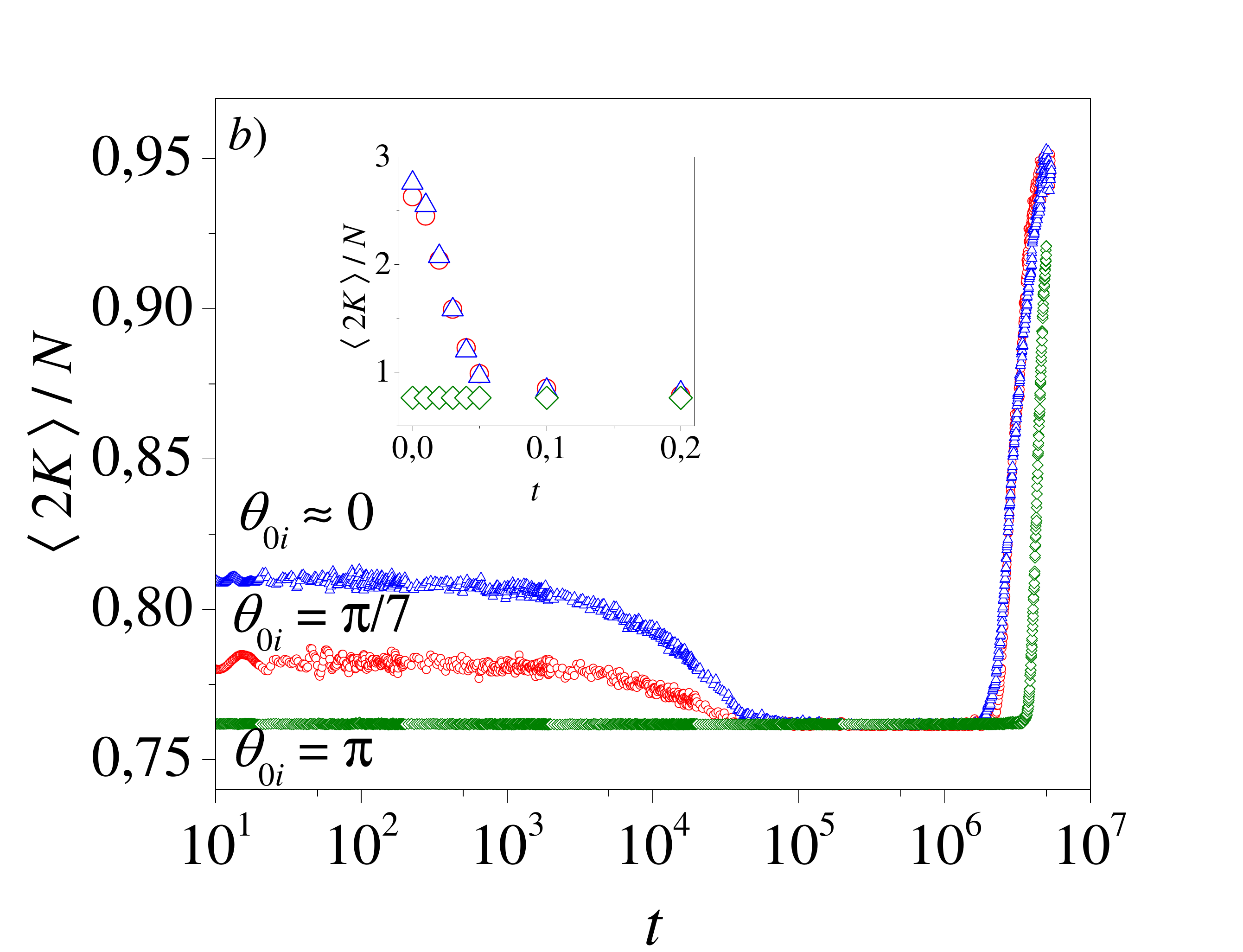}}
\caption{En \emph{a}) se muestra la evolución temporal de la energía cinética promedio por partícula para diferentes valores de $N$ con las condiciones iniciales water-bag inhomogéneas. Se aprecian dos QSS previos al equilibrio de BG. En \emph{b}) se muestra la evolución temporal de la energía cinética promedio por partícula para diferentes water-bag con $N=25000$. Se observa una desaparición del primer QSS al pasar del caso inhomogéneo al caso homogéneo. En el panel pequeño de la derecha se evidencia la relajación violenta del sistema al inicio de la simulación Ref. \cite{ATENAS3,ATENAS4}.}\label{tvst}
\end{figure}

\begin{figure}
    \centering
  \includegraphics[width=0.5\textwidth]{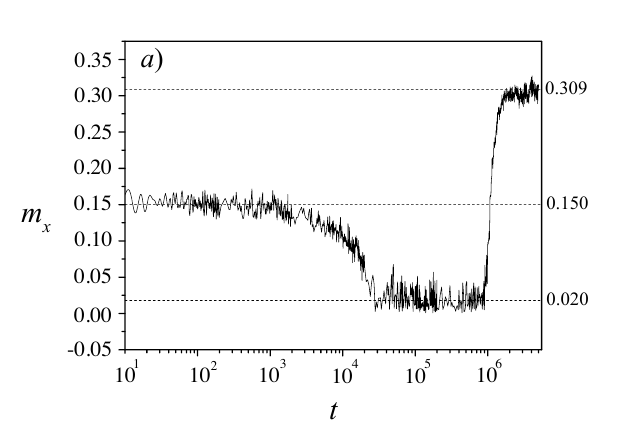}\includegraphics[width=0.445\textwidth]{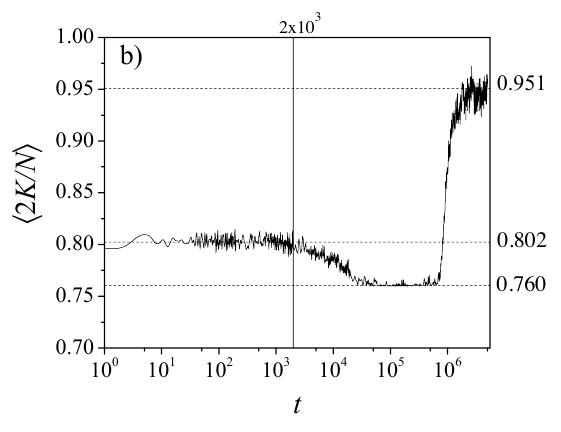}
   \caption{En $a)$ el valor de magnetización observado para el primer QSS es $m_x=0$.$15$, para el segundo es $m_x=0$.$020\approx0$, y para el equilibrio $m_x=0$.$309$. En $b)$, la evolución de la energía cinética promedio, el valor observado para el primer QSS es $\langle 2K/N\rangle=0$.$802$, para el segundo $\langle 2K/N\rangle=0$.$706$ y para el equilibrio $\langle 2K/N\rangle=0$.$951$.}\label{Fig2}
\end{figure}

En la Fig. \ref{tvst} \emph{a}) se muestra la evolución temporal de la energía cinética promedio por partícula para diferentes valores de $N$ con las condiciones iniciales water-bag $\theta_{0k}\approx0$, es decir el caso inhomogéneo (ver Fig. \ref{WBIC})\footnote{En el apéndice \ref{ApendiceB} se encuentra un desarrollo detallado para el cálculo de las condiciones iniciales water-bag  asociadas a los estados mostrados en la Fig. \ref{tvst}}. Se observa que a medida que nos acercamos al caso homogéneo el primer QSS se confunde con el segundo. Esto nos dice que la naturaleza de los QSS es mucho más rica de lo que se pensaba, pues bajo ciertas condiciones iniciales el sistema puede pasar por diferentes QSS. Este es el primer modelo que presenta más de un QSS. Tal característica ha sido evidenciada en sistemas atmosféricos \cite{MONAHAN, ITOH}.

Un estudio más detallado de los QSS se muestra en la Fig. \ref{Fig2}. En el panel izquierdo se muestra la evolución de la magnetización, donde se observa que el primer QSS tiene una magnetización distinta de cero. El segundo QSS tiene una magnetización cero, lo que implica, que en promedio las fuerzas se anulan, esto se ve claramente en la ec. (\ref{eqmot}). Los motivos por los que el primer QSS presenta magnetización no nula aún son desconocidos, sin embargo se sospecha que puede deberse a un equilibrio hidrostático.

En \cite{DAUXOIS} se discute acerca del caos que se produce bajo ciertas condiciones iniciales. En ella se menciona que el modelo HMF es caótico bajo un water-bag inhomogéneo.

Para el modelo HMF, se ha reportado que bajo un water-bag inhomogéneo el sistema es caótico, sin embargo en el modelo d-HMF, vemos que esto no ocurre, pues al aumentar el número de partículas la duración del primer QSS se extiende como se aprecia en el panel \emph{a}) de la Fig. \ref{tvst}.

\begin{figure}
\centering
  \subfloat{
    \includegraphics[width=0.45\textwidth]{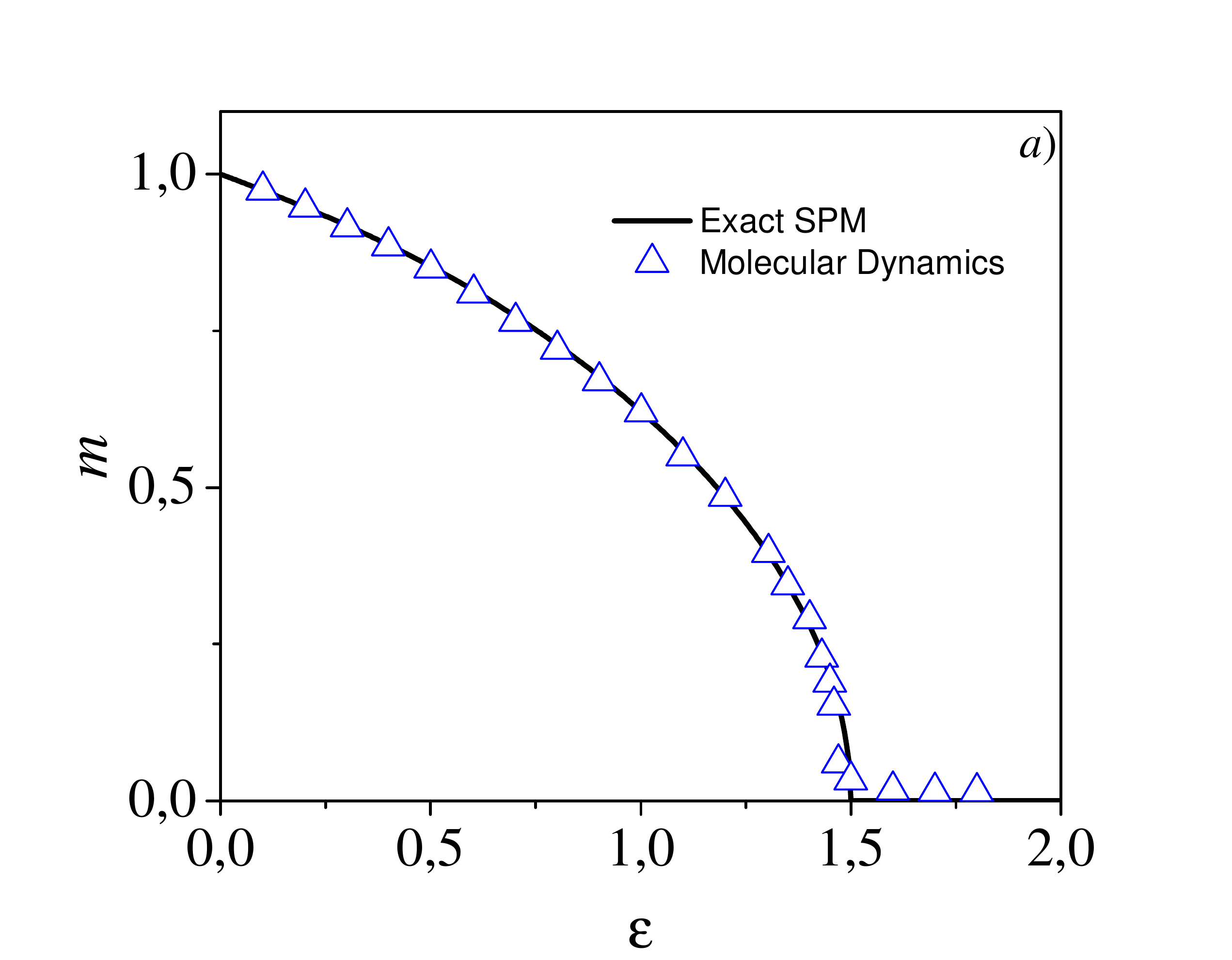}}
  \subfloat{
    \includegraphics[width=0.45\textwidth]{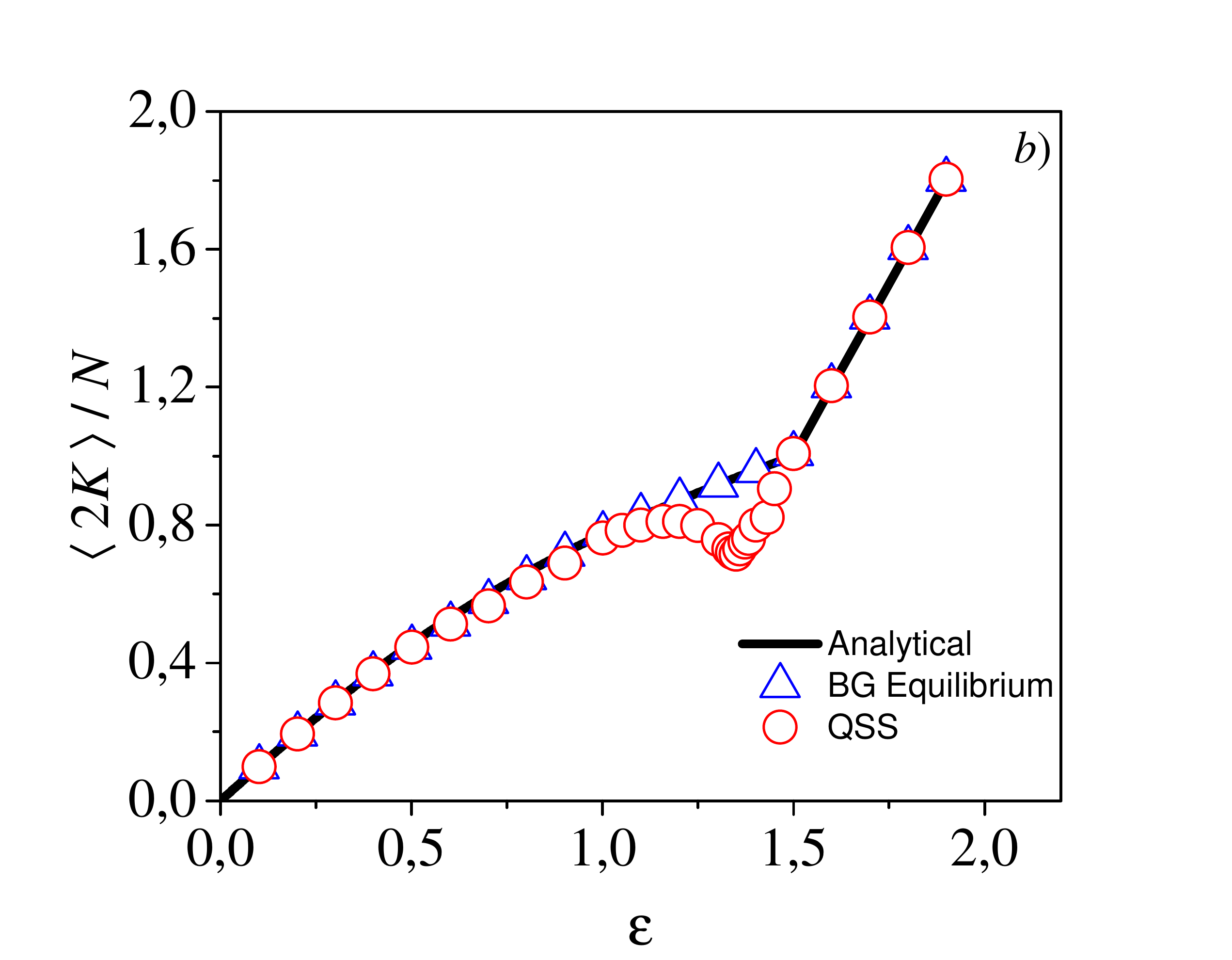}}
\caption{En el panel izquierdo, se muestra la magnetización con los datos del equilibrio BG de la simulación de color azul. De color Negro la curva analítica. En el panel derecho, se muestra la curva calórica. De color negro la solución analítica, de color azul, los datos del equilibrio BG de la simulación y de color rojo, los datos encontrados para el segundo QSS.}\label{magtvsu}
\end{figure}

En la Fig. \ref{magtvsu} \emph{a}) se muestra la magnetización obtenida por dinámica molecular de color azul, en contraste con la solución analítica de color negro. Los valores de la simulación se hicieron con $N=4000$. En el panel \emph{b}) se muestra la energía cinética promedio \textit{versus} la energía interna por partícula del sistema. De color azul se muestran los datos del equilibrio BG, de color negro la solución analítica, y de color rojo los valores obtenidos del segundo QSS.

\begin{figure}
\centering
  \subfloat{
    \includegraphics[width=0.45\textwidth]{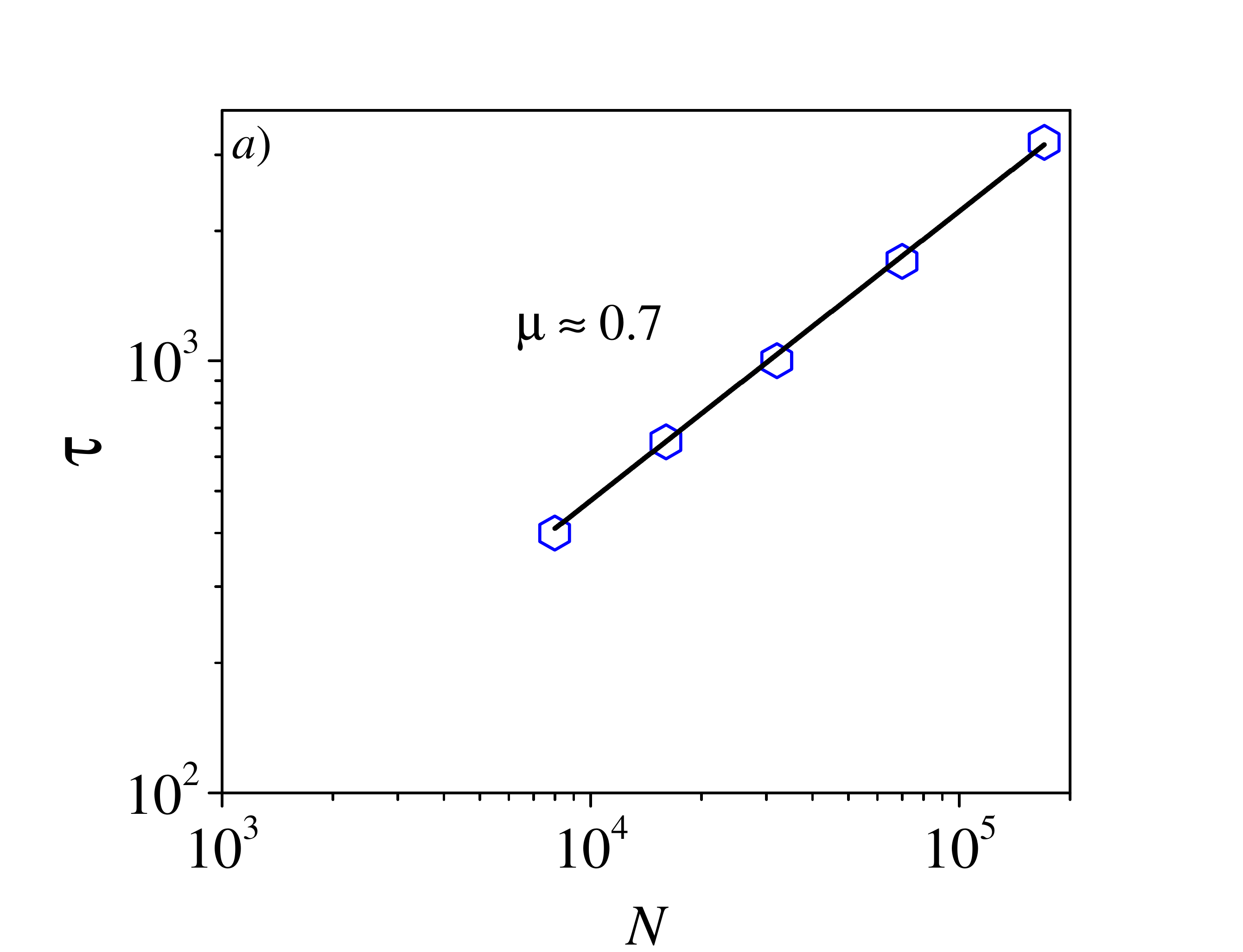}}
  \subfloat{
    \includegraphics[width=0.45\textwidth]{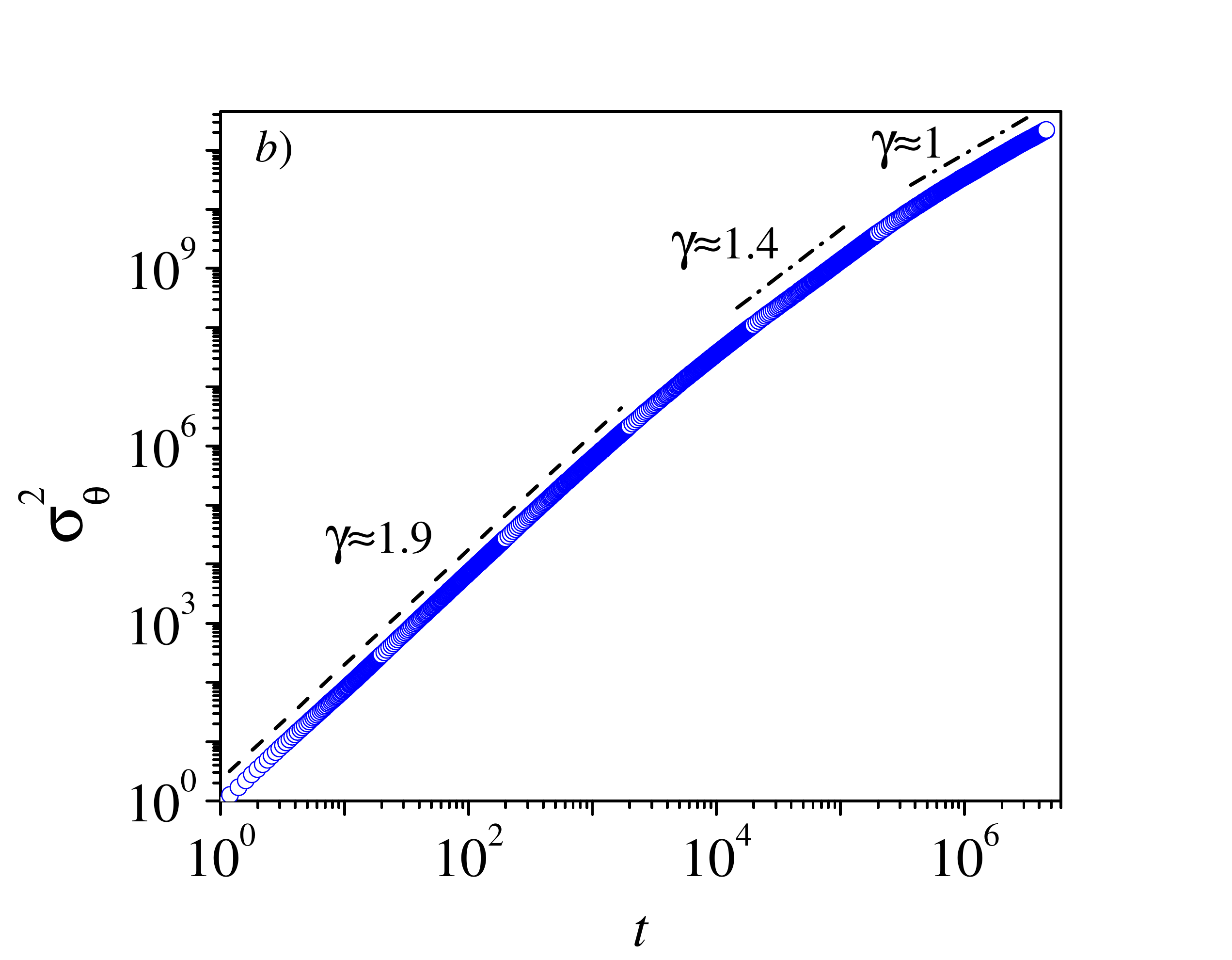}}
\caption{En \emph{a}) una ley de escala ($\tau\propto N^{\gamma}$) para el primer QSS. en \emph{b}) la ley difusiva con los tres regímenes, los primeros corresponden a los dos QSS y el último es el equilibrio BG.}\label{escaladim}
\end{figure}

En la Fig. \ref{escaladim} \emph{a}) se muestra la ley de escala $\tau\propto N^{\gamma}$ obtenida para la duración del primer QSS. En el panel \emph{b}) la ley difusiva que caracteriza los dos regímenes cuasi-estacionarios y el equilibrio BG. Los primeros dos estados son superdifusivos con valores de $\gamma>1$, mientras que en el equilibrio BG $\gamma=1$ lo que nos dice que efectivamente el sistema se ha difundido por completo.

\begin{figure}
\centering
  \subfloat{
    \includegraphics[width=0.45\textwidth]{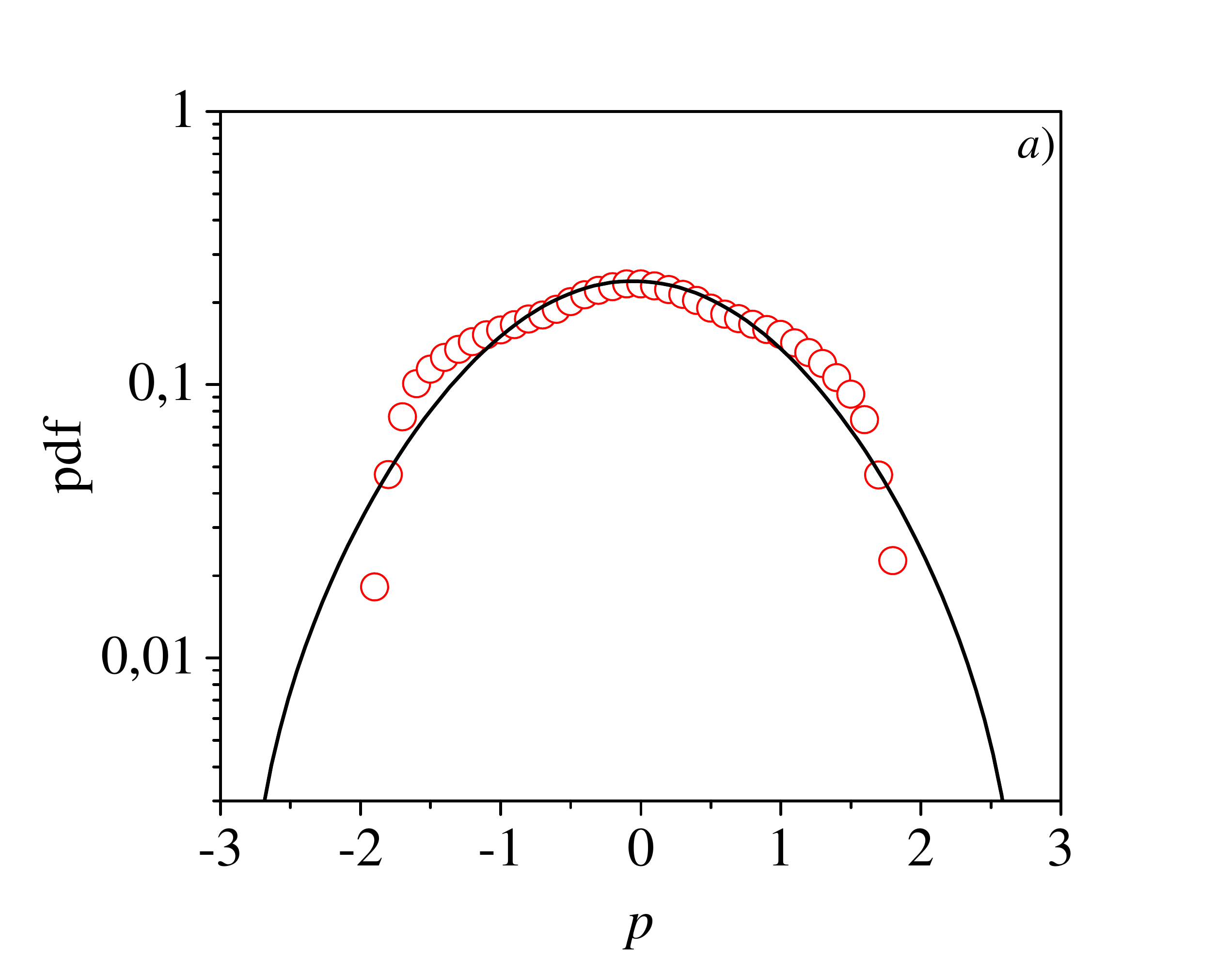}}
  \subfloat{
    \includegraphics[width=0.45\textwidth]{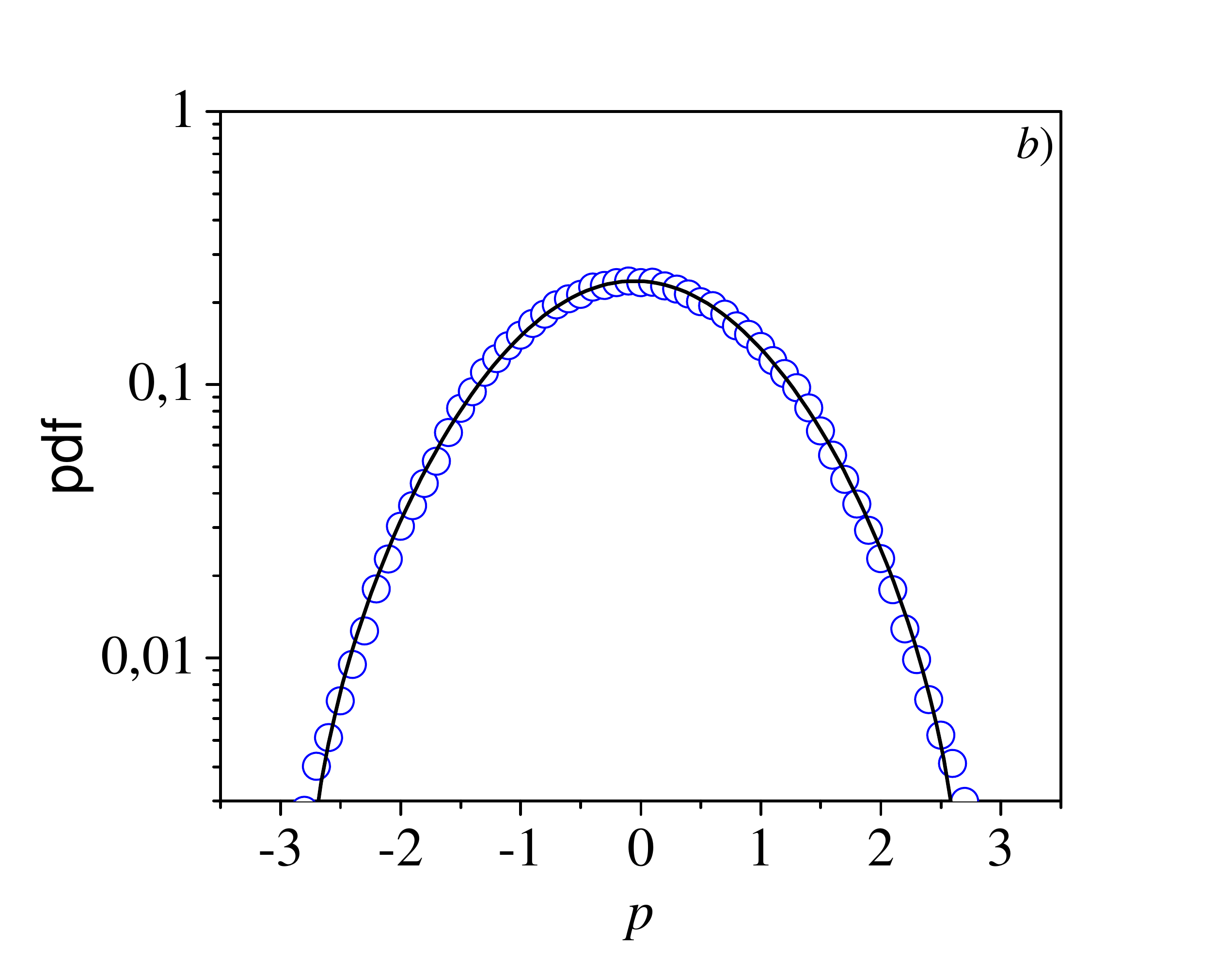}}
\caption{En el panel \emph{a}) se muestra el perfil de distribución (pdf) de momentum $p$ en el segundo QSS. Se puede apreciar que no es un perfil gaussiano. En el panel \emph{b)} el perfil de distribución del equilibrio BG claramente gaussiano. Estos datos fueron tomados para $N=500$, con energía $\varepsilon=1$.$38$.}\label{gaussdim}
\end{figure}

\begin{figure}
\centering
    \includegraphics[width=0.6\textwidth]{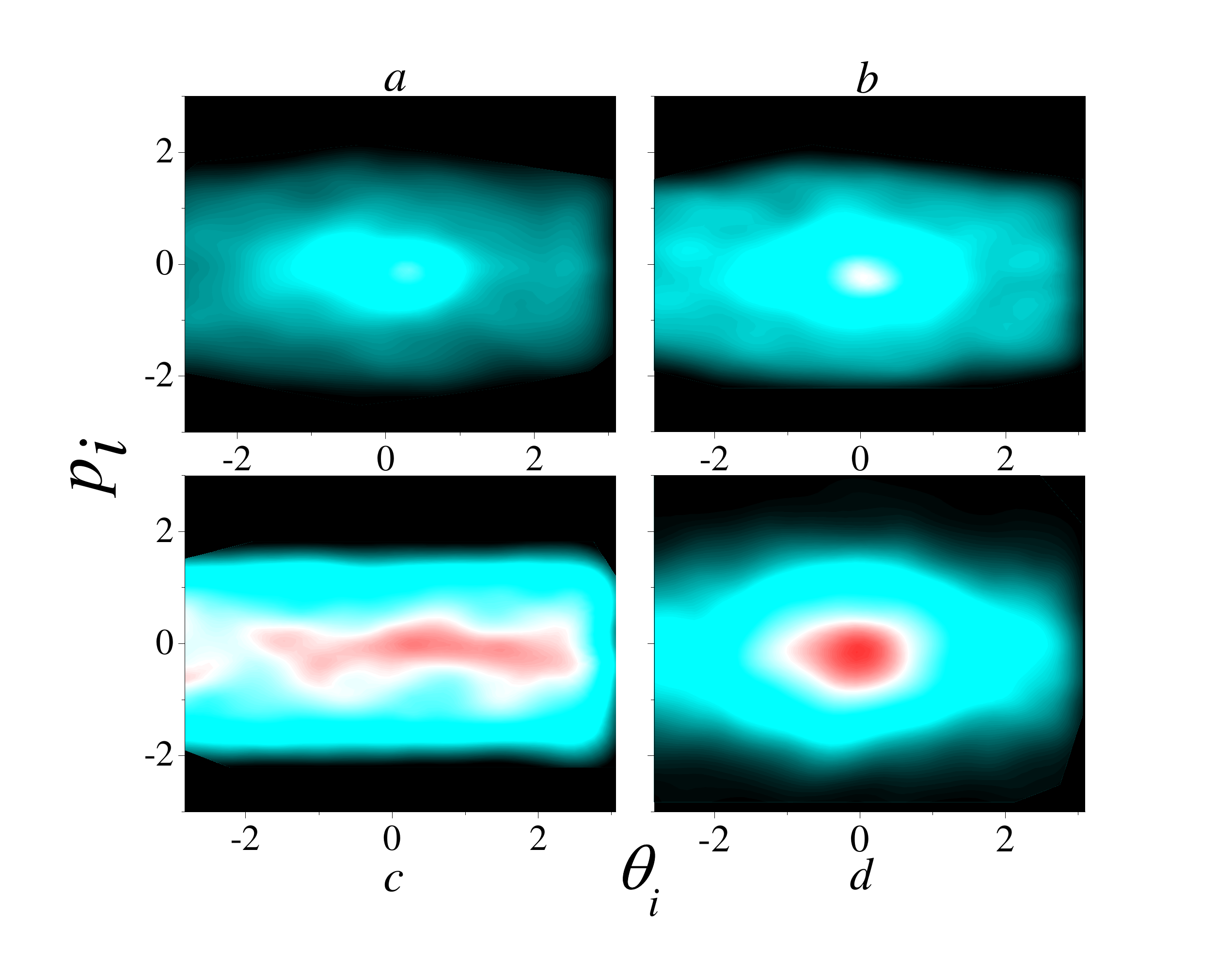}
\caption{Espacio de fases del modelo d-HMF. Cada instantánea es una representación del espacio de fases en los tiempos $t=20s$, $t=200s$, $t=6\times10^5s$ y $t=5\times10^6s$.}\label{snapdim}
\end{figure}

\begin{figure}
\centering
\includegraphics[width=0.45\textwidth]{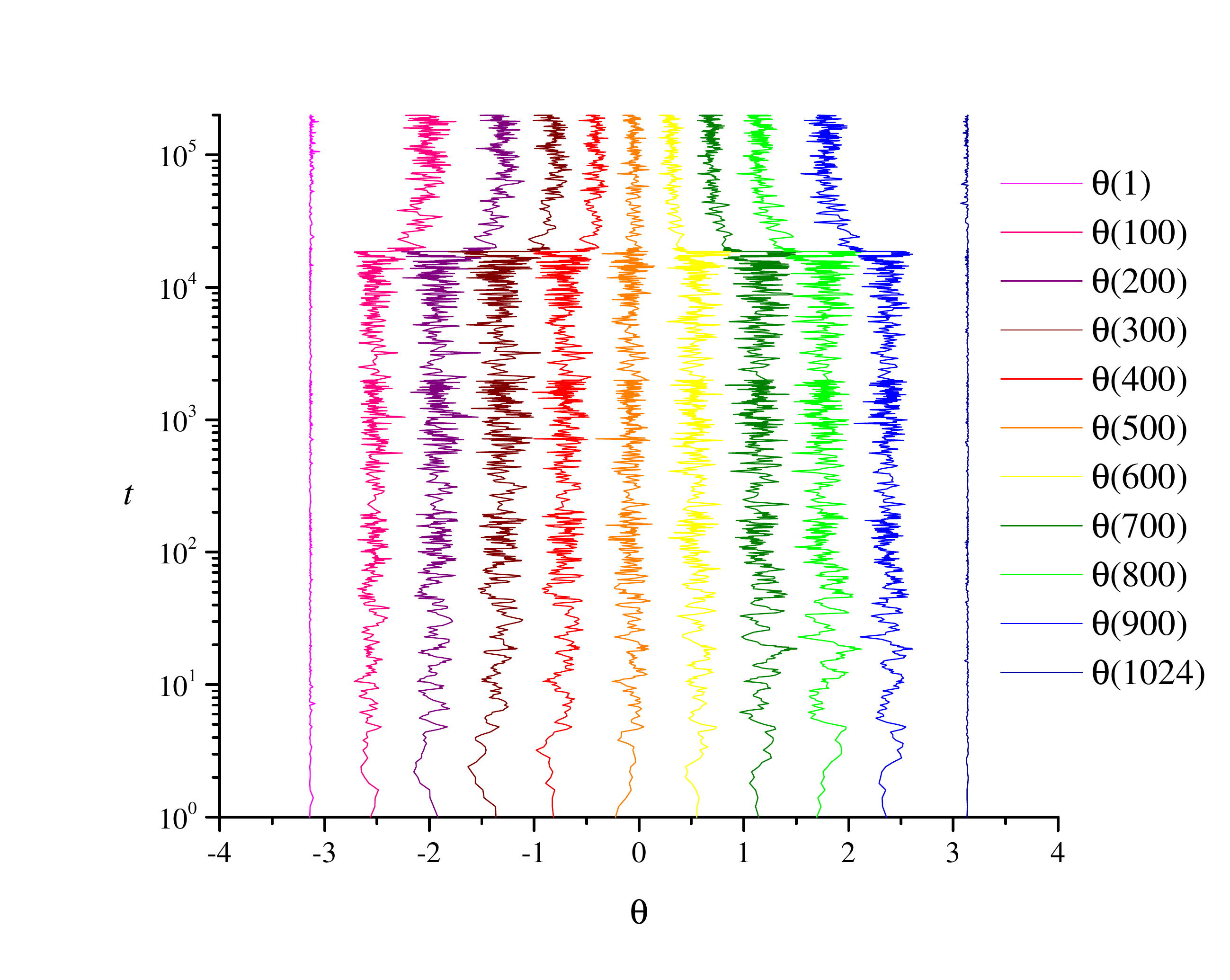}\includegraphics[width=0.45\textwidth]{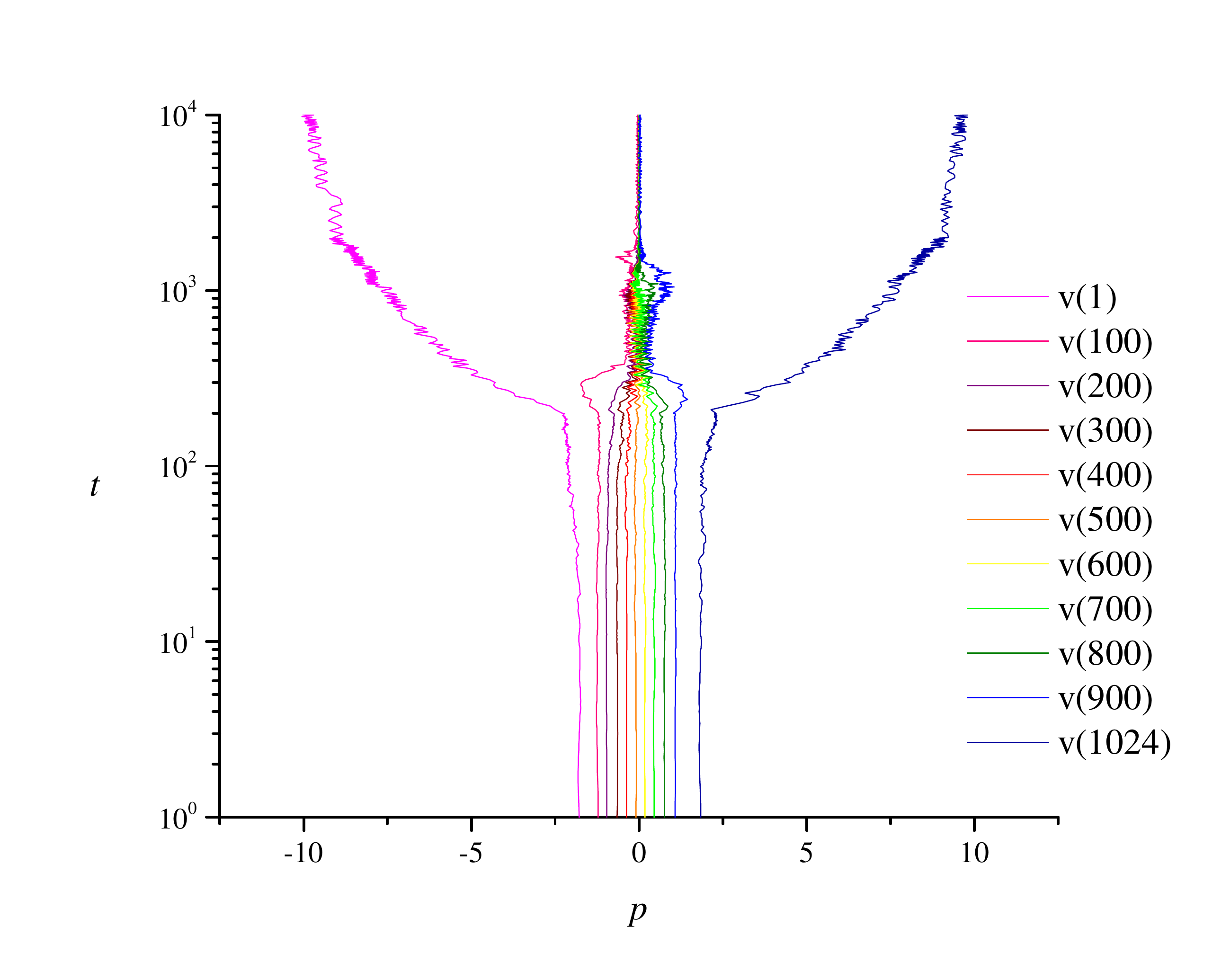}
    
\includegraphics[width=0.45\textwidth]{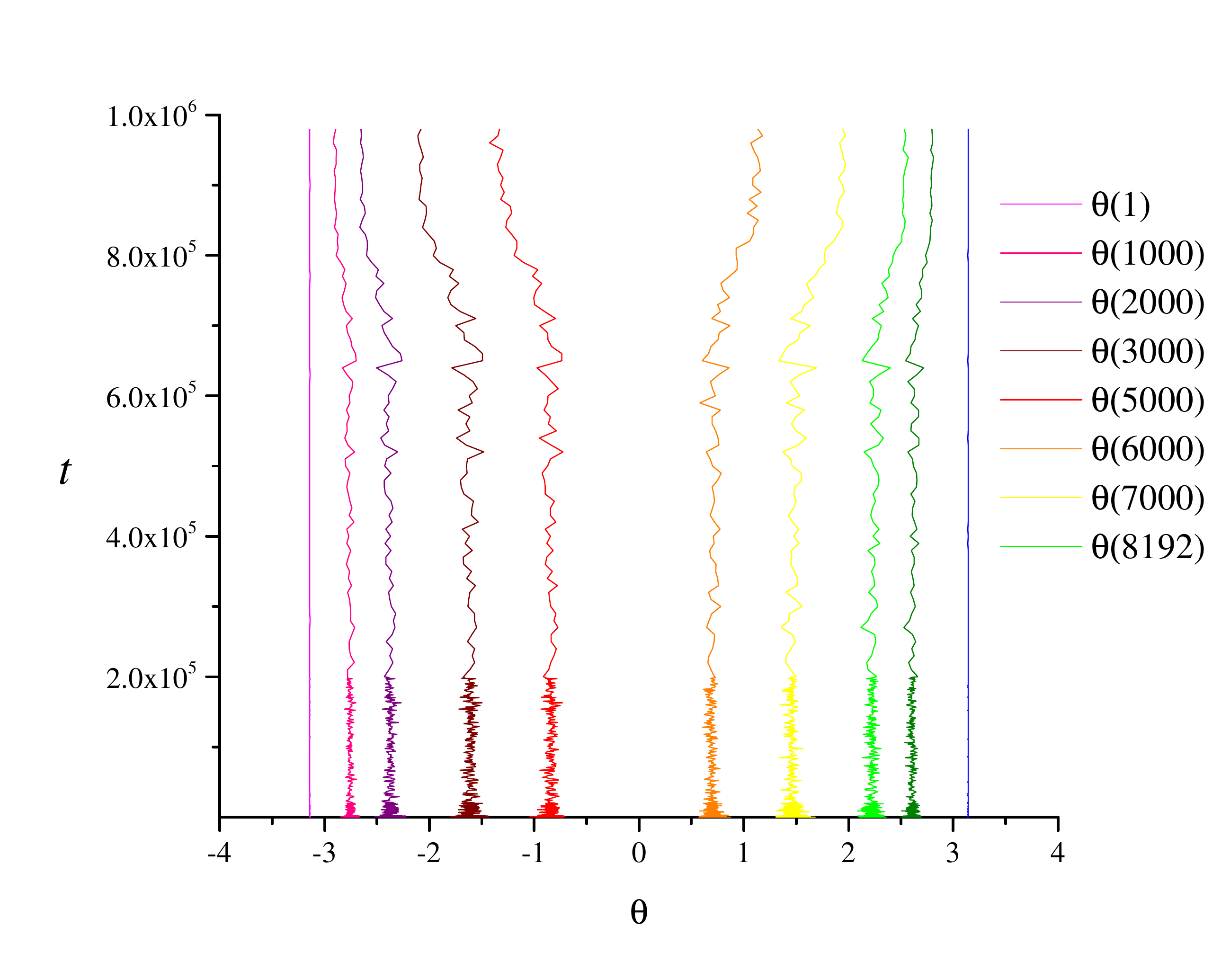} 
\includegraphics[width=0.45\textwidth]{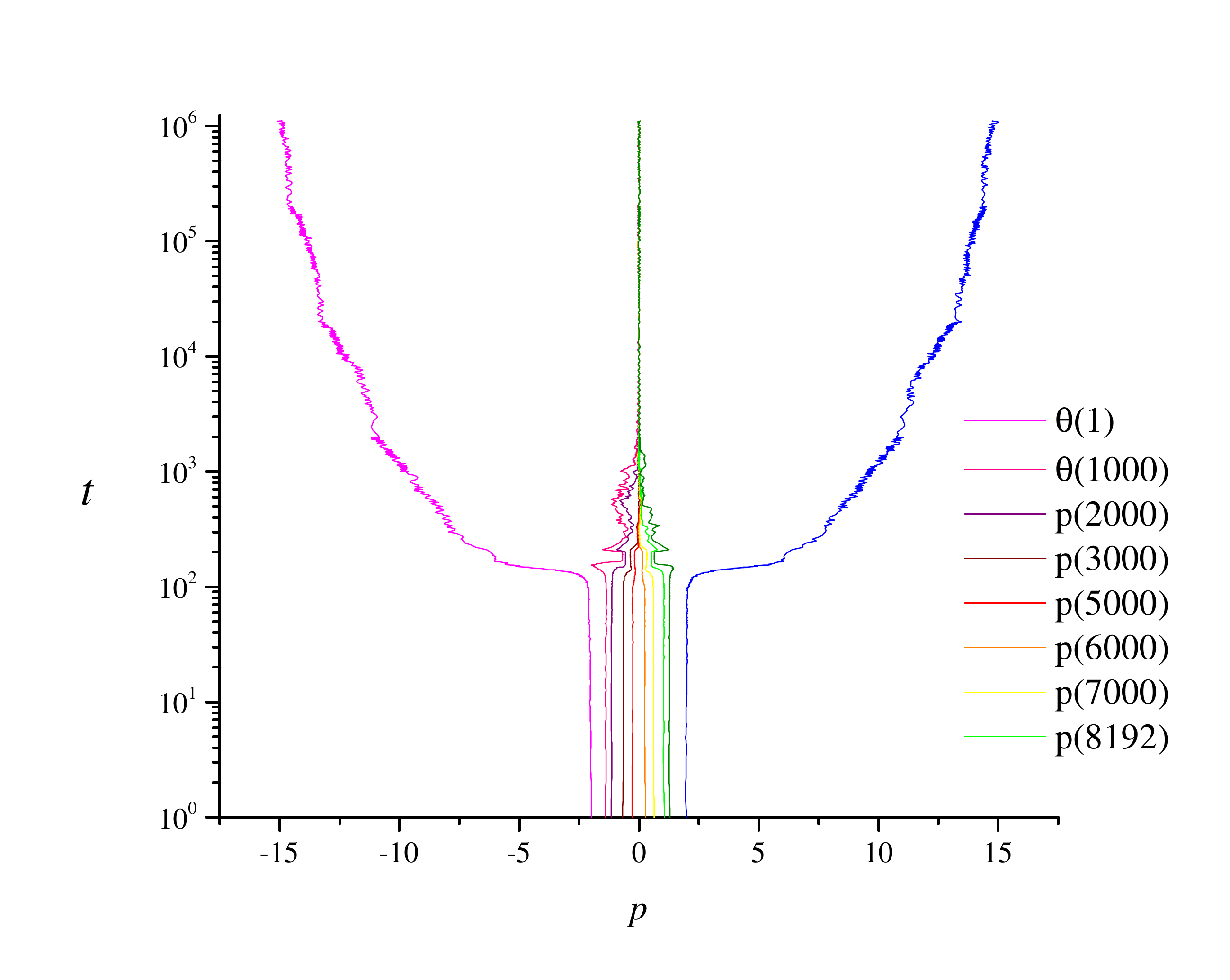}
\caption{Seguimiento de las orientaciones y velocidades de algunas partículas. En la parte superior para 1024 partículas en la parte inferior para 8192.}\label{timeEv}
\end{figure}

En la Fig. \ref{snapdim} se muestran instantáneas del espacio de fases. En el panel \emph{a}), con $t=20s$, se muestra al sistema poco tiempo después de iniciar la dinámica, como se puede apreciar rápidamente el sistema se relaja desde las condiciones iniciales water-bag inhomogéneas a una distribución similar a una gaussiana. En \emph{b}) con $t=200s$, el tiempo en el que ocurre el primer QSS, en \emph{c}), con $t=6\times10^5s$, el tiempo en el que sistema se encuentra en el segundo QSS y finalmente en \emph{d}) con $t=5\times10^6s$, cuando el sistema ha llegado al equilibrio BG, donde se ve claramente las elipses características del perfil gaussiano. La corrida fue hecha para $N=8000$ y $\varepsilon=1.38$. En la Fig. \ref{timeEv}, se muestra la evolución temporal de las orientaciones y los momentos. En la parte superior para 1024 partículas, y en la inferior para 8192 con condiciones iniciales inhomogéneas. En la parte superior, se corta el tiempo hasta el segundo QSS. En la parte inferior, se deja evolucionar hasta el equilibrio, como se puede observar la forma de las distribuciones en el equilibrio y los QSS son consistentes con las mostradas en la Fig. \ref{snapdim}.

\newpage
Los resultados realizados al modelo d-HMF, han revelado una rica fenomenología en cuanto a los estados QSS que este presenta \cite{ATENAS4}. Sin embargo, la información obtenida mediante la dinámica molecular nos lleva a hacer un segundo estudio mediante la ecuación cinética de Vlasov para descartar los efectos de tamaño finito que pueden estar relacionados con la aparición del primer QSS. 

\chapter{Soluciones analíticas del modelo d-HMF}\label{cap2}

\section{Cálculos en el ensamble canónico}\label{canonico}

En el ensamble canónico estamos interesados en el cálculo de la función de partición del sistema. A partir de ésta, es posible calcular la energía libre por partícula, energía interna y la curva calórica, entre otras cantidades termodinámicas \cite{ATENAS3,ATENAS5}.

La parte interactuante de la ec.( \ref{hamiltoniano}), es comúnmente expresada en términos del vector de espín $\overrightarrow{m}_i=(\cos\theta_i,\sin\theta_i)$.
Luego podemos introducir el vector de espín total,
\begin{eqnarray}\label{totalspin}
\overrightarrow{M} &=& \frac{1}{N}\sum_{i=1}^N\overrightarrow{m}_i \label{defM} \\
&=&(M_x,M_y) \label{defMxy} \\
&=& m\exp(i\phi),
\end{eqnarray}
donde $(M_x,M_y)$ y $m$ son las componentes y el módulo del vector $\overrightarrow{M}$, respectivamente, y $\phi$ denota la fase del parámetro de orden. Luego, las ecuaciones de movimiento son,
\begin{equation}\label{eqmot}
\dot{p}_i=-\lambda\left(2M_x\sin\theta_i +M_y\cos\theta_i\right)
\end{equation}
y la energía potencial puede ser escrita como,
\begin{equation}
U = - N\frac{\lambda}{2} \left(2M_x^2-M_y^2-2\right). \label{epot}
\end{equation}
Como se mencionó anteriormente esta definición es crucial en la definición del nivel de energía. En consecuencia, cuando $t=0$ la energía potencial es cero y la energía cinética es máxima y coincide con el valor de energía total. 

\begin{figure}
 \centering
  \subfloat{
    \includegraphics[width=0.4\textwidth]{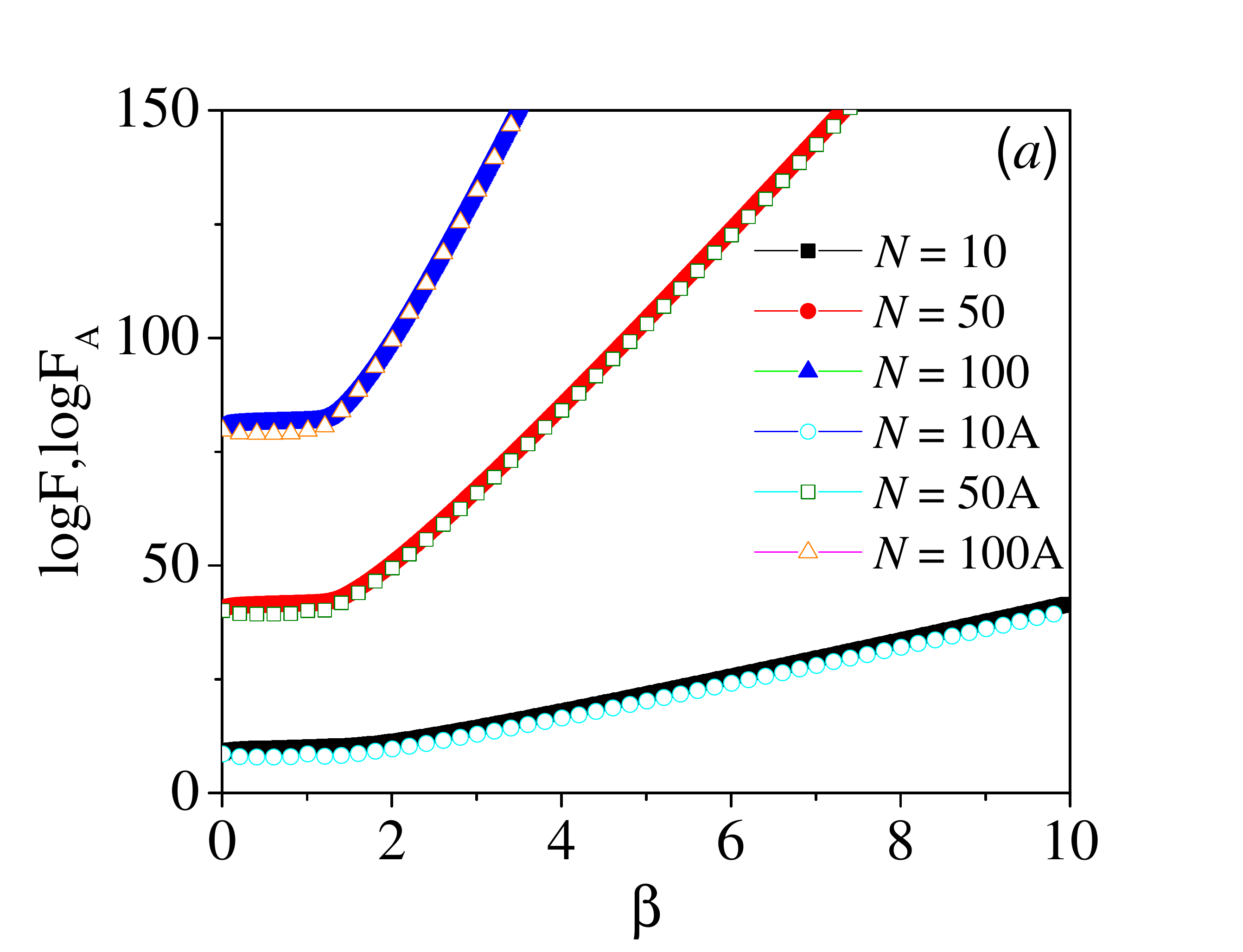}}
  \subfloat{
    \includegraphics[width=0.4\textwidth]{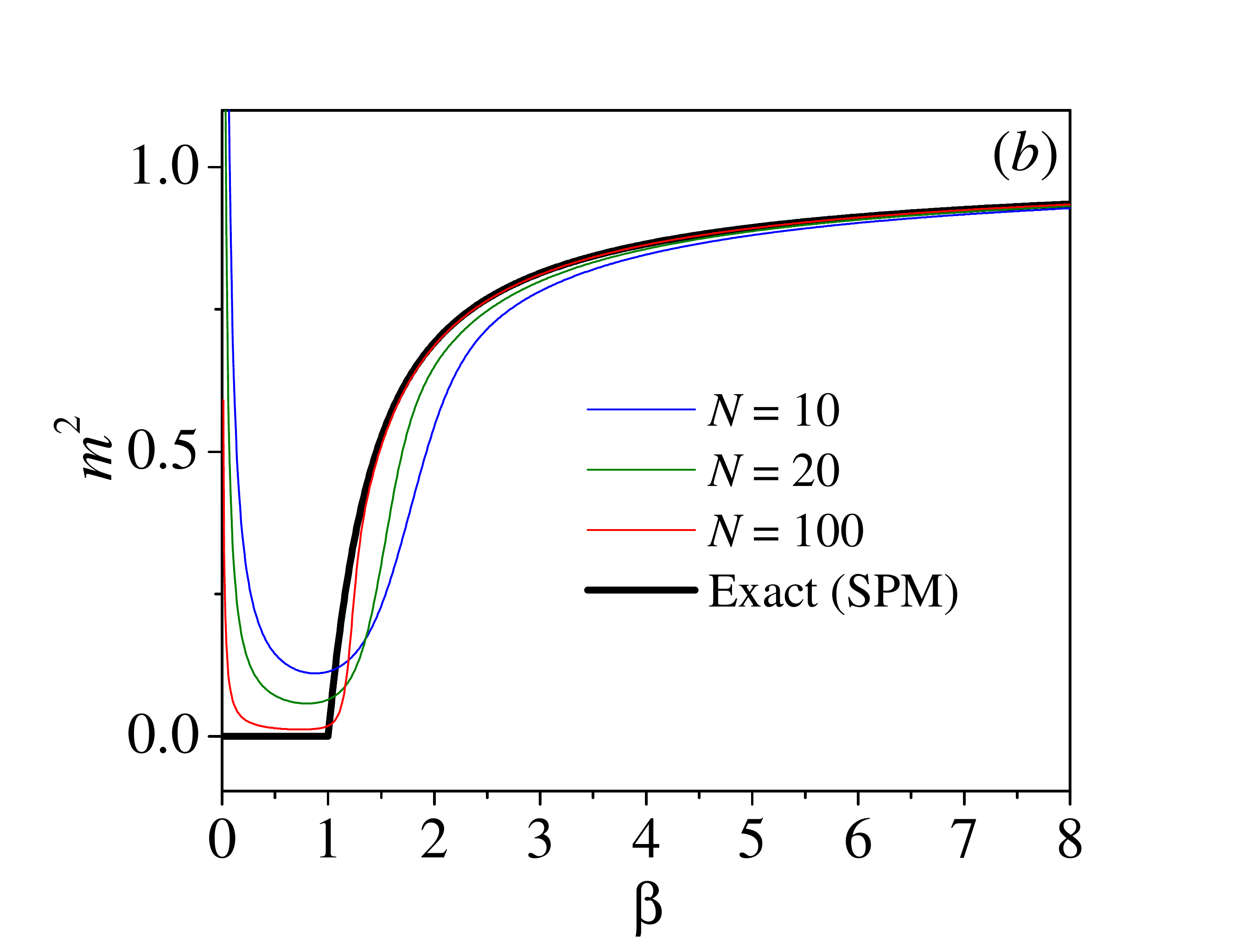}}
\caption{La figura muestra las funciones; exacta ($F_N(\beta \lambda)$) ($F$ en la gráfica izquierda) y aproximada ${\cal F}(N,\beta \lambda)$ ($F_A$ en la gráfica izquierda) por el método de punto silla (SPM), las cuales son cercanas a medida que $N$ crece. En ({\emph{a}}) ${\cal F}_N(\beta \lambda)$  se compara con ${\cal F}(N,\beta \lambda)$ para varios valores de $N$. En ({\emph{b}}), $dX(N,\beta \lambda)/d\beta$ para diferentes valores de $N$ es comparada con $dX_{N\rightarrow\infty}(\beta \lambda)/d\beta$ (exacta), la cual coincide con el cuadrado de la magnetización como función de la temperatura inversa $m^2(\beta \lambda)$. }\label{approx}
\end{figure}

La función partición puede ser expresada como sigue,
\begin{eqnarray}
Z(\beta,N)\!&=&\!\!\int \texttt{d}^{\!N}\!p_i\, \int  \texttt{d}^{\!N}\!\theta_i e^{-\beta H}\\ &=&Z_K(\beta,N)Z_U(\beta,N),
\end{eqnarray}
donde $Z_K(\beta,N)$ es la parte cinética y $Z_U(\beta,N)$ la parte interactuante.
La parte cinética bien conocida \cite{PATHRIA} y equivale a,
\begin{eqnarray}
Z_K(\beta,N)&=&\!\int \texttt{d}^N\!p_i  \exp\left(\!-\frac{\beta}{2} \sum_i p_i^2\right)\\&=&\left(\frac{2\pi}{\beta}\right)^{N/2}.
\end{eqnarray}
Por otro lado, la parte interactuante puede ser expresada como,
\begin{eqnarray}
Z_U(\beta,N)=\int\mathrm{d}^N\theta_i \mathrm{exp}\left(\beta N\frac{\lambda}{2}(2M^2_x-M^2_y-2)\right),
\end{eqnarray}
en la que se puede notar que se tienen dos integrales gaussianas,
\begin{eqnarray}
Z_U(\beta,N)= e^{-\beta N\lambda} \int\mathrm{d}^N\theta_i e^{\beta N{\lambda}M^2_x} e^{-\beta N\frac{\lambda}{2}M^2_y}.\label{zv}
\end{eqnarray}
Estas integrales gaussianas pueden ser reescritas mediante las transformaciones de Hubbard-Stratonovich \cite{STRATONOVICH, HUBBARD},
\begin{eqnarray}
\sqrt{\frac{\pi}{b}}&=&\int^{\infty}_{-\infty} \mathrm{d}x\: \mathrm{exp}\left(-b(x-M_x)^2\right)\\
\mathrm{exp}\left(b M^2_x\right)&=&\sqrt{\frac{b}{\pi}}\int^{\infty}_{-\infty} \mathrm{d}x\: \mathrm{exp}\left(-bx^2+2bM_xx\right)\label{emx}
\end{eqnarray}
y
\begin{eqnarray}
\sqrt{2 \pi b}&=&\int^{\infty}_{-\infty} \mathrm{d}y\: \mathrm{exp}\left(-\frac{1}{2}\left(\frac{y}{\sqrt{b}}+i\sqrt{b}{M_y}\right)^2\right)\\
\mathrm{exp}\left(-\frac{b M^2_y}{2}\right)&=&\sqrt{\frac{1}{2\pi b}}\int^{\infty}_{-\infty} \mathrm{d}y \: \mathrm{exp}\left(-\frac{y^2}{2b}-iM_y y\right),\label{emy}
\end{eqnarray}
entonces, aplicando al problema con $b=\beta \lambda N$. En el equilibrio, se espera una distribución simétrica de las orientaciones $\rho(\theta)$; luego para $N$ grande tenemos que
\begin{equation}
M_y=\lim_{N\rightarrow\infty}\frac{\sum_i \sin\theta_i}{N} \approx \int_{0}^{2\pi} \mbox{d}\theta \sin \theta \rho(\theta)=\langle \sin\theta\rangle=0,
\end{equation}
entonces
\begin{equation}
\exp\left(\beta \lambda N M^2_x\right)= \sqrt{\frac{\beta \lambda N }{\pi}}\int^{\infty}_{-\infty} \mathrm{d}x \exp{( -\beta \lambda N x^2+2\beta  \lambda  x \sum_i\cos\theta_i)},\label{emxx}
\end{equation}
y
\begin{equation}
 \exp\left(-\beta N\frac{\lambda}{2}M^2_y\right)= \frac{1}{\sqrt{2 \pi \beta \lambda N }}\int^{\infty}_{-\infty} \mathrm{d}y \exp{\left( -\frac{ y^2}{2\beta \lambda N}\right)}=1.\label{emyy}
\end{equation}
Si substituimos las ecs. (\ref{emxx}) y (\ref{emyy}) en (\ref{zv}), obtenemos
\begin{equation}
Z_U(\beta,N)=e^{-\beta \lambda N}\sqrt{\frac{1}{2\pi^2}}\int^{\infty}_{-\infty}\mathrm{d}y e^{-\frac{y^2}{2\beta \lambda N}} \int^{\infty}_{-\infty} \: \mathrm{d}x e^{-\beta \lambda N x^2 } \int\mathrm{d}^N\theta_i e^{2\beta  \lambda x \sum_i\cos\theta_i}.
\end{equation}
Luego escribimos
\begin{equation}
Z_U(\beta,N)\!=\!\sqrt{\frac{\beta \lambda N}{\pi}}e^{-\beta \lambda N}\!\!\int_{-\infty}^\infty \!\! \!\texttt{d}x\, e^{-\beta \lambda N x^2} (2\pi\textrm{I}_0(2\beta \lambda x))^N,
\end{equation}
donde $\textrm{I}_k(y)$ es la función modificada de Bessel de k-ésimo orden.
Luego la función de partición puede ser expresada como,
\begin{equation}
Z\!=\!\sqrt{\frac{\beta \lambda N}{\pi}}e^{-\beta \lambda N}\! \left(\frac{2\pi}{\beta}\right)^{\! \! \frac{N}{2}}\!\! {\cal F}_N(\beta \lambda), \label{function}
\end{equation}
donde ${\cal F}_N(\beta \lambda)$ representa la integral
\begin{equation}
{\cal F}_N(\beta \lambda)= \int_{-\infty}^\infty \!\texttt{d}x\, e^{-N(\beta \lambda x^2\!-\ln(2\pi\textrm{I}_0(2\beta \lambda x)))}. \label{intF}
\end{equation}
Si tomamos ahora la función  $f(x)=N (\beta \lambda x^2-\ln(2\pi\textrm{I}_0(2\beta \lambda x)))$, podemos definir un extremo de la función en $x_0=M_x =m=\frac{\textrm{I}_1(2\beta \lambda x_0)}{\textrm{I}_0(2\beta \lambda x_0)}$ que corresponde a la magnetización. La derivada de segundo orden de la función $f(x)$, evaluada en $x_0$, está dada por
\begin{eqnarray}
f^{''}(x_0)=4N\beta \lambda\left(1+\beta \lambda(m^2-1)\right)).
\end{eqnarray}
Luego la función $f(x)\simeq f(x_0)+\frac{1}{2}(x-x_0)^2f^{''}(x_0)+...$ es explícitamente escrita como
\begin{equation}
f(x)\simeq N\beta \lambda x_0^2-N\ln(2\pi\textrm{I}_0(2\beta \lambda x_0))+\frac{1}{2}(x-x_0)^24N\beta \lambda\left(1+\beta \lambda(m^2-1)\right)),\label{approxf}
\end{equation}
la cual puede ser calculada por la siguiente aproximación,
\begin{eqnarray}
\int_{-\infty}^{\infty}\texttt{d}x \: e^{-f(x)}
&\approx&\int_{-\infty}^{\infty} \texttt{d}x \: e^{-f(x_0)-\frac{1}{2}(x-x_0)^2f^{''}(x_0)}\\&\approx&e^{-f(x_0)}\sqrt{\frac{2\pi}{f^{''}(x_0)}}\;\;\;\;=\;\;\;{\cal F}(N,\beta \lambda),
\end{eqnarray}
donde ${\cal F}(N,\beta \lambda)$ corresponde a la aproximación de la función ${\cal F}_N(\beta \lambda)$, la cual es obtenida de la evaluación de la ec.( \ref{intF}) usando la aproximación dada por la ec.( \ref{approxf}). Luego,
 ${\cal F}_N(\beta \lambda)$ coincide con ${\cal F}(N,\beta \lambda)$ cuando  $N\rightarrow\infty$.
En la Fig. \ref{approx} (a), se describe ${\cal F}_N(\beta \lambda)$ comparada con ${\cal F}(N,\beta \lambda)$ como una función de $\beta \lambda$  para ilustrar la validez de la aproximación (ver apéndice \ref{app2}). Se muestran los valores exactos y aproximados los cuales se vuelven más cercanos a medida que $N$ crece. El mismo efecto se muestra en el panel (b) de la Fig. \ref{approx} para las funciones definidas como,
\begin{eqnarray}
X_N(\beta \lambda) &=&\frac{1}{N}\ln \left( {\cal F}_N(\beta \lambda)\right)\\
X(N,\beta \lambda) &=&\frac{1}{N}\ln \left( {\cal F}(N,\beta \lambda)\right).
\end{eqnarray}
Entonces, $X_N(\beta \lambda)$ y $X(N,\beta \lambda) $ coinciden cuando $N\rightarrow\infty$.

Luego la función de partición puede ser expresada como,
\begin{eqnarray}
Z\!=\sqrt{\frac{\beta \lambda N}{\pi}}e^{-\beta \lambda N}\left(\frac{2\pi}{\beta}\right)^{N/2}\!\:\:e^{-N\beta \lambda x_0^2+N\ln(2\pi\textrm{I}_0(2\beta \lambda x_0))}\frac{\sqrt{\pi}}{\sqrt{2N\beta\lambda\left(1+\beta \lambda(m_0^2-1)\right)}},\nonumber\\
\end{eqnarray}
donde $x_0$ es el extremo de la función $f(x)$.

Ahora procederemos a obtener las magnitudes termodinámicas del sistema.
Calculando $\ln Z$, tenemos
\begin{eqnarray}
\ln Z&=&\frac{N}{2}\ln\!\!\left(\frac{2\pi}{\beta}\right)+\frac{1}{2}\ln\left(\frac{\beta \lambda N}{\pi}\right)- \beta \lambda N-N\beta \lambda x_0^2\nonumber\\
&+&N\ln(2\pi\textrm{I}_0(2\beta \lambda x_0)\!)\!+\!\frac{1}{2}\ln\left(\frac{\pi}{2N\beta \lambda\left(\!1\!+\!\beta \lambda(m^2\!-\!1)\right)}\!\right).
\end{eqnarray}
El caso límite de esta cantidad por partícula es 
\begin{eqnarray}\label{fi}
\displaystyle\lim_{N\longrightarrow \infty}\frac{\ln Z}{N}=\frac{1}{2}\ln\left(\frac{2\pi}{\beta}\right)- \beta \lambda-\beta \lambda x_0^2+\ln(2\pi\textrm{I}_0(2 \beta \lambda x_0)).
\end{eqnarray}
La última expresión se evalúa en el límite termodinámico, $N\rightarrow \infty$ \cite{STANLEY}, para $x=x_0$.
Sea $\varphi=F\beta=-\displaystyle\lim_{N\longrightarrow \infty}\ln Z/N$ donde $F$ es la energía libre por partícula, luego tenemos
\begin{equation}\label{problemaext}
\varphi(\beta)\!=-\!\frac{1}{2}\ln\frac{\beta}{2\pi} \!+\!\lambda\beta
-\!\inf_{x\geq 0}\![-\!\beta \lambda x^2\!+\!\ln (2\pi\textrm{I}_0(2\beta \lambda x))].
\end{equation}
Como mencionamos anteriormente la solución del problema extremal se obtiene de
\begin{equation}\label{XSolution}
x =\frac{\texttt{I}_1(2\beta \lambda x)}{\texttt{I}_0(2\beta \lambda x)}.
\end{equation}
La temperatura inversa crítica es $\beta_c=1$.

Si $\lambda <0$, la ecuación tiene como solución trivial $x=0$. En contraste, si $\lambda >0$, la ecuación tiene un conjunto de valores para $x$ y $\beta$, los cuales definen la solución del problema. Finalmente, la energía interna por partícula se obtiene como una función de la temperatura inversa y la magnetización
\begin{equation}\label{ASolution}
\varepsilon=\frac{\partial \varphi(\beta,N)}{\partial \beta} = \frac{1}{2\beta}-\lambda\left( m^2-1\right),
\end{equation}
donde $m$ es la solución del problema extremal ec.( \ref{problemaext}) y corresponde a la solución del ensamble canónico. De ahora en adelante usaremos $\lambda=1$.
En la  Fig.\ref{approx} describimos la magnetización de equilibrio $m$ como una función de la energía interna $\varepsilon$. La solución analítica se obtiene del ensamble canónico dada por la ec.( \ref{ASolution}). El punto crítico está localizado en $\varepsilon_c=3/2$.
\begin{figure}[t]
\centering
  \subfloat{
    \includegraphics[width=0.4\textwidth]{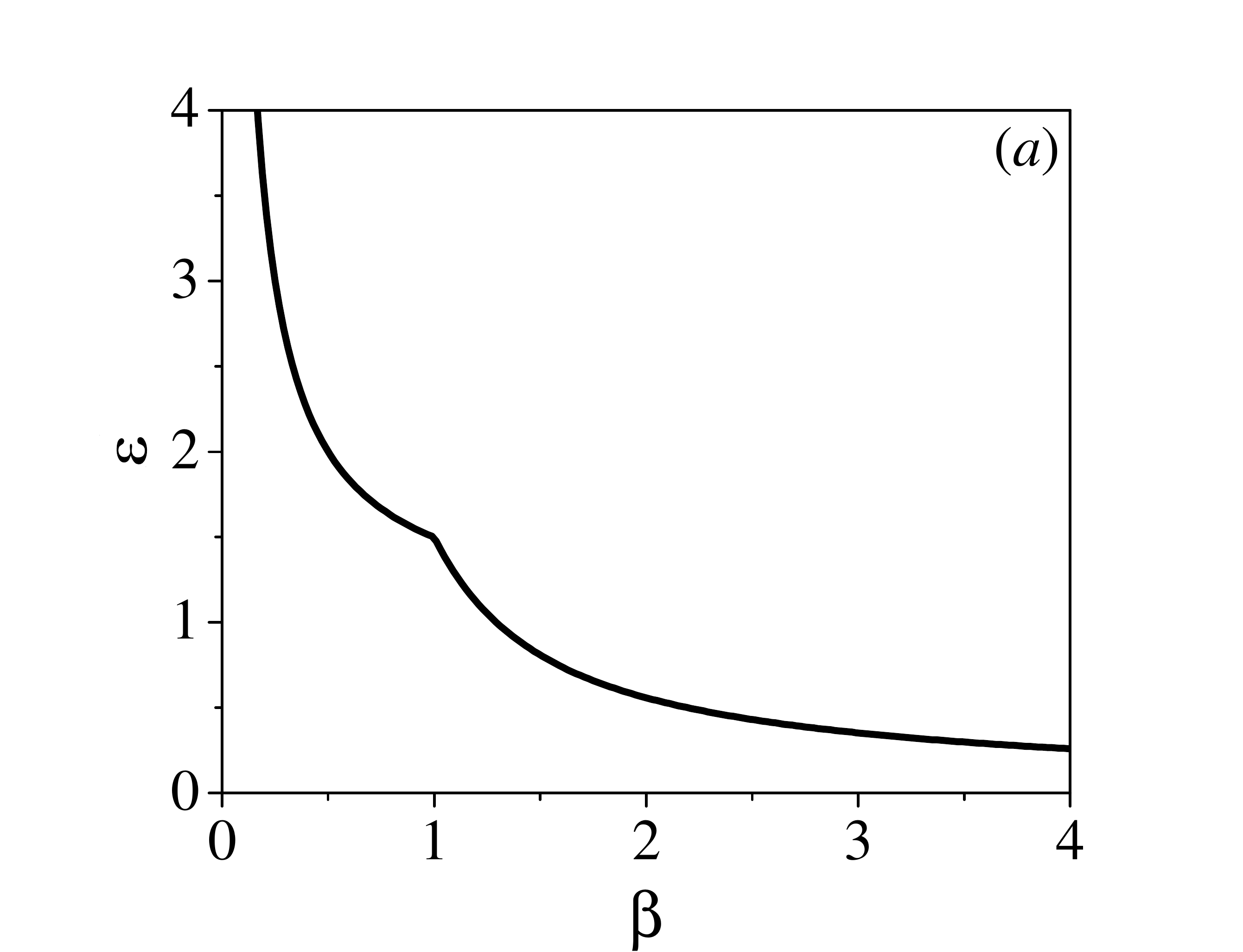}}
  \subfloat{
    \includegraphics[width=0.4\textwidth]{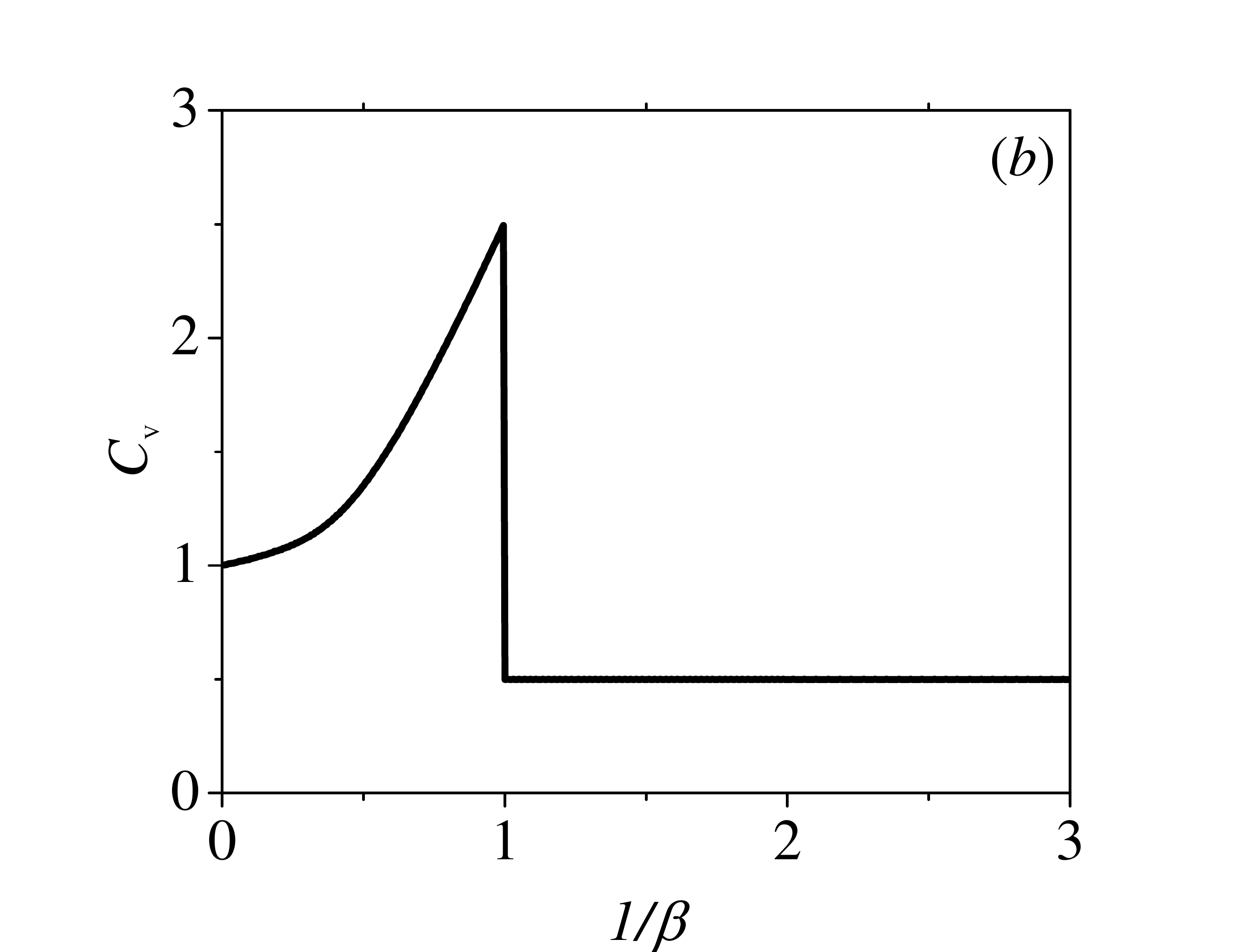}}
\caption{Describimos en ({\emph{a}}) la energía interna $\varepsilon$ como función de $\beta$. El punto crítico está en $\varepsilon_c = 3/2$, $\beta_c= 1$ para $\lambda=1$. Adicionalmente, describimos en  ({\emph{b}}) el calor específico $c_v$ como una función de $1/\beta$, con $\beta_c=1$.}\label{Energy}
\end{figure}
Además, en la Fig. \ref{Energy} mostramos en ({\emph{a}}) la energía interna $\epsilon$ como una función de $\beta$ y en ({\emph{b}}) el calor específico $C_v$ como una función de $1/\beta$.  El comportamiento de las funciones termodinámicas $\varepsilon$ y $C_v$ definen dos regiones diferentes. Esto refleja que el sistema presenta una transición de fase continua.
En particular, la Fig. \ref{Energy} ({\emph{b}}) muestra, por un lado, que el calor específico crece cuando la temperatura aumenta y tras el punto crítico cae y se mantiene constante con el valor $c_v=1/2$, el cual corresponde a un gas ideal en una dimensión.

\section{Cálculos en el ensamble microcanónico}\label{microcanonico}

En esta sección nos centraremos en el cálculo microcanónico de la entropía a partir de la densidad de estados, y con éste la curva calórica del sistema. Estos resultados han sido publicados en \cite{ATENAS5}.

En el ensamble microcanónico, a partir del hamiltoniano ec.( \ref{hamiltoniano}) es posible encontrar el número de microestados
\begin{eqnarray}
\Omega(E,N)=\int\displaystyle\prod_{i=1}^N \mbox{d}p_i\mbox{d}\theta_l\delta(E-H_N(\theta_i,p_i)),
\end{eqnarray}
donde podemos introducir una identidad de Dirac en $K$, luego
\begin{eqnarray}
\Omega(E,N)=\int \mbox{d}K\displaystyle\prod_{i=1}^N \mbox{d}p_i\mbox{d}\theta_i\delta\left(K-\sum_{j=1}^N\frac{p^2_j}{2}\right)\delta(E-K-U\{\theta_i\}).
\end{eqnarray}
Esta expresión puede ser separada en dos partes, la cinética y la configuracional, esto es,
\begin{eqnarray}
\Omega_{\text{kin}}(K)&=&\int \displaystyle\prod_{i=l}^N \mbox{d}p_i\delta\left(K-\sum_{j=1}^N\frac{p^2_j}{2}\right),\\
\Omega_{\text{conf}}(E-K)&=&\int \displaystyle\prod_{i=1}^N \mbox{d}\theta_i \delta(E-K-U\{\theta_i\}),
\end{eqnarray}
donde $\Omega(E,N)=\int \mbox{d}K \Omega_{\text{kin}}(E,N)\Omega_{\text{conf}}(E-K)$. Nuevamente, la parte cinética \cite{PATHRIA} está dada por,
\begin{eqnarray}
\Omega_{\text{kin}}(K)=\frac{\pi(2\pi K)^{N/2-1}}{\Gamma(N/2)}.
\end{eqnarray}
Luego usando la propiedad $\ln\Gamma(N)=\left(N-\frac{1}{2}\right)\ln N-N+\frac{1}{2}\ln(2\pi)$, y despreciando unos términos sobre otros para $N$ grande, tenemos,
\begin{eqnarray}
\ln\Omega_{\text{kin}}(K)\simeq\frac{N}{2}\left(1+\ln(2\pi)+\ln \frac{2K}{N}\right).
\end{eqnarray}
Definiendo $E=K+U$, $u=2K/N$, $\tilde{u}=U/N$, y $\varepsilon=E/N$, $\Omega_{\text{kin}}(K)$ puede ser expresado como
\begin{equation}
\Omega_{\text{\text{kin}}}(K)\simeq \exp\left(\frac{N}{2}\left(1+\ln(2\pi)+\ln u\right)\right).
\end{equation}

\begin{figure}
\centering
  \subfloat{
    \includegraphics[width=0.4\textwidth]{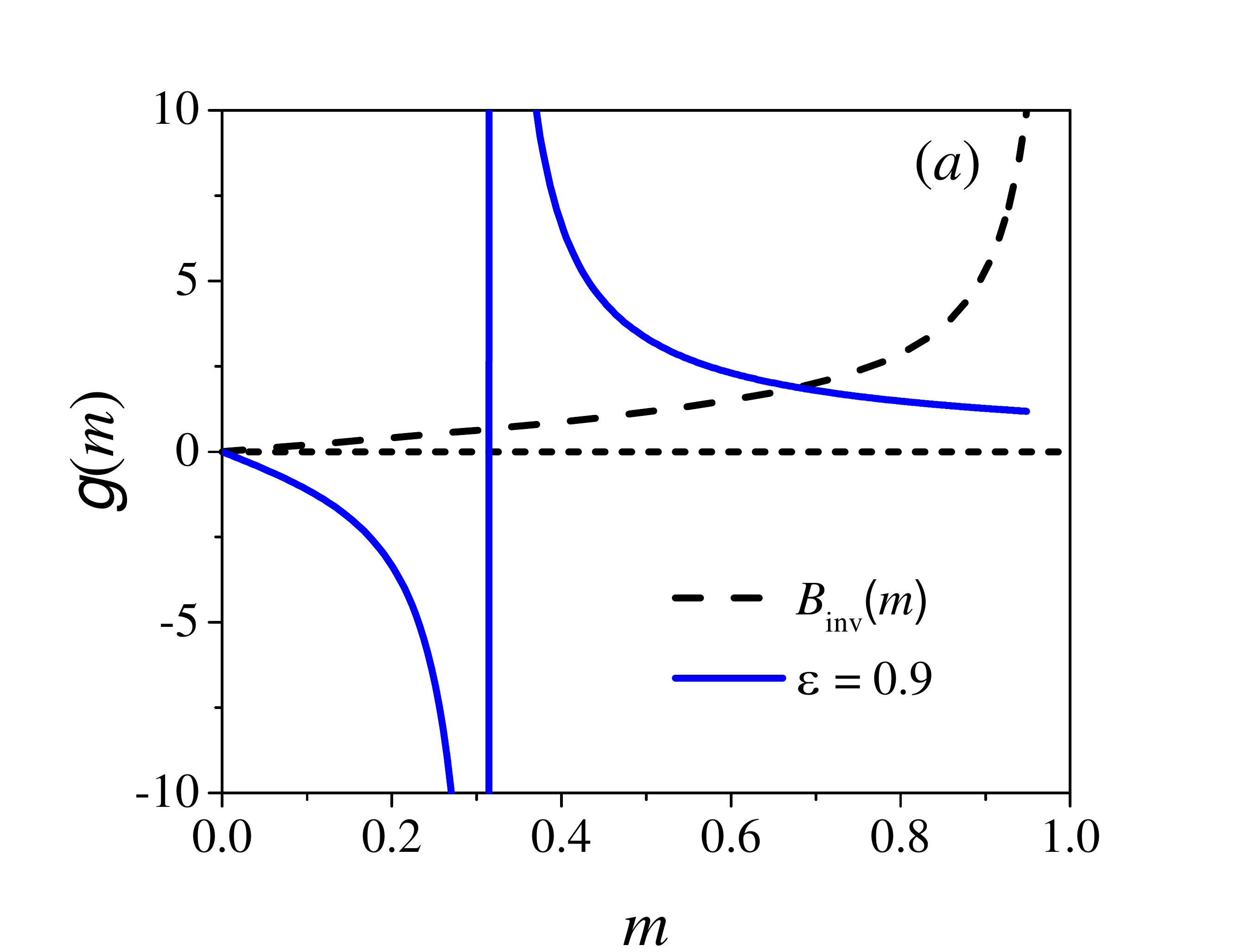}}
  \subfloat{
    \includegraphics[width=0.4\textwidth]{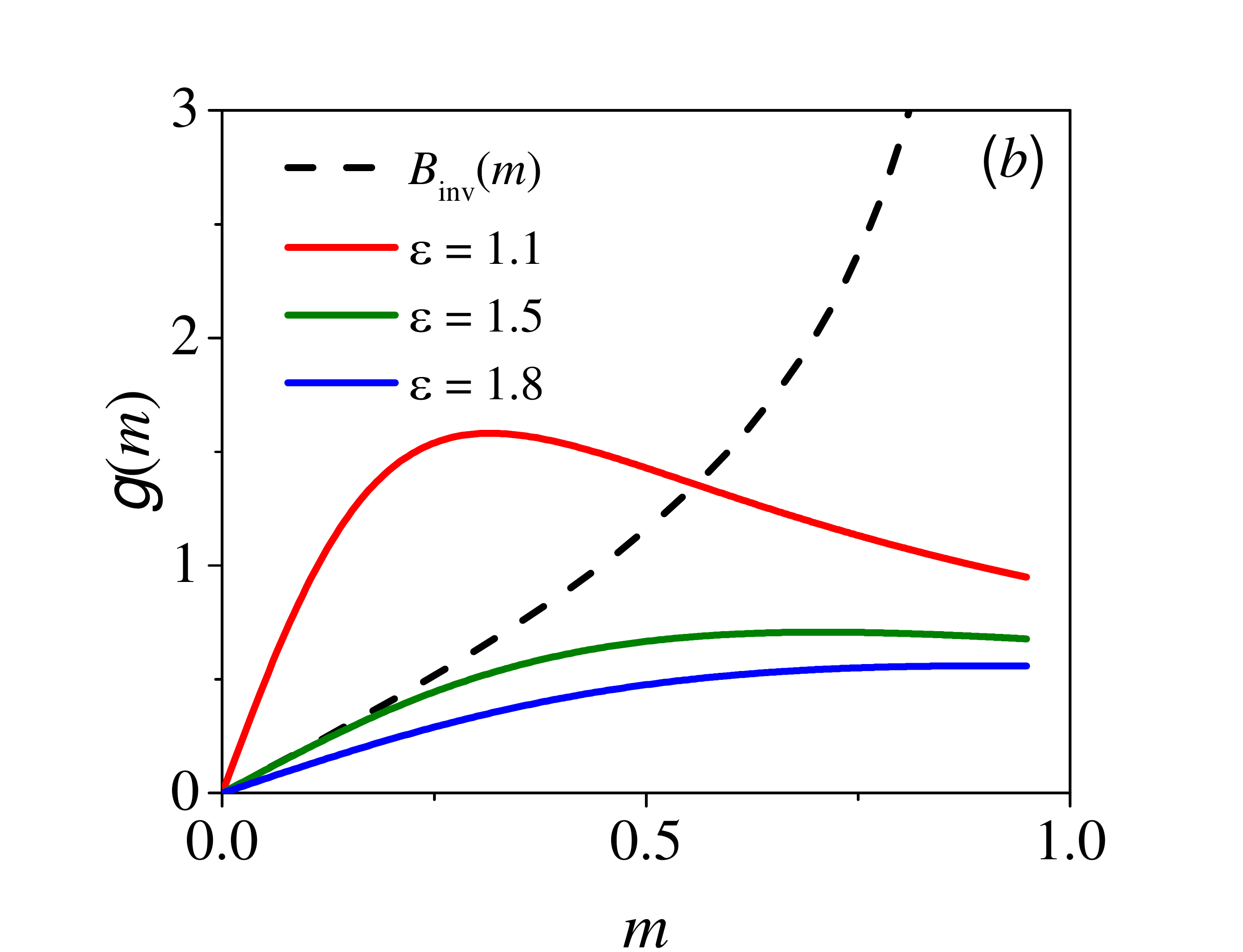}}
\caption{Nosotros describimos $g(m)$, como una función de $m$ para valores de ({\emph{a}}) $\varepsilon$ =0.9 y ({\emph{b}}) $\varepsilon$ =1.1, 1.5, 1.8, con  $\lambda=1$. El valor $\varepsilon =1$ cambia la tendencia de la función $g(m)$ en el intervalo $0\leq m<1$.}\label{Energy2}
\end{figure}
La parte configuracional viene dada por
\begin{eqnarray}
\Omega_{\text{conf}}(E-K)\simeq \exp{\left( \ln \Omega_{\text{conf}}\right)}.
\end{eqnarray}
Luego, la entropía es $s=\frac{1}{N}\ln \Omega$, por lo tanto, $\Omega(E,N)$ puede ser expresada como
\begin{eqnarray}
\Omega(E,N)=\frac{N}{2}\int \mbox{d}u  \;\text{e}^{N\left(\frac{1}{2}+\frac{1}{2}\ln(2\pi)+\frac{1}{2}\ln u+s_{\text{conf}}(N \tilde{u})\right)},
\end{eqnarray}
donde $\tilde{u}$ es la energía potencial por partícula, luego $\Omega_{\text{conf}}(E-K)=\Omega_{\text{conf}}(N\tilde{u})$. Como se realizó anteriormente, la integral se puede transformar en un problema extremal dado por
\begin{eqnarray}
s&\!\!=\!\!&\frac{1}{N}\ln \Omega(E,N) \\
\!\!&=\!\!&\frac{1}{N}\ln\!\left(\!\frac{N}{2}\!\int \!\mbox{d}u  \;\exp\!\left(\!N\!\left(\!\frac{1}{2}\!+\!\frac{1}{2}\ln(2\pi)+\frac{1}{2}\ln u+s_{\text{conf}}(N \tilde{u})\!\!\right)\!\!\right)\!\!\right).\label{entr}
\end{eqnarray}
Resolviendo la integral (\ref{entr}) de la misma manera que para el ensamble canónico
\begin{eqnarray}
s&=&\!\!\frac{1}{N}\ln\left(\frac{N}{2}\exp{\!\left(\!\frac{N}{2}\!+\!\frac{N}{2}\ln(2\pi)\right)}\!\right)\displaystyle \!\sup_{u}\!\left(\!\frac{N}{2}\ln u\!+\!Ns_{\text{conf}}(N\tilde{u}(u))\!\!\right)\nonumber\\
\!\!&=&\!\!\frac{1}{N}\ln\frac{N}{2}\!+\!\frac{1}{2}+\frac{1}{2}\ln 2\pi+\displaystyle\sup_u\!\left(\!\frac{1}{2}\ln u+ s_{\text{conf}}(N\tilde{u}(u))\right).
\end{eqnarray}
Además, la energía potencial por partícula puede ser expresada de la ec.( \ref{epot}), para $M_x\approx m$ and $M_y\approx 0$ como,
\begin{equation}
\tilde{u}=U/N=-\lambda\left(m^2-1\right).
\end{equation}
De la expresión $U=E-K$, $\tilde{u}=\varepsilon-u/2$, y $u=2(\varepsilon-\tilde{u})=2(\varepsilon+\lambda(m^2-1))$; luego, en el límite termodinámico, la entropía se puede escribir como,
\begin{equation}
s=\frac{1}{2}+\frac{1}{2}\!\ln 2 \pi+\frac{1}{2} \ln 2+ \displaystyle\sup_m\left[\frac{1}{2}\ln (\varepsilon+\lambda(m^2-1))+s_{\text{conf}}(N\tilde{u}(u))\right].
\end{equation}
Ahora, calculamos la entropía configuracional $s_{\text{conf}}$. Como se mostró antes, el término $M_y$ es despreciable comparado con $M_x$, esta información puede ser introducida en $s_{\text{conf}}$, como sigue
\begin{equation}
\Omega_{\text{conf}}=\int \displaystyle\prod_{l=1}^N \mbox{d}\theta_l \delta\left(\sum_j \cos \theta_j-Nm\right)\delta\left(\sum_j\sin\theta_j\right),
\end{equation}
esto es $M_x\simeq m$, y $M_y\simeq 0$. Luego, podemos calcular, expresando en la representación de Fourier
\begin{equation}
\Omega_{\text{conf}}\!=\!\left(\!\frac{1}{2\pi}\!\right)^2\!\!\int\!\! \mbox{d} q_1\!\int\!\mbox{d}q_2\!\int\! \displaystyle{\prod_{l=1}^N} \mbox{d}\theta_l \exp\!\left(\!iq_1\!\sum_j \!\cos\! \theta_j\!-\!Nm \!\right)
\!\exp\!\!\left(\!iq_2\sum_j\!\sin\theta_j\right),
\end{equation}
la cual corresponde a la función de Bessel de primera especie $J_0(z)$,
\begin{equation}
\Omega_{\text{conf}}=\left(\frac{1}{2\pi}\right)^2\int \mbox{d}q_1\int \mbox{d}q_2 \exp\left({N \left(
-iq_1 m+\ln(2\pi J_0(z))\right)}\right),
\end{equation}
donde el módulo de $z$ is $(q_1^2+q_2^2)^{1/2}$
Seguidamente, resolvemos la última integral de la misma forma que en el canónico, se satisfacen las siguientes ecuaciones
\begin{eqnarray}
-im-\frac{J_1(z)}{J_0(z)}\frac{q_1}{z}&=&0,\\
-\frac{J_1(z)}{J_0(z)}\frac{q_2}{z}&=&0,
\end{eqnarray}
donde las soluciones son $q_2=0$, y $q_1=-i\gamma$, y $\gamma$ es la solución de la ecuación
\begin{equation}
\frac{I_1(\gamma)}{I_0(\gamma)}=m.\label{Binv}
\end{equation}
Denotando por $B_{\text{inv}}(m)$ la inversa de la ec.( \ref{Binv}), obtenemos en el límite termodinámico
\begin{equation}
s_{\text{conf}}=\displaystyle\lim_{N\rightarrow \infty} \frac{1}{N}\ln \Omega_{\text{conf}}=-mB_{\text{inv}}(m)+\ln I_0( B_{\text{inv}}(m)).
\end{equation}
Usando $u=2(\varepsilon+\lambda(m^2-1))$, se tiene
\begin{equation}
s\!=\!\frac{1}{2}\!+\!\frac{1}{2}\!\ln 2 \pi\!+\!\frac{1}{2}\! \ln 2\!
+\!\sup_m\! \left[\!\frac{1}{2}\!\ln(\varepsilon+\lambda(m^2-1))\!-\!mB_{\text{inv}}(m)\!+\!\ln \!I_0( B_{\text{inv}}(m))\!\right]\!\!.
\end{equation}
El problema extremal está dado por la solución de la ecuación
\begin{eqnarray}
\frac{\lambda \; m}{\varepsilon+\lambda(m^2-1)}-B_{\text{inv}}(m)=0,\label{gm}
\end{eqnarray}
donde definimos, 
\begin{equation}
g(m)=\frac{m}{\varepsilon+\lambda(m^2-1)},    
\end{equation}
la cual se muestra en la Fig. \ref{Energy2} como una solución de la ec.( \ref{gm}). La solución para $\varepsilon<1$ se grafica en el panel ($a$). En el panel ($b$) la función $g(m)$ para $\varepsilon > 1$, donde la magnetización es siempre $m=0$. La única solución para $m\neq0$ ocurre cuando $\varepsilon<1$. La Fig. \ref{Energy2} muestra la función $g(m)$ para diferentes valores de $\varepsilon$.\\
\begin{figure}[h!]\centering
  \subfloat{
    \includegraphics[width=0.4\textwidth]{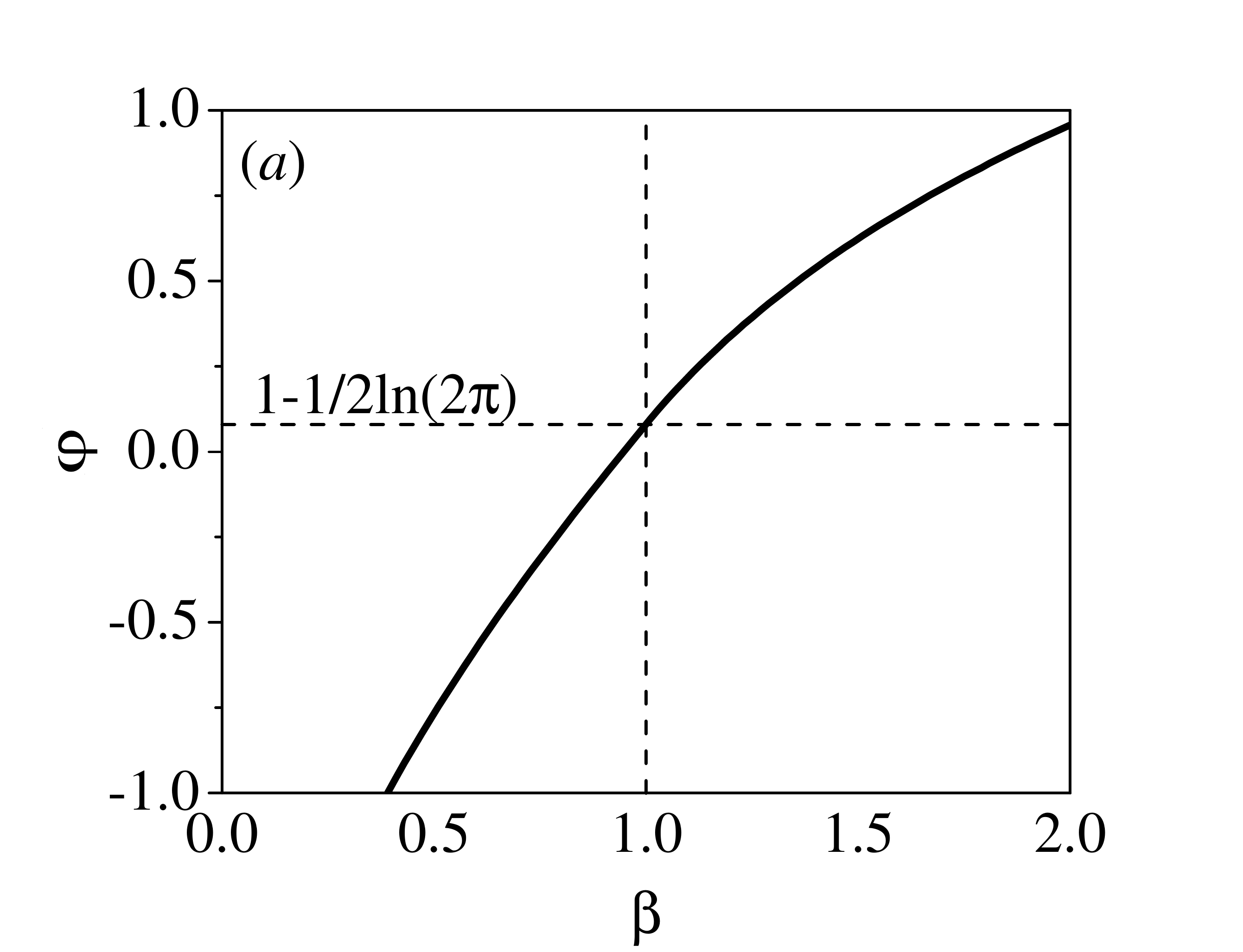}}
  \subfloat{
    \includegraphics[width=0.4\textwidth]{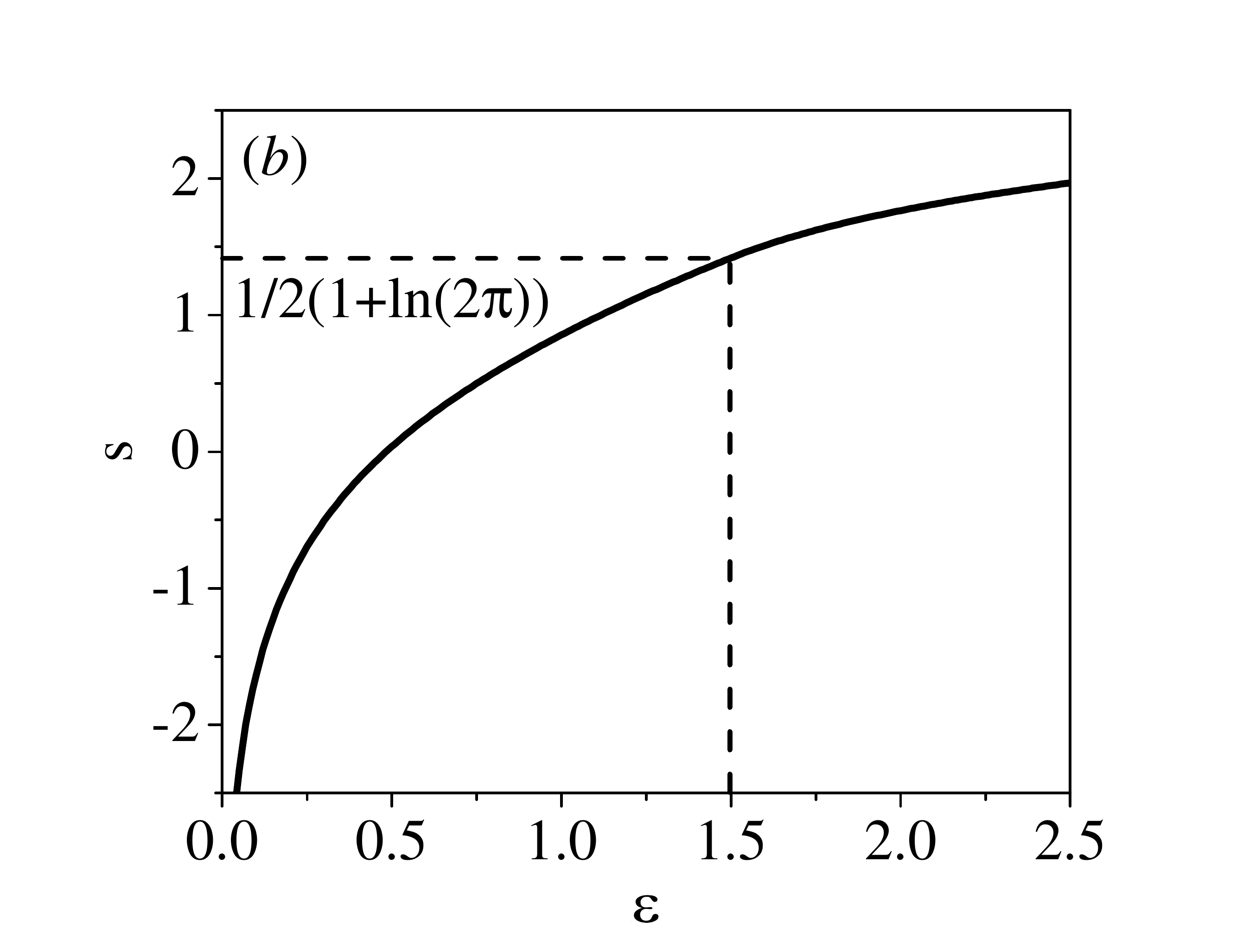}}
\caption{Describimos en ({\textit a}) la energía libre como una función de $\beta$, en ({\textit b}) la entropía $s$ como una función de la energía interna $\varepsilon$. El punto crítico está localizado en $\beta_c= 1$ y $\varepsilon_c = 3/2$ para $\lambda=1$.
}\label{Entropy}
\end{figure}
Llamando $m=m(\beta)$ a la solución del problema extremal para $m$, finalmente la entropía puede expresarse como
\begin{equation}
s\!=\!\frac{1}{2}\!+\!\frac{1}{2}\ln 2 \pi\!+\!\frac{1}{2} \ln 2\!+\! \frac{1}{2}\ln(\varepsilon\!+\!\lambda(m^2(\varepsilon)\!-\!1))\!-\!xB_{\text{inv}}(x)\!+\!\ln I_0( B_{\text{inv}}(m)).
\end{equation}
La Fig. \ref{Entropy} muestra la entropía microcanónica del problema, y si tomamos la derivada respecto a $\varepsilon$, recuperamos la solución canónica, esto es
\begin{eqnarray}
\beta=\frac{ds}{d\varepsilon}=\frac{1}{2(\varepsilon+\lambda(m^2-1))},\label{ccurve}
\end{eqnarray}
que coincide con la solución canónica de la ec.( \ref{ASolution}).
Vemos que en el equilibrio de BG, en el límite termodinámico hay equivalencia de ensambles entre el canónico y microcanónico; sin embargo, como veremos en la siguiente sección, al estudiar el sistema fuera del equilibrio mediante dinámica molecular, es posible encontrar estados QSS, que producen capacidades caloríficas negativas en la zona cercana al punto crítico.

\chapter{Dinámica de Vlasov del modelo d-HMF}\label{cap3}

En esta sección se derivan las ecuaciones para la dinámica de Vlasov, del modelo d-HMF. 

A continuación a partir de la ec.( \ref{hamiltoniano}) es posible derivar una expresión para la la energía potencial de una partícula, cuya expresión viene dada por
\begin{eqnarray}
e&=&\frac{\lambda}{N}\sum_{j=1}^N\left(\cos(\theta)\cos\theta_j-3\cos\theta\cos\theta_j+2)\right)\nonumber\\
&=&\frac{\lambda}{N}\sum_{j=1}^N\left(\sin\theta\sin\theta_j-2\cos\theta\cos\theta_j+2\right)\nonumber\\
&=&\lambda\left(M_y\sin\theta-2M_x\sin\theta+2)\right),\label{eppdhmf}
\end{eqnarray}
donde $\theta$ es la orientación de la partícula. En la ec.( \ref{hamiltoniano}) hemos dejado el factor $1/2$ fuera de nuestro modelo ya que estamos considerando la energía que ejercen todas las demás partículas sobre una sola.

Pasando al continuo la ec.( \ref{eppdhmf}) obtenemos
\begin{eqnarray}
\langle U(\theta,t)\rangle&=&\int f(\theta',p',t)(2+\cos(\theta'-\theta)-3\cos\theta'\cos\theta)d\theta' dp'\nonumber\\
&=&2-2M_x\cos\theta +M_y\sin\theta.\label{potdhmfvlas}
\end{eqnarray}
Derivamos la ec.( \ref{potdhmfvlas}) para obtener las fuerzas de campo medio,
\begin{eqnarray}
-\frac{\partial \langle U(\theta,t)\rangle}{\partial \theta}=-2M_x\sin\theta -M_y\cos\theta.\label{campomediodhmfvlas}
\end{eqnarray}
Luego como $\dot{\vec{p}}=\vec{F}_{mf}=-\nabla \langle U\rangle$, la ecuación de Vlasov puede expresarse como
\begin{eqnarray}
\frac{\partial f}{\partial t}+p
\frac{\partial f}{\partial \theta}
+(-2M_x\sin\theta-M_y\cos\theta)\frac{\partial f}{\partial p} =0\label{vlasoveccdHMF}
\end{eqnarray}
y la energía total específica o densidad de energía en el tiempo $t$ es
\begin{eqnarray}
e(t)&=&\frac{1}{2}\int f(\theta,p,t)p^2d\theta dp -\frac{1}{2}\int f(\theta,p,t)(2-2M_x\cos\theta+M_y\sin\theta)d\theta dp\nonumber\\
&=&\frac{1}{2}\int f(\theta,p,t)p^2d\theta dp+\frac{1}{2}\left(2-2M_x^2+M_y^2\right).\label{energyespdhmf}
\end{eqnarray}

\begin{figure}
    \centering
   \includegraphics[width=0.5\textwidth]{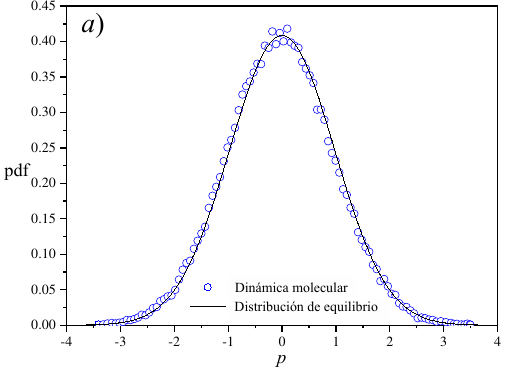}\includegraphics[width=0.487\textwidth]{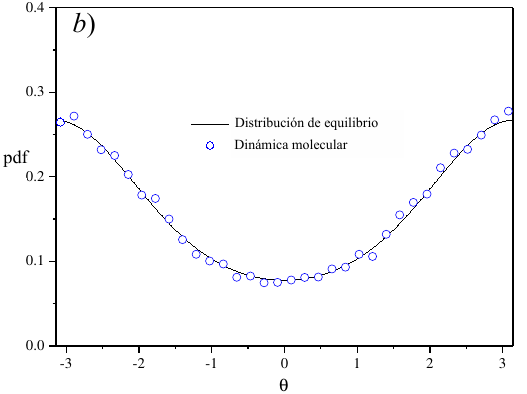}
    \caption{Distribuciones de equilibrio de las ec. \ref{distrib} y \ref{distrib2} comparadas con los datos de dinámica molecular. En $a)$ la distribución en los momentos y en $b)$ la distribución en las orientaciones.}
    \label{DistributionsBG-Eq}
\end{figure}

Por otro lado de la ec.( \ref{hamiltoniano}) la energía de una partícula queda,
\begin{eqnarray}\label{IndividualEnergy}
e(\theta,p)=\frac{p^2}{2}+ 2-2m_x\cos\theta+m_y\sin\theta.
\end{eqnarray}


\section{Función densidad del equilibrio}\label{VlasovEquilibrio}

Como se muestra en la referencia \cite{YAMAGUCHI}, las soluciones estacionarias de la ecuación de Vlasov vienen dadas por una función que depende sólo de la energía de una partícula individual, esto es
\begin{equation}
f(\theta,p)=\phi(e(\theta,p)).
\end{equation}
Por lo tanto, para el equilibrio la densidad es
\begin{eqnarray}\label{eqdistrib}
f(\theta,p)=C\exp\left(-\beta_{eq}\left(p^2/2+2-2m_x\cos\theta+m_y\sin\theta\right)\right).
\end{eqnarray}
Nosotros enfatizamos aquí la temperatura inversa para el equilibrio con subíndice, esto es $\beta_{eq}$, para distinguir del parámetro $\beta$ que usaremos para describir los estados QSS fuera del equilibrio mediante Tsallis.

Normalizando la densidad obtenemos
\begin{eqnarray}
C=\frac{\sqrt{\beta_{eq}}\exp(2\beta_{eq})}{(2\pi)^{3/2}I_0(2m_x\beta_{eq})}.
\end{eqnarray}
A partir de este resultado, integramos la ec. (\ref{eqdistrib}) en $p$ y $\theta$, respectivamente, obteniendo las distribuciones en los momentos y las orientaciones,
\begin{eqnarray}\label{distrib}
\varrho(p)&=&\int f(\theta,p)\:d\theta=\sqrt{\frac{\beta_{eq}}{2\pi}}e^{-\beta_{eq}p^2/2}, \\
\rho(\theta)&=&\int f(\theta,p)\:dp=\frac{e^{2\beta_{eq}m_x\cos\theta}}{2\pi I_0(2m_x\beta_{eq})}.\label{distrib2}
\end{eqnarray}
Como se aprecia en la Fig. \ref{DistributionsBG-Eq}, las soluciones analíticas coinciden con los resultados de dinámica molecular.

\section{Soluciones estacionarias de Vlasov fuera del equilibrio}

\begin{figure}[!h]
\centering
\includegraphics[width=0.8\textwidth]{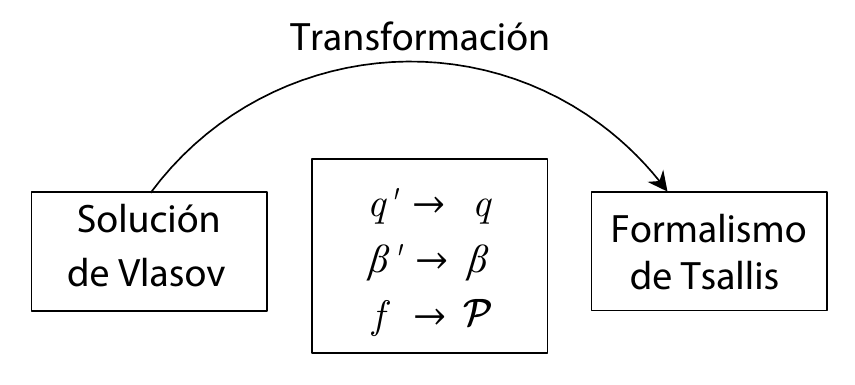}
\caption{Esquema de la metodología utilizada para abordar el problema.}\label{resum}
\end{figure}

En esta sección, se presenta una transformación que permite describir de manera adecuada los QSS del modelo d-HMF. Esta transformación permite conectar la estadística de Tsallis con la formulación de Vlasov según el esquema de la Fig. \ref{resum}

Primero daremos una pequeña introducción al formalismo de Tsallis, el cual se basa en el principio de máxima entropía de Jaynes \cite{JAYNES} sujeto a restricciones, el cual permite obtener la conocida forma $q$-exponencial de Tsallis. Estas restricciones, serán en nuestro caso, los valores de energía cinética de los QSS hallados por dinámica molecular. Cabe mencionar que esta metodología constituye una variación de la transformación de Lima-Penna \cite{LIMA1999}.

\section{Formalismo de Tsallis y  soluciones estacionarias de Vlasov} \label{Tsallis-like}

La propuesta inicial de Tsallis \cite{TSALLIS}, ha sufrido diversas variaciones con el objetivo de hacer compatible su formulación con los principios de la termodinámica y las leyes de la probabilidad. En aras de realizar una conexión adecuada entre el formalismo de Tsallis y la dinámica de Vlasov es que utilizaremos la forma más reciente dada por
\begin{eqnarray}\label{TsallisConstraints}
s_q&=&-k_B\frac{1-\displaystyle\int p^q (x)dx}{1-q}\\    
u_q&=&\frac{\displaystyle\int p^q(x) e(x) dx}{\displaystyle\int p^q(x) dx}\\
1&=&\displaystyle\int p(x) dx,
\end{eqnarray}
donde $s_q$ es la q-entropía, $u_q$ es el valor de esperado 
$q$ de la energía y $\varepsilon(x)$ es la energía del estado $x$. 

La distribución de Tsallis se obtiene al aplicar el principio de máxima entropía de Jaynes, sujeto a las restricciones de las ecs. (\ref{TsallisConstraints}),
\begin{eqnarray}
\delta (\alpha_1 \mathbb{I}+\alpha_2  u_q+s_q)=0,
\end{eqnarray}
\begin{eqnarray}
\delta \left(\alpha_1\int p(x)dx+\alpha_2 \frac{\int p^q(x)e(x)dx}{\int p^q(x)dx}-k\frac{\int p(x)dx-\int dx p^q(x)dx}{1-q}\right)=0,
\end{eqnarray}
como la energía total se conserva, luego
\begin{eqnarray}
\int dx\left(\alpha_1+\alpha_2q \frac{ p^{q-1}(x)e(x)}{\int p^q(x)dx}-\alpha_2q \frac{u_q p^{q-1}(x)}{\int p^q(x)dx}
-k\frac{1-q p^{q-1}(x)}{1-q}\right)\delta p(x)=0,
\end{eqnarray}

\begin{eqnarray}\label{TsallisMulti}
\int dx\left(\alpha_1-\frac{k}{1-q}+p^{q-1}(x)q\left[\alpha_2\frac{e(x)-u_q}{\int p^q(x)dx}+\frac{k}{1-q}\right]\right)\delta p(x)=0,
\end{eqnarray}
entonces, la probabilidad viene dada por
\begin{eqnarray}
p(x)=\left(\frac{q\frac{k}{1-q}\left[1+\frac{\alpha_2(1-q)}{k\int p^q(x)dx}\left(e(x)-u_q\right)\right]}{\frac{k}{1-q}\left(1-\alpha_1\frac{1-q}{k}\right)}\right)^{\frac{1}{1-q}},
\end{eqnarray}
después de una manipulación algebraica obtenemos
\begin{eqnarray}\label{TsallisMulti2}
p(x)=\left(\frac{q}{1-\alpha_1\frac{1-q}{k}}\right)^{\frac{1}{1-q}}\left[1+\frac{\alpha_2(1-q)}{k\int p^q(x)dx}\left(e(x)-u_q\right)\right]^{\frac{1}{1-q}}.
\end{eqnarray}
Por lo tanto, la normalización corresponde a la función de partición 
\begin{eqnarray}
Z_{q}(\beta)=\left(\frac{q}{1-\alpha_1\frac{1-q}{k}}\right)^{-\frac{1}{1-q}}.
\end{eqnarray}

Para construir las distribuciones escolta (escort distributions), elevamos la probabilidad al parámetro $q$ e integramos en todo el espacio
\begin{eqnarray}
p^q(x)=\left(\frac{q}{1-\alpha_1\frac{1-q}{k}}\right)^{\frac{q}{1-q}}\left[1+\frac{\alpha_2(1-q)}{k\int p^q(x)dx}\left(e(x)-u_q\right)\right]^{\frac{q}{1-q}},
\end{eqnarray}
finalmente, obtenemos
\begin{eqnarray}\label{EscortDistTsallis}
f(x)= \frac{p^q(x)}{\int p^q(x) dx}=\frac{\left[1+\frac{\alpha_2(1-q)}{k\int p^q(x)dx}\left(e(x)-u_q\right)\right]^{\frac{q}{1-q}}}{\displaystyle\int \left[1+\frac{\alpha_2(1-q)}{k\int p^q(x)dx}\left(e(x)-u_q\right)\right]^{\frac{q}{1-q}}dx}.
\end{eqnarray}
Esta última distribución, la ec. (\ref{EscortDistTsallis}), es la que se utilizará para hacer la conexión con la dinámica de Vlasov.

Uno de los principales objetivos planteados en esta tesis es describir los QSS del modelo d-HMF mediante un esquema teórico. Cuando intentamos utilizar el software vmf$90$ de Buyl \cite{BUYL} para describir los QSS hallados en el modelo d-HMF, sólo logramos describir el segundo QSS. Como este software no  permite estudiar en detalle todos los aspectos teóricos, es que optamos por este esquema analítico. Los resultados mostrados en esta sección se encuentran en proceso de revisión \cite{ATENAS6}.
Partimos testeando una distribución de Tsallis q-exponencial que optimizan los valores de $q'$ y $\beta'$ para las distribuciones halladas por las simulaciones. Luego, mediante un algoritmo numérico variacional hallamos los  valores de $q'$ y $\beta'$ que describen de manera más adecuada estos estados  \footnote{En el apéndice \ref{app2} se describe el procedimiento de optimización utilizado para hallar los valores óptimos de $q'$ y $\beta'$ para los dos QSS.}.
Una vez optimizados estos valores, procedemos a contrastar los perfiles (ver Fig. \ref{DistributionsQSS}).

La forma $q$-exponencial escogida como solución estacionaria de Vlasov, para describir los QSS, viene dada por

\begin{eqnarray}\label{VlasovSolutionEq}
f(x)=C(1-(1-q')\beta'e(x))^{\frac{1}{1-q'}},
\end{eqnarray}
donde $q'$ y $\beta'$ son parámetros de la solución de Vlasov, $C$ es una constante de normalización y $e(x)$ es la energía de una partícula individual. Por consiguiente, el cálculo del promedio de las cantidades termodinámicas  viene dado por
\begin{eqnarray}\label{MeanValues}
\langle O \rangle =\int O(x) f(x)dx.
\end{eqnarray}
Como se mencionó anteriormente, esta simple expresión calcula los valores de expectación de cantidades físicas de un sistema generalizado en el formalismo de Tsallis a través del $q$-valor esperado (en inglés $q$ expectation value) definido previamente en la ec.( \ref{TsallisConstraints}). Tomando en cuenta las ecs.( \ref{EscortDistTsallis}) y \ref{VlasovSolutionEq}, escribimos
\begin{eqnarray}
\frac{p^q(x)}{\int p^q(x)dx}=C(1-(1-q')\beta'e(x))^{\frac{1}{1-q'}}.
\end{eqnarray}
Después de manipular algebraicamente el argumento de la solución de Vlasov, llegamos a
\begin{eqnarray}\label{Relationship}
\frac{\left[1-\frac{(1-q)\beta}{\int p^q(x)dx}\left(e(x)-u_q\right)\right]^{\frac{q}{1-q}}}{\int p^q(x)dx}=C\left(1-(1-q')\beta'u_q\right)^{\frac{1}{1-q'}}\left(1-\frac{(1-q')\beta'(e(x)-u_q)}{1-(1-q')\beta'u_q}\right)^{\frac{1}{1-q'}}\nonumber\\
\end{eqnarray}
Primero comparando los lados izquierdo y el derecho de la ec. (\ref{Relationship}) obtenemos una relación para la constante de normalización
\begin{eqnarray}\label{Normalization}
\frac{1}{\int p^q(x)dx}=C\left(1-(1-q')\beta'u_q\right)^{\frac{1}{1-q'}}.
\end{eqnarray}

\begin{figure}[!h]
    \centering
   \includegraphics[width=1.0\textwidth]{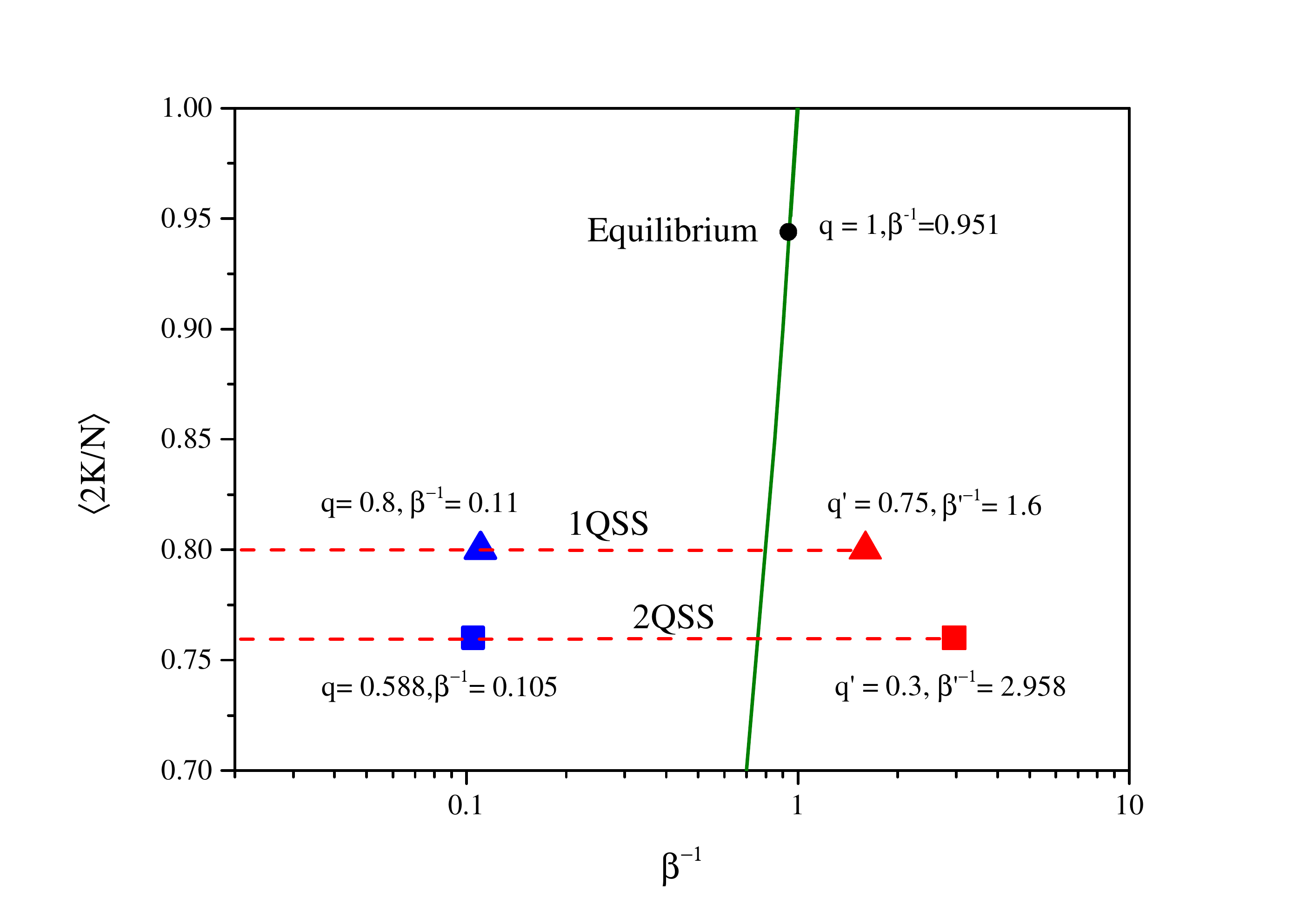}\
    \caption{Energía cinética promedio por partícula. A la izquierda, en símbolos azules, los parámetros de Tsallis $q$ y $\beta$. En el lado derecho, símbolos en rojo, los parámetros de Vlasov $q'$ y $\beta'$. La línea verde corresponde al equilibrio, esto es válido sólo si $q=1$}
    \label{TsallisBeta}
\end{figure}
A continuación conectamos ambas distribuciones, esto es
\begin{eqnarray}
\left[1-\frac{(1-q)\beta}{\int p^q(x)dx}\left(e(x)-u_q\right)\right]^{\frac{q}{1-q}}=\left(1-\frac{(1-q')\beta'(e(x)-u_q)}{1-(1-q')\beta'u_q}\right)^{\frac{1}{1-q'}}.
\end{eqnarray}
En consecuencia, la relación entre los parámetros $q$ de Tsallis y $q'$ de Vlasov $q'$ es

\begin{eqnarray}
q=\frac{1}{2-q'},
\end{eqnarray}
y la relación entre los parámetros $\beta$ y $\beta'$ es,
\begin{eqnarray}
\frac{(1-q)\beta}{\int p^q(x)dx}=\frac{(1-q')\beta'}{1-(1-q')\beta'u_q}
\end{eqnarray}
usando la ec. (\ref{Normalization}) encontramos

\begin{eqnarray}
\beta=\frac{\beta'(2-q')}{C\left(1-(1-q')\beta'u_q\right)^{\frac{2-q'}{1-q'}}}.
\end{eqnarray}
Por lo tanto, empleando las soluciones de Vlasov y la probabilidad $q$-exponencial, se propone una conexión formal entre ambos esquemas; más precisamente, se presentan soluciones estacionarias de Vlasov como distribuciones $q$-exponenciales adecuadas para ambos QSS.
\begin{figure}
    \centering
    \includegraphics[width=0.5\textwidth]{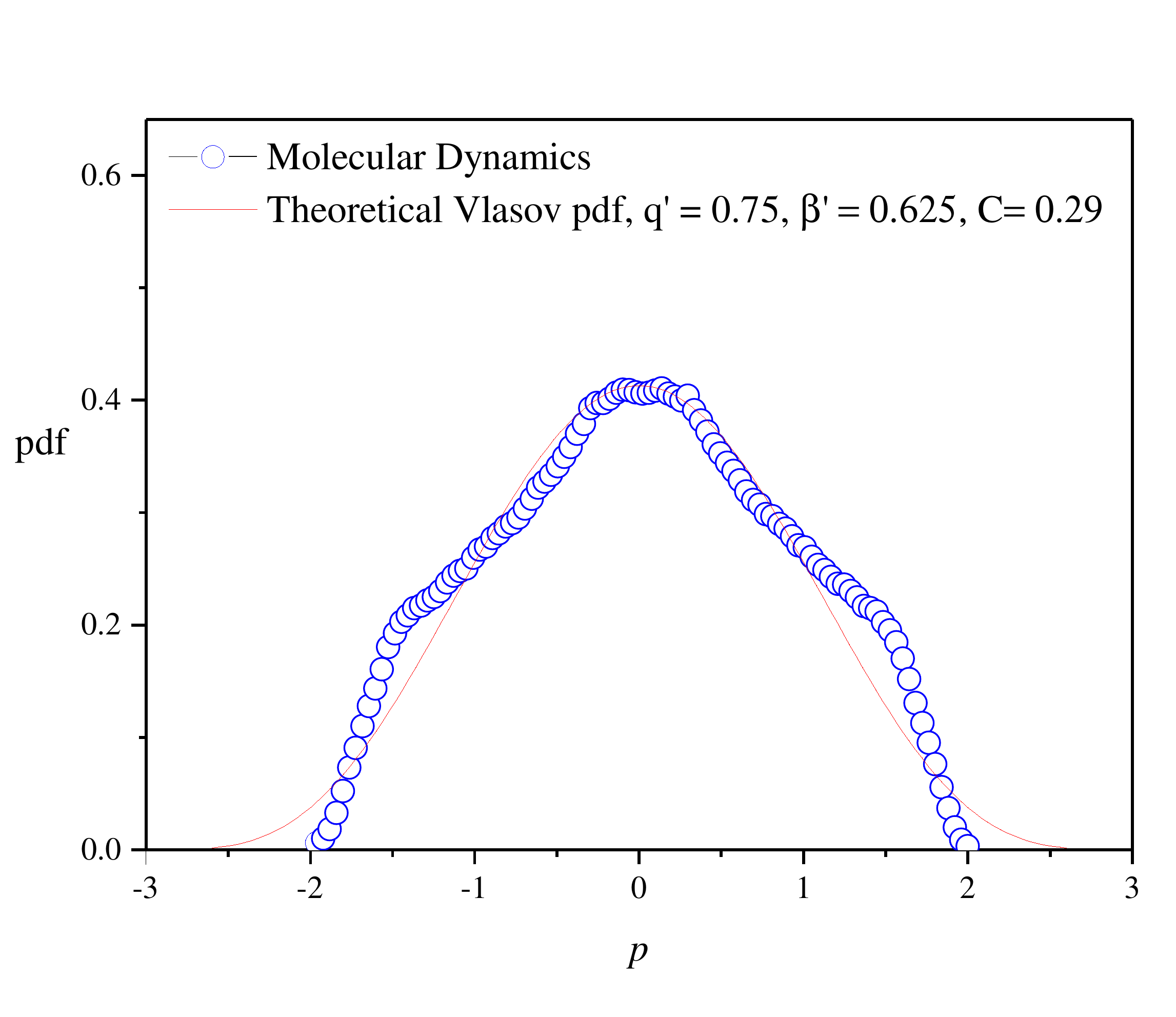}\includegraphics[width=0.49\textwidth]{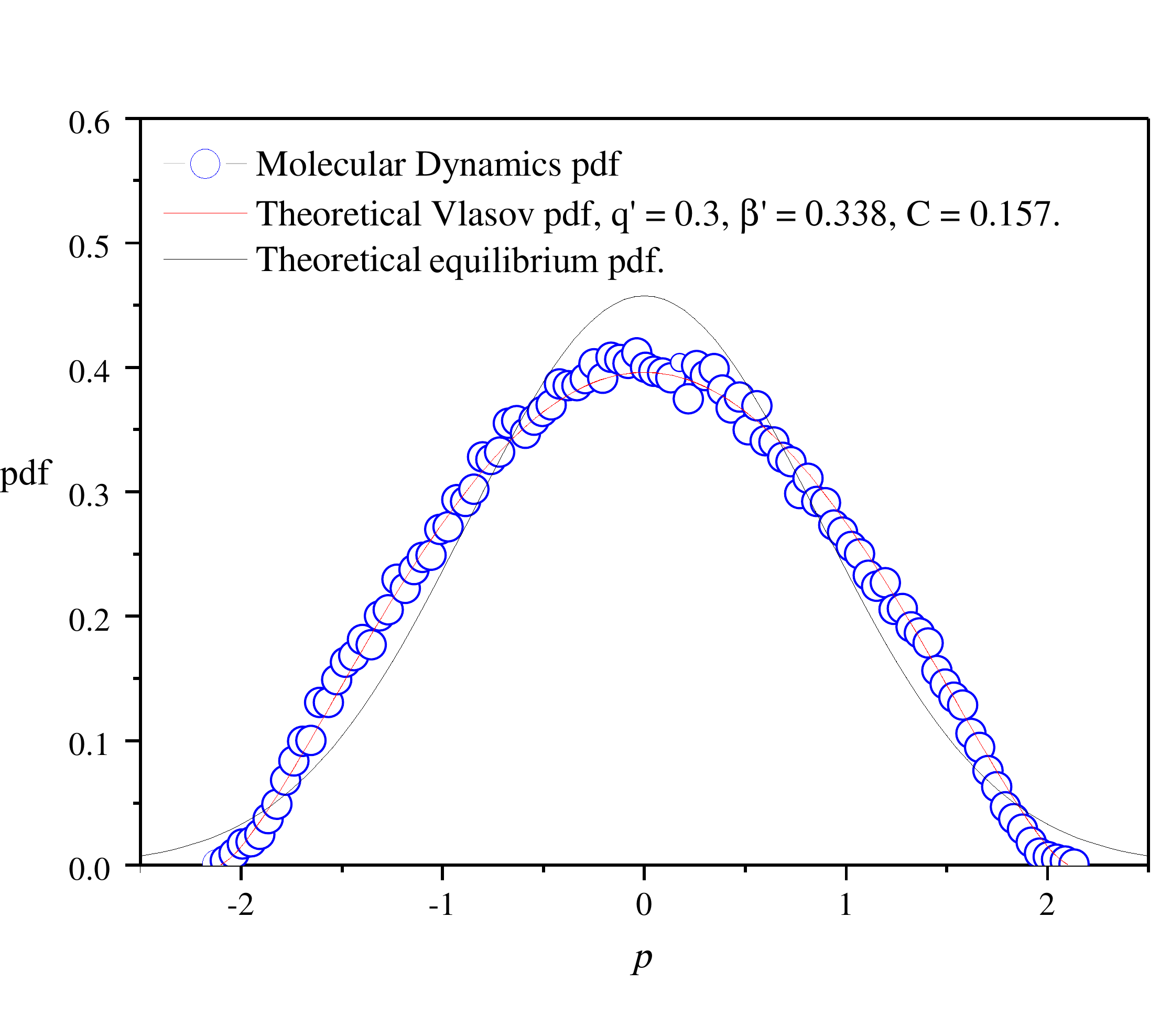}
   \\

   \caption{Comparación entre los resultados de dinámica molecular y la solución de Vlasov. En el panel izquierdo, el primer QSS y en la derecha el segundo QSS. La línea negra es un ajuste gaussiano, de color rojo, la solución de Vlasov, y en círculos azules los datos de dinámica molecular.}\label{DistributionsQSS}
\end{figure}
En la Fig. \ref{TsallisBeta}, se muestran los resultados obtenidos para los dos QSS. Se observa, cuando el valor de energía cinética promedio se acerca al valor del equilibrio, $q\rightarrow1$. Observamos en la Fig. \ref{DistributionsQSS} que la mejor descripción ocurre para el segundo QSS, mientras que para el primer QSS la solución de Vlasov si bien tiene los mismos valores de energía cinética y magnetización, la aproximación es óptima, pero no es buena. En la tabla 1 se resumen los valores de los parámetros tanto de Vlasov como Tsallis.

\begin{center}
Tabla 1.Resumen parámetros de Tsallis y Vlasov para los QSS y el Equilibrio.\\

\begin{tabular}{ccccccc}
   
  Estado & $q'$ &  $\beta'$ & $q$ & $\beta$ &  $\langle 2K/N\rangle$ & $m_x$ \\
     \hline
   1QSS & 0.750  & 0.625 & 0.800 & 9.091 & 0.800 & 0.150\\
      \hline
    2QSS & 0.300 & 0.338 &  0.588 & 9.524 &0.760 & 0.000 \\
       \hline
    Equilibrio &1.000 & 1.050 & 1.000 & 1.050 & 0.951 & 0.309 \\
       \hline
\end{tabular}
\end{center}

\section{Distribuciones analíticas en las orientaciones y en los momentos}

Los resultados mostrados en la sección anterior \ref{Tsallis-like} (Fig. \ref{DistributionsQSS}) corresponden a integraciones numéricas de la solución de Vlasov propuesta en la ec. (\ref{VlasovSolutionEq}). En esta sección revisaremos la integración analítica que permite obtener las distribuciones en las orientaciones y en los momentos.

Las distribuciones en las orientaciones y los momentos se obtienen a partir de la integración de la ec. (\ref{VlasovSolutionEq}), esto es,
\begin{eqnarray}
\rho(\theta)=\int_{-\infty}^{\infty}f(\theta,p)dp,\:\:\:\varrho(p)=\int_{-\pi}^{\pi}f(\theta,p)d\theta,
\end{eqnarray}
donde se obtiene $C$  de la condición
\begin{eqnarray}
\int_{-\infty}^{\infty}\int_{-\pi}^{\pi}f(\theta,p)d\theta dp=1. \label{normal}
\end{eqnarray}
Utilizaremos la transformación de Cahen-Mellin \cite{MELLIN,MELLIN2}, que para el caso de la función exponencial resulta ser la función gamma, dada por
\begin{eqnarray}
\Gamma(z)=\int_0^\infty x^{z-1} e^{-x}dx
\end{eqnarray}
que puede ser invertida mediante la transformación $x\rightarrow xy$, esto es
\begin{eqnarray}
x^{-z}=\frac{1}{\Gamma(z)}\int_0^\infty y^{z-1} e^{-xy}dy.
\end{eqnarray}\label{Mellin}
A partir del ansatz dado por la ecuación \ref{VlasovSolutionEq}, tenemos que $z=1/(q'-1)$ y
\begin{eqnarray}
x=(1-(1-q')\beta'(p^2+V(\theta))),
\end{eqnarray}
inmediatamente, podemos integrar la ecuación (\ref{Mellin}) en los momentos para obtener la distribución en las orientaciones
\begin{eqnarray}
\int_{-\infty}^{\infty}x^{-z}dp=\frac{C}{\Gamma(z)}\int_{-\infty}^{\infty}\int_0^{\infty} y^{z-1} \exp\left(-(1+(q'-1)\beta'(p^2/2+V(\theta)))y\right)dydp.\nonumber\\
\end{eqnarray}\label{Mellinqexp}
Invirtiendo el orden de integración al lado derecho de la ecuación \ref{Mellinqexp}, tenemos el siguiente resultado
\begin{eqnarray}
\rho(\theta)&=&\frac{C}{\Gamma(z)}\int_{0}^{\infty}y^{z-1}\int_{-\infty}^{\infty} \exp(-(1+(q'-1)\beta'(p^2/2+V(\theta)))y)dpdy, \nonumber\\
\end{eqnarray}
si $q'>1$, entonces
\begin{eqnarray}
\rho(\theta)&=&\frac{C}{\Gamma(z)}\int_{0}^{\infty}y^{z-1} \exp(-y(1+(q'-1)\beta' V(\theta))) \sqrt{\frac{2\pi}{(q'-1)\beta' y}}dy\nonumber\\
&=&\frac{C}{\Gamma(z)}\sqrt{\frac{2\pi}{(q'-1)\beta' }}\int_{0}^{\infty}y^{z-1/2-1} \exp(-y(1+(q'-1)\beta' V(\theta))) dy, \nonumber\\
\end{eqnarray}
haciendo el cambio de variables $u=y(1+(q'-1)\beta' V(\theta))$, obtenemos
\begin{eqnarray}
\rho(\theta)=\frac{C}{\Gamma(z)}\sqrt{\frac{2\pi}{(q'-1)\beta' }}\Gamma(z-1/2)\left(1+(q'-1)\beta' V(\theta)\right)^{-z+1/2}.\label{ec2108}
\end{eqnarray}
Esta distribución es especial ya que presenta un cambio importante de comportamiento cuando el exponente se vuelve a cero, esto es:
\begin{eqnarray}
-z+\frac{1}{2}=\frac{1}{1-q'}+\frac{1}{2}=0
\end{eqnarray}
\begin{eqnarray}
\frac{1}{1-q'}=-\frac{1}{2}
\end{eqnarray}

\begin{eqnarray}
2=q'-1
\end{eqnarray}

\begin{eqnarray}
q'=3
\end{eqnarray}

Al integrar nuevamente la ec.( \ref{ec2108}), podemos encontrar una expresión para la constante de normalización $C$
\begin{eqnarray}
\int_{-\pi}^{\pi}\rho(\theta)d\theta=1
\end{eqnarray}
En lo que viene, vamos a especificar el potencial del problema, el modelo d-HMF, donde
\begin{eqnarray}
V(\theta)=2-2m_x\cos\theta+m_y\sin\theta
\end{eqnarray}
Deseamos estudiar las soluciones de la ecuación de Vlasov para el caso $\varepsilon=1$.$38$, del que se obtiene $m_x=0$.$309$ y $m_y\approx0$.
\begin{eqnarray}
\int_{-\pi}^{\pi}\frac{C}{\Gamma(z)}\sqrt{\frac{2\pi}{(q'-1)\beta' }}\Gamma(z-1/2)\left(1+(q'-1)\beta' (2-2m_x\cos\theta)\right)^{-z+1/2}d\theta=1, \nonumber\\
\end{eqnarray}

\begin{eqnarray}
\frac{C}{\Gamma(z)}\sqrt{\frac{2\pi}{(q'-1)\beta' }}\Gamma(z-1/2)\int_{-\pi}^{\pi}\left(1+(q'-1)\beta' (2-2m_x\cos\theta)\right)^{-z+1/2}d\theta=1, \nonumber\\
\end{eqnarray}

\begin{eqnarray}
C=\frac{\Gamma(z)}{\Gamma(z-1/2)}\sqrt{\frac{(q'-1)\beta' }{2\pi}}\left/\int_{-\pi}^{\pi}\left(1+(q'-1)\beta' (2-2m_x\cos\theta)\right)^{-z+1/2}d\theta\right.\nonumber\\
\end{eqnarray}

A continuación, se muestran algunas distribuciones para diferentes valores de $q'$.

\begin{figure}[t]
\centering
 \includegraphics[scale=0.42]{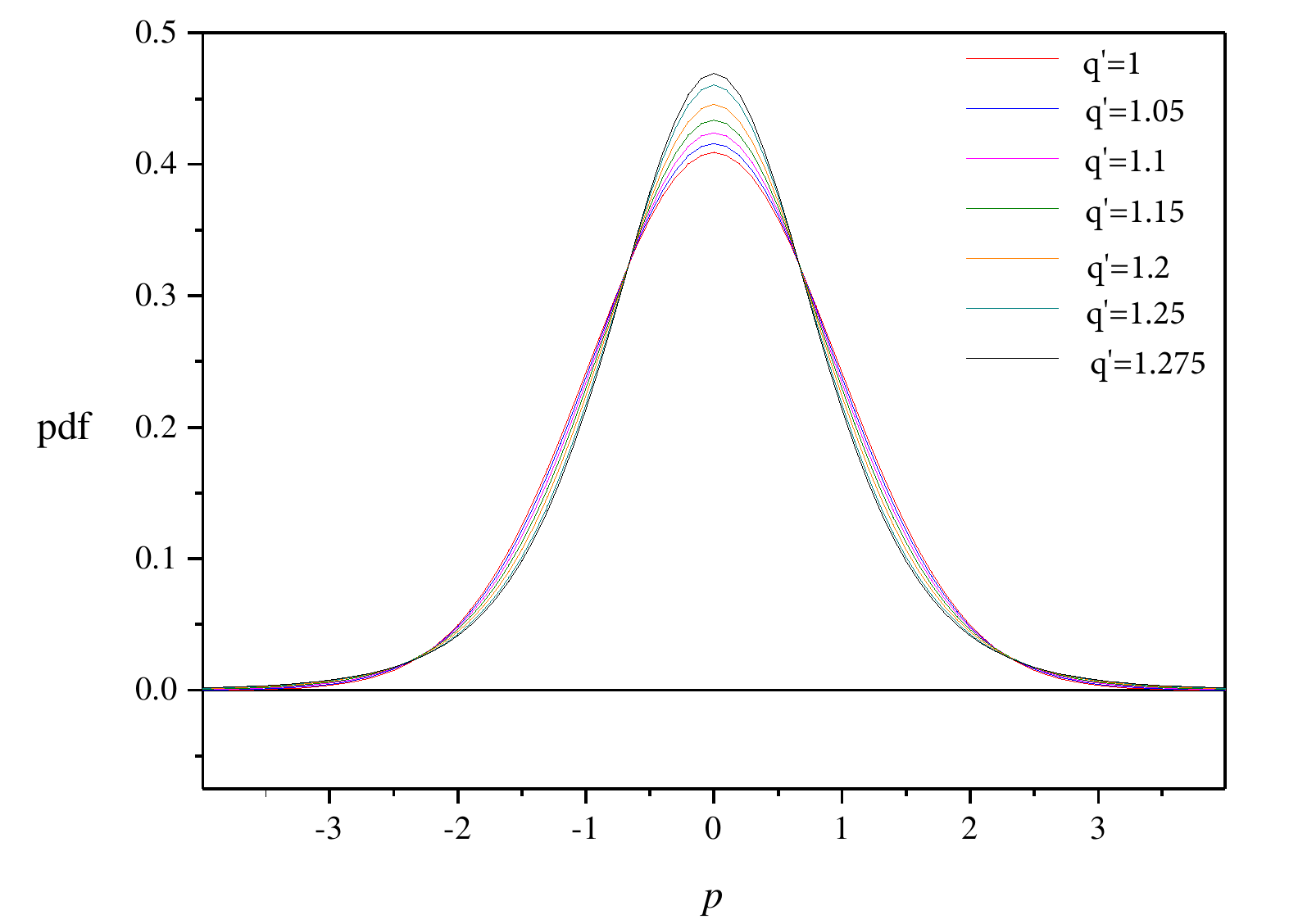} \includegraphics[scale=0.42]{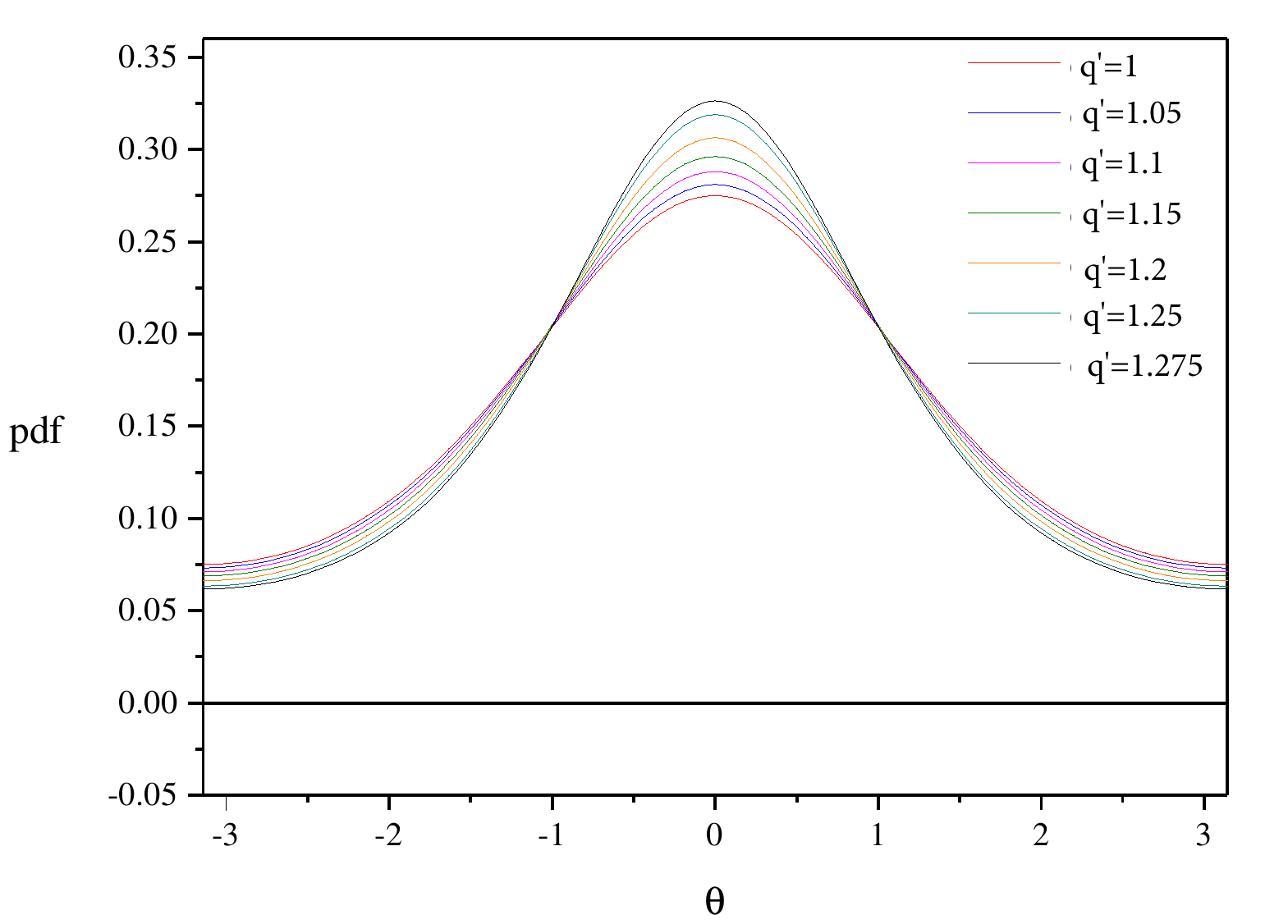}
      \caption{Distribuciones en las orientaciones $\rho(\theta)$ y en los momentos $\varrho(p)$, en función del parámetro $q'$.}\label{ComparaDistVlasov}
\end{figure}

\begin{figure}[t]
\centering
 \includegraphics[scale=0.5]{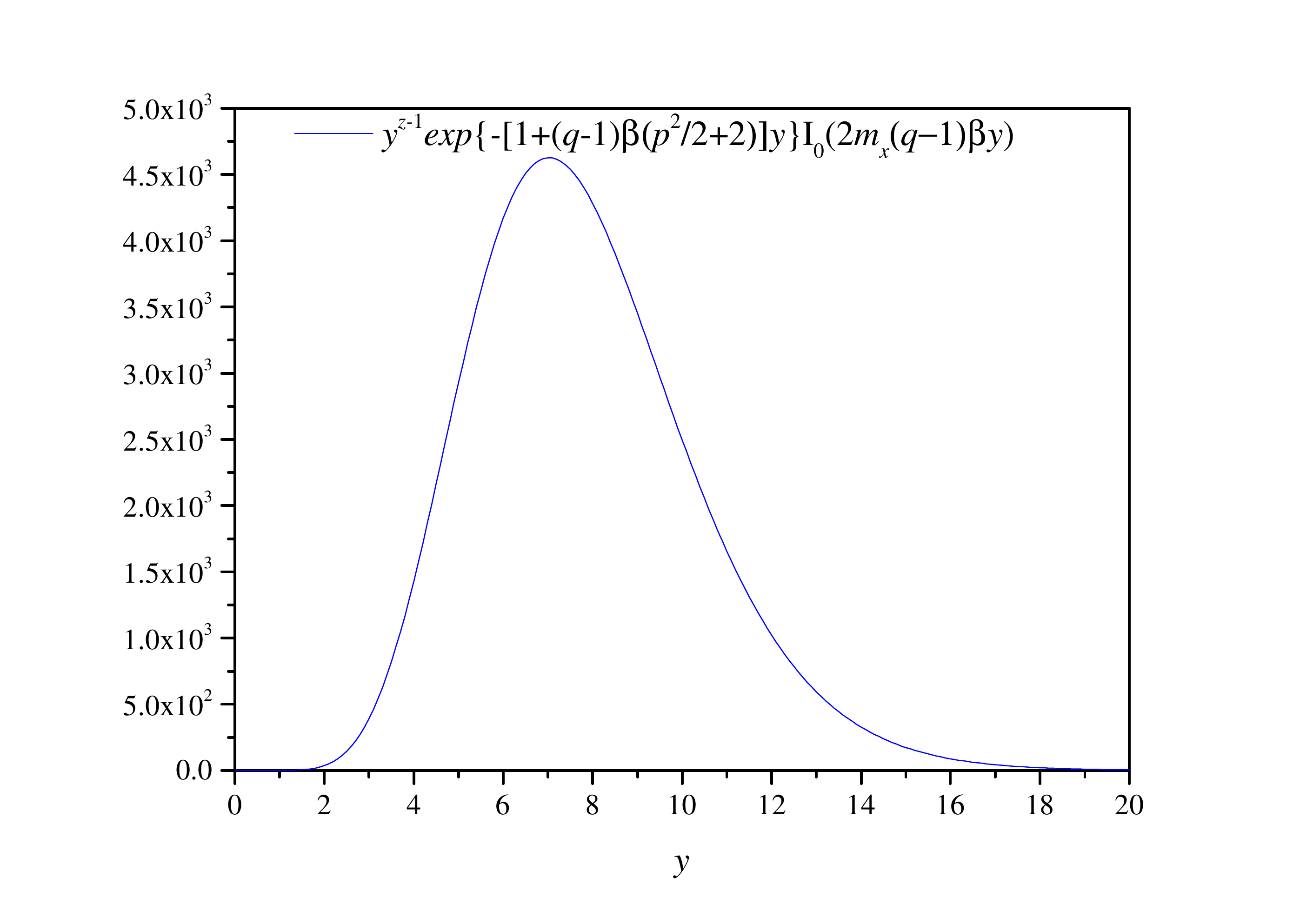}
      \caption{Integrando de la expresión \ref{varrho(p)} para $q'=1$.$1$, con parámetros $\beta'=1$.$5603$ y $m_x=0$.$309$ que garantizan $\varepsilon=1$.$38$.}\label{integrando}
\end{figure}

\begin{center}
Tabla 2. Algunos valores de $q'$ y $\beta'$, para los que se obtiene $\varepsilon=1$.$38$.
\begin{tabular}[t]{|l |c|c|c|c|c| |r|}
\hline
$q'$ & $\beta'$ & $\langle K \rangle$ & $\langle U \rangle$ & $\varepsilon$\\
\hline
1 & 1.0515 & 0.4755 & 0.9045 & 1.38 \\
\hline
1.050 & 1.2568 & 0.4788 & 0.9012 & 1.38\\
\hline
1.100 & 1.5603 & 0.4826 & 0.8975 & 1.38\\
\hline
1.150 & 2.0545 & 0.4869 & 0.8932 & 1.38\\
\hline
1.200 & 3.0001 & 0.4919 & 0.8882 & 1.38\\
\hline
1.250 & 5.5270 & 0.4977 & 0.8824 & 1.38\\
\hline
1.275 & 9.5100 & 0.5009 & 0.8791 & 1.38\\
\hline
\end{tabular}\label{table}\\
\end{center}
La tabla \ref{table} muestra la dependencia del parámetro $\beta'$ en función en $q'$ y la variación de la energía cinética y potencial a medida que $q'$ crece. El valor $q'=1$ es el caso gaussiano mostrado en la sección anterior. \\
Vamos ahora a obtener $\varrho(p)$,
\begin{eqnarray}
\varrho(p)=\int_{-\pi}^{\pi}f(\theta,p)d\theta,
\end{eqnarray}
utilizaremos la integral tipo Mellin ec.( \ref{Mellin}),
donde, $x=1+(q'-1)\beta'(p^2/2+V(\theta))$ y $z=1/(q'-1)$. De esta manera tenemos,

\begin{eqnarray}
\varrho(p)&=&\int_{-\pi}^{\pi}x^{-z}d\theta\nonumber\\
&=&\frac{C}{\Gamma(z)}\int_{-\pi}^{\pi}\int_{0}^{\infty}y^{z-1}\exp\left\{-\left[1+(q'-1)\beta'(p^2/2+2-2m_x\cos(\theta))\right]y\right\}dyd\theta\nonumber\\
\end{eqnarray}
separando tenemos,
\begin{eqnarray}
\varrho(p)=\frac{C}{\Gamma(z)}\int_{0}^{\infty}y^{z-1}\exp\left\{-\left[1+(q'-1)\beta'(p^2/2+2)\right]y\right\} \nonumber\\
\cdot\int_{-\pi}^{\pi}\exp\left\{2m_x(q'-1)\beta'\cos(\theta)y\right\}d\theta dy
\end{eqnarray}
luego,
\begin{eqnarray}
\varrho(p)=\frac{2\pi C}{\Gamma(z)}\int_{0}^{\infty}y^{z-1}\exp\left\{-\left[1+(q'-1)\beta'(p^2/2+2)\right]y\right\} I_0(2m_x(q'-1)\beta' y) dy. \nonumber\\\label{varrho(p)}
\end{eqnarray}
La última expresión no tiene primitiva conocida, excepto para ciertos valores de $p$, $q'$, $\beta'$ y $m_x$, que no son los que determinan el valor de energía $\varepsilon=1.38$. La convergencia de esta integral, depende del valor del parámetro $q'$. Si $q'<1$ la integral diverge. Haciendo un análisis de la convergencia, tenemos que la exponencial compite con la función modificada de Bessel; el producto $e^{-ax}I_0(bx)$ converge sólo si $a>b$.
\begin{eqnarray}
1+(q'-1)\beta'(p^2/2+2)>2m_x(q'-1)\beta' .
\end{eqnarray}
Para las condiciones que pretendemos modelar, se tiene que $m_x=0$.$309$ de lo que obtenemos,
\begin{eqnarray}
1+(q'-1)\beta'(p^2/2+2)>2\cdot0{.}309(q'-1)\beta' 
\end{eqnarray}
Claramente para el valor más bajo del lado izquierdo tendríamos que $p=0$. 
\begin{eqnarray}
1+2(q'-1)\beta'>2\cdot0{.}309(q'-1)\beta' 
\end{eqnarray}
con lo que tendríamos siempre convergencia, al menos en este producto. 
La función potencia $y^{z-1}$, también juega un rol importante en la convergencia de esta integral, la cual diverge si $z-1<-1$, lo que nos lleva al resultado $q'>1$ (condición necesaria para la convergencia). En la Fig. \ref{integrando}, se muestra un ejemplo de esta función.

Por otro lado podemos calcular la energía cinética promedio en función de $\theta$, esto es:
\begin{eqnarray}
\langle K\rangle (\theta)=\int_{-\infty}^{\infty}\frac{p^2}{2}f(\theta,p)dp,
\end{eqnarray}
\begin{eqnarray}
\langle K\rangle (\theta)=C\int_{-\infty}^{\infty}\frac{p^2}{2}\left[1+(q'-1)\beta'(p^2/2+V(\theta))\right]^{-\frac{1}{q'-1}}dp,
\end{eqnarray}
De la misma manera anterior, utilizamos la integral de Mellin, con los mismos parámetros, esto es,
\begin{eqnarray}
\langle K\rangle (\theta)=\frac{C}{2\Gamma(z)}\int_{-\infty}^{\infty}\int_{0}^{\infty}p^2 y^{z-1}e^{-\left[1+(q'-1)\beta'(p^2/2+V(\theta))\right]y}dydp,
\end{eqnarray}
\begin{eqnarray}
\langle K\rangle (\theta)=\frac{C}{2\Gamma(z)}\int_{0}^{\infty} y^{z-1}e^{-\left[1+(q'-1)\beta' V(\theta)\right]y}\int_{-\infty}^{\infty}p^2e^{-\left[\beta'(q'-1)p^2/2\right]y}dpdy,
\end{eqnarray}
La integral en los momentos es fácilmente integrable, pues,
\begin{eqnarray}
\int_{-\infty}^{\infty}x^2 e^{-ax^2}dx=\frac{\sqrt{\pi}}{2a^{3/2}}
\end{eqnarray}
\begin{eqnarray}
\langle K\rangle (\theta)=\frac{C}{2\Gamma(z)}\int_{0}^{\infty} y^{z-1}e^{-\left[1+(q'-1)\beta' V(\theta)\right]y}\frac{\sqrt{\pi}}{2\left(y(q'-1)\beta'/2\right)^{3/2}}dy,
\end{eqnarray}
\begin{eqnarray}
\langle K\rangle (\theta)=\frac{C\sqrt{\pi}}{4\Gamma(z)\left((q'-1)\beta'/2\right)^{3/2}}\int_{0}^{\infty} y^{z-5/2}e^{-\left[1+(q'-1)\beta' V(\theta)\right]y}dy,
\end{eqnarray}
Nuevamente esta integral, se resuelve fácilmente, 
\begin{eqnarray}
\int_{0}^{\infty}x^{z-5/2} e^{-bx}dx=b^{3/2-z}\:\Gamma(z-3/2)
\end{eqnarray}
finalmente,
\begin{eqnarray}
\langle K\rangle (\theta)=\frac{C\sqrt{\pi}\Gamma(z-3/2)\left[1+(q'-1)\beta' V(\theta)\right]^{3/2-z}}{4\Gamma(z)\left((q'-1)\beta'/2\right)^{3/2}}.
\end{eqnarray}
Ahora debemos obtener $\langle K\rangle$ integrando en $\theta$, numéricamente.\\

\begin{eqnarray}
\langle K\rangle=\frac{C\sqrt{\pi}\Gamma(z-3/2)}{4\Gamma(z)\left((q'-1)\beta'/2\right)^{3/2}}\int_{-\pi}^{\pi}\left[1+(q'-1)\beta'V(\theta)\right]^{3/2-z}d\theta,
\end{eqnarray}
Realizando la transformación de Mellin con $x=[1+(q'-1)\beta' V(\theta)]$ y $w=z-3/2$, tenemos,

\begin{eqnarray}
\langle K\rangle &=& \int_{-\pi}^{\pi}x^{-w}d\theta=\frac{C\sqrt{\pi}}{4\Gamma(z)\left((q'-1)\beta'/2\right)^{3/2}}\int_{-\pi}^{\pi}\int_{0}^{\infty}y^{w-1} e^{-\left[1+(q'-1)\beta' V(\theta)\right]y}dy d\theta\nonumber\\
 &=&\frac{C\sqrt{\pi}}{4\Gamma(z)\left((q'-1)\beta'/2\right)^{3/2}}\int_{-\pi}^{\pi}\int_{0}^{\infty}y^{w-1} e^{-\left[1+(q'-1)\beta' (2-2m_x\cos\theta)\right]y}dy d\theta\nonumber\\
 &=&\frac{C}{2\Gamma(z)}\left(\frac{2\pi}{(q'-1)\beta'}\right)^{3/2}\int_{0}^{\infty}y^{z-5/2} e^{-\left[1+2(q'-1)\beta' \right]y}I_0(2m_x\beta'(q-1) y)dy. \nonumber\\
\end{eqnarray}
Esta última integral nuevamente no tiene primitiva conocida para los valores de $q'$, $\beta'$ y $m_x$ compatibles con $\varepsilon=1$.$38$.

De la misma manera, podemos encontrar la expresión analítica para calcular la energía potencial promedio por partícula.

\begin{eqnarray}
\langle U\rangle (\theta)=\int_{-\infty}^{\infty}(2-2m_x \cos\theta)f(\theta,p)dp,
\end{eqnarray}

\begin{eqnarray}
\langle U\rangle (\theta)=C\int_{-\pi}^{\pi}(1-m_x \cos\theta)(1+(q'-1)\beta' (p^2/2+2-2m_x\cos\theta))^{-\frac{1}{q'-1}}dp,\nonumber\\
\end{eqnarray}
si integrando en las orientaciones, como la función $f(\theta,p)$ está normalizada tenemos,
\begin{eqnarray}
\langle U\rangle =1-Cm_x\int_{-\pi}^{\pi}\int_{-\infty}^{\infty} \cos\theta(1+(q'-1)\beta' (p^2/2+2-2m_x\cos\theta))^{-\frac{1}{q'-1}}dpd\theta,\nonumber\\
\end{eqnarray}
usando la transformación de Mellin,
\begin{eqnarray}
\langle U\rangle =1-\frac{Cm_x}{\Gamma(z)}\int_{0}^{\infty}\int_{-\pi}^{\pi}\int_{-\infty}^{\infty} \cos\theta y^{z-1}e^{-(1+(q'-1)\beta' (p^2/2+2-2m_x\cos\theta))y}dpd\theta dy,\nonumber\\
\end{eqnarray}

\begin{eqnarray}
\langle U\rangle =1-\frac{Cm_x}{\Gamma(z)}\int_{0}^{\infty}\int_{-\infty}^{\infty}y^{z-1}e^{-(1+(q'-1)\beta' (p^2/2+2))y}\int_{-\pi}^{\pi} \cos\theta e^{2\beta'(q'-1)m_x\cos\theta y}d\theta dpdy,\nonumber\\
\end{eqnarray}
la integral, en las orientaciones, se puede calcular y corresponde a una función de Bessel modificada de primera especie $I_1$, luego tenemos,
\begin{eqnarray}
\langle U\rangle =1-\frac{Cm_x}{\Gamma(z)}\int_{0}^{\infty}\int_{-\infty}^{\infty}y^{z-1}e^{-(1+(q'-1)\beta' (p^2/2+2))y}2\pi I_1(2\beta'(q'-1)m_xy) dpdy, \nonumber\\
\end{eqnarray}
la integral en los momentos también se puede calcular, es una gaussiana,
\begin{eqnarray}
\langle U\rangle =1-\frac{2\pi C m_x}{\Gamma(z)}\int_{0}^{\infty}y^{z-1}e^{-(1+2(q'-1)\beta' )y}\int_{-\infty}^{\infty}e^{-(q'-1)\beta' p^2/2y} dpI_1(2\beta'(q'-1)m_xy) dy,\nonumber\\
\end{eqnarray}

\begin{eqnarray}
\langle U\rangle =1-\frac{2\pi C m_x}{\Gamma(z)}\int_{0}^{\infty}y^{z-1}e^{-(1+2(q'-1)\beta' )y}\sqrt{\frac{2\pi}{(q'-1)\beta' y}}I_1(2\beta'(q'-1)m_xy) dy,\nonumber\\
\end{eqnarray}

\begin{eqnarray}
\langle U\rangle =1-\frac{(2\pi)^{3/2} C m_x}{\Gamma(z)\sqrt{(q'-1)\beta'}}\int_{0}^{\infty}y^{z-3/2}e^{-(1+2(q'-1)\beta' )y}I_1(2\beta'(q'-1)m_xy) dy,\nonumber\\
\end{eqnarray}

Nuevamente esta integral como hemos mencionado anteriormente no tiene primitiva conocida. Sin embargo, es posible obtener los valores de la tabla \ref{table} a partir de esta expresión realizando una evaluación numérica.

\chapter{Conclusiones}\label{conclu}

En esta tesis se investigó la dinámica y termodinámica del modelo d-HMF, describiendo los estados de equilibrio y los estados QSS presentes.

El problema se resolvió analíticamente en el conjunto canónico mediante los procedimientos estándar de la mecánica estadística calculando la función de partición y la posterior derivación de la energía libre de Helmholtz para finalmente obtener la temperatura del equilibrio y la magnetización. En el proceso se emplearon las transformaciones reales y complejas de Hubbard-Stratonovich. Asimismo, el problema se resolvió analíticamente en el ensamble microcanónico calculando directamente el número de microestados accesibles para obtener la entropía del sistema. Comparando ambos resultados se demostró la equivalencia de ensambles entre el canónico y microcanónico.

Desde la perspectiva de la teoría cinética, se encontró la función de distribución de equilibrio BG del modelo d-HMF dada por la ec.(\ref{eqdistrib}) y las distribuciones marginales en las orientaciones y los momentos dadas por las ecs.( \ref{distrib}) y \ref{distrib2}, las cuales se corresponden con los datos obtenidos por los métodos de la dinámica molecular, como se observa en la Fig. \ref{DistributionsBG-Eq}.

Mediante los métodos de la dinámica molecular, se encontró que si dejamos evolucionar el sistema desde condiciones iniciales uniformes (water-bag), el sistema queda atrapado en estados QSS descritos por diferentes valores de magnetización y energía cinética promedio y cuya duración crece conforme crece el número de partículas como se observa en las figuras \ref{tvst} $a$) y \ref{Fig2}. Los resultados fueron interpretados como la existencia de dos estados QSS diferentes, un primer estado QSS de energía cinética promedio mayor que el segundo QSS. Además se encontró una ley de potencia Ref. \cite{ATENAS3,ATENAS4} para el tiempo de duración del segundo QSS como se observa en la Fig. \ref{escaladim} $a$), lo que sugirió el estudio del sistema basándonos en la dinámica de Vlasov. Por otro lado, se encontró que el sistema es altamente sensible a las condiciones iniciales como se aprecia en la Fig. \ref{tvst} $b$), lo que sugiere futuros trabajos respecto a la caoticidad del modelo. 

La hipótesis central del problema consistió en la aceptación de la existencia de estados QSS presentes en la dinámica fuera del equilibrio que resultó en la observación de dos estados QSS en el modelo d-HMF y que éstos pueden ser descritos mediante la dinámica de Vlasov. Bajo este supuesto se llevaron a cabo simulaciones mediante los métodos de la dinámica molecular para obtener información adicional sobre las distribuciones en los momentos y las orientaciones de los estados QSS y posteriormente intentar describirlos mediante una función de distribución analítica. Esta hipótesis resultó ser acertada, pues mediante la dinámica de Vlasov, se encontraron (de forma analítica)  funciones de distribución del tipo q-exponencial para describir los estados QSS del modelo, siendo más adecuado el procedimiento para describir el segundo estado QSS y presentando limitaciones para describir el primer QSS como se aprecia en la Fig. \ref{DistributionsQSS}. Los resultados sugieren  valores de los parámetros $q'$ y $\beta'$ óptimos (mostrados en la tabla $1$) para la descripción de estos estados mediante la función de distribución de la ec.( \ref{VlasovSolutionEq}). A raíz de esta investigación se logró publicar recientemente un artículo en Physica A Ref. \cite{ATENAS6}.

Las soluciones encontradas del tipo q-exponencial sugieren la existencia de una relación estrecha con la estadística de Tsallis. Como se mostró en la sección \ref{Tsallis-like}, se encontró una transformación entre los parámetros $q$ y $\beta$ de Tsallis con los parámetros $q'$ y $\beta'$ propuestos en las soluciones estacionarias de Vlasov. Este resultado nos permite mirar la termodinámica y al menos sugerir una posible conexión entre la estadística de Tsallis y las soluciones de Vlasov.

Por último, se encontraron (de forma analítica) las distribuciones marginales $\rho(\theta)$ y $\varrho(p)$ para estados estacionarios fuera del equilibrio, mediante la transformación Cahen-Mellin, tomando como función la forma q-exponencial propuesta en la ec.( \ref{VlasovSolutionEq}). La solución exacta para la distribución marginal en las orientaciones $\rho(\theta)$ viene dada por la ec.( \ref{ec2108}), sin embargo, la distribución marginal en los momentos $\varrho(p)$ resultó no tener primitiva, por lo que sólo fue posible obtener una expresión integral dada por la ec.( \ref{varrho(p)}), cuyo cálculo debe realizarse numéricamente.

El d-HMF es un modelo interesante, porque está construido a partir de un hamiltoniano no simétrico, lo que representa un ejemplo de un modelo con una transición de fase que probablemente no provenga de una ruptura espontánea de la simetría, hecho se será investigado en futuros trabajos, así como también la verificación de la hipótesis de ergodicidad y la caoticidad del modelo. Adicionalmente un posible trabajo futuro sería la implementación de un método Monte Carlo para la descripción del equilibrio y también posibles aplicaciones del modelo al estudio de rotores moleculares y/o dispositivos de cortina eléctrica.

Por último esta tesis puede ser de interés para estudiantes que estén iniciando sus trabajos de investigación en este campo, ya que entrega una perspectiva general de las técnicas teóricas analíticas y computacionales que permiten describir la dinámica y la termodinámica de sistemas con interacciones de largo alcance.

\appendix
\chapter{Algoritmos computacionales}\label{app2}

\textbf{Solución analítica canónica}

Para comparar los cálculos de la función de partición completa, \textit{versus} la aproximación del SPM de la ecuación \ref{approxf}, esto es ${\cal F}_N(\beta \lambda)$ comparada con ${\cal F}(N,\beta \lambda)$, se utiliza el método del trapecio con $n=1000$ rectángulos.
\begin{eqnarray}
\int_a^b f(x)dx\approx \frac{b-a}{n}\left[\frac{f(a)+f(b)}{2}+\sum_{k=1}^{n-1}f\left(a+k\frac{b-a}{n}\right)\right]
\end{eqnarray}
Para la Fig. \ref{approx}, solo se calcula hasta $N=100$, ya que ${\cal F}(N,\beta \lambda)\sim 10^{4096}$ para $N>100$.

\textbf{Dinámica Molecular}

La dinámica molecular es una herramienta poderosa para resolver la dinámica de sistemas en los que se conocen las ecuaciones de movimiento de cada partícula. La desventaja es el tiempo de simulación porque la integración de las ecuaciones de movimiento se vuelve computacionalmente más demandante conforme crece el número de partículas. Existen diversos mecanismos para integrar las ecuaciones de movimiento, ejemplos de ellos son el algoritmo Método de Euler, Runge-Kutta, Verlet, coeficientes simplécticos, etc.

\begin{figure}[!h]
\centering
 \includegraphics[scale=1.0]{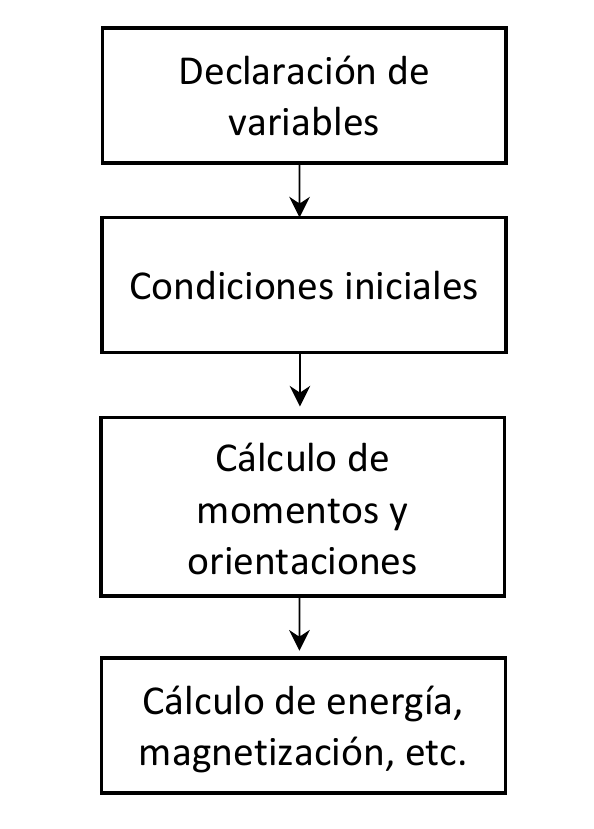}
      \caption{Estructura del código del programa de dinámica molecular. En la etapa de condiciones iniciales, se generan números aleatorios que siguen una distribución uniforme (water-bag initial conditions). La parte central del código está en el cálculo de los momentos y las orientaciones, aquí se encuentra la rutina simpléctica para el cálculo de integración de las ecuaciones de movimiento. En esta etapa es donde se almacenan los datos de interés en los archivos.}\label{codemap}
\end{figure}

Para esta tesis se utilizó un algoritmo simpléctico de cuarto orden \cite{RUTH} para integrar las ecuaciones de movimiento. Además se implementó una paralelización MPI del código en FORTRAN90. En la Fig. \ref{codemap} se ilustra la estructura general del código.

\textbf{Soluciones estacionarias de la ecuación de Vlasov}

Para describir las soluciones más cercanas a las distribuciones encontradas por los métodos de la dinámica molecular, se realiza un proceso de optimización que permite obtener los parámetros $q'$ y $\beta'$. Este procedimiento considera los valores de energía cinética promedio de los estados QSS hallados, es decir $\langle 2K/N\rangle=0$.$800$ para el $1$QSS y $\langle 2K/N\rangle=0$.$760$ para el $2$QSS. Se implementó con el software mathematica, dejando libre los parámetros de magnetización $M_x$, $q'$ y $\beta'$. Con este procedimiento se logran determinar los tres estados que se aprecian en la tabla $1$, es decir los dos estados QSS y el equilibrio. Con menor probabilidad se encuentran los estados con la magnetización  $M_x=0$.$15$ ($1$QSS), luego, con mayor probabilidad se encuentran los estados con valores de magnetización $M_x=0$ ($2$QSS) y $M_x=0$.$309$ (equilibrio),  en este último cabe resaltar que se obtienen los valores esperados de $q'=1$ y $\beta'=\beta_{eq}$.

\chapter{Función de distribución uniforme (Water-Bag)}\label{ApendiceB}

Ahora, veamos como obtener la función de distribución en los momentos y en las orientaciones para una distribución uniforme (water-bag), y que corresponde a un valor constante de la distribución en un cierto rango de los momentos y de las orientaciones, esto es $\theta<|\theta_0|$ y $p<|p_0|$. Al normalizar se tiene,
\begin{eqnarray}
f(\theta,p)=\frac{1}{4\theta_0 p_0}
\end{eqnarray}
Entonces, la distribución en las orientaciones es:
\begin{eqnarray}
\rho(\theta)=\int_{-p_0}^{p_0}
f(\theta,p) dp=\frac{1}{2\theta_0}
\end{eqnarray}
Mientras que la distribución en los momentos es:
\begin{eqnarray}
\varrho(p)=\int_{-\theta_0}^{\theta_0}
f(\theta,p) dp=\frac{1}{2p_0}
\end{eqnarray}
Recordemos que para la distribución uniforme $f(\theta,p)=1/4\theta_0 p_0$, se tiene que la energía total es:
\begin{eqnarray}
\varepsilon=\int_{-p_0}^{p_0}\int_{-\theta_0}^{\theta_0}\frac{p^2}{2}f(\theta,p)d\theta dp+\frac{1}{2}\int_{-p_0}^{p_0}\int_{-\theta_0}^{\theta_0}(2-2m_x \cos\theta + m_y\sin\theta)f(\theta,p)d\theta dp, \nonumber\\
\end{eqnarray}
donde 
\begin{eqnarray}
m_x=\int_{-p_0}^{p_0}\int_{-\theta_0}^{\theta_0}f(\theta,p)\cos\theta d\theta dp,\nonumber\\
m_y=\int_{-p_0}^{p_0}\int_{-\theta_0}^{\theta_0}f(\theta,p)\sin\theta d\theta dp,
\end{eqnarray}
donde $m_x=\sin\theta_0/\theta_0$ y $m_y=0$, la energía total es,
\begin{eqnarray}
\varepsilon=\frac{p_0^2}{6}+1-\frac{\sin^2\theta_0}{\theta^2_0}=\frac{p_0^2}{6}+1-m^2.
\end{eqnarray}
Para el caso homogéneo $\theta_0=\pi$, $m=0$, se obtiene que
\begin{eqnarray}
p_0=\sqrt{6 (\varepsilon-1)}\approx1{.}51.
\end{eqnarray}
Consecuentemente, la energía cinética promedio del water-bag homogéneo es:

\begin{eqnarray}
\langle K \rangle=\int_{-p_0}^{p_0}
\int_{-\theta_0}^{\theta_0}\frac{p^2}{2}f(\theta,p) d\theta dp=\frac{p_0^2}{6},
\end{eqnarray}

\begin{eqnarray}
\langle K \rangle=0{.}38,
\end{eqnarray}
así, las distribuciones en $\theta$ y $p$ son,
\begin{eqnarray}
\rho(\theta)=\frac{1}{2\theta_0}=0{.}159155,
\end{eqnarray}
\begin{eqnarray}
\varrho(p)=\frac{1}{2p_0}=0{.}331133.
\end{eqnarray}

Para un water-bag inhomogéneo, por ejemplo $\theta_0=0$.$01$, la magnetización por partícula es alta (cercana a la unidad), por lo que casi toda la energía del sistema es energía cinética,
\begin{eqnarray}
p_0=\sqrt{6\left(e-1+\frac{\sin^2 \theta_0}{\theta_0^2}\right)}\approx2{.}877464162,
\end{eqnarray}
\begin{eqnarray}
\langle K \rangle=\frac{p_0^2}{6}=1{.}379966667,
\end{eqnarray}

\begin{eqnarray}
f(\theta,p)=\frac{1}{4\theta_0 p_0}=\frac{1}{4\cdot0{.}01\cdot2{.}2978}=8{.}688205514,
\end{eqnarray}

Para un water-bag $\theta_0=\pi/2$ tenemos,
\begin{eqnarray}
p_0=\sqrt{6\left(e-1+\frac{\sin^2 \theta_0}{\theta_0^2}\right)}\approx2{.}170647002,
\end{eqnarray}

\begin{eqnarray}
\langle K \rangle=\frac{p_0^2}{6}=0{.}7852847346.
\end{eqnarray}

Para un water-bag $\theta_0=\pi/4$ tenemos,
\begin{eqnarray}
p_0=\sqrt{6\left(e-1+\frac{\sin^2 \theta_0}{\theta_0}\right)}\approx 2{.}672717122,
\end{eqnarray}

\begin{eqnarray}
\langle K \rangle=\frac{p_0^2}{6}=1{.}190569469.
\end{eqnarray}

Para un water-bag $\theta_0=\pi/3$ tenemos,
\begin{eqnarray}
p_0=\sqrt{6\left(e-1+\frac{1\sin^2 \theta_0}{\theta_0}\right)}\approx2{.}6603843313
\end{eqnarray}

\begin{eqnarray}
\langle K \rangle=\frac{p_0^2}{6}=1{.}13
\end{eqnarray}  

Los valores obtenidos de $p_0$ para los diferentes tipos de condiciones iniciales, fueron utilizados para fijar las condiciones iniciales en el programa de dinámica molecular. Con esto se realizó el estudio mostrado en la Fig. \ref{Fig2} $b$).

\end{document}